\documentclass[%
 reprint,
 amsmath,amssymb,
 aps,
prb,
]{revtex4-2}

\usepackage{graphicx}
\usepackage{array}%
\usepackage{dcolumn}%
\usepackage{bm}%
\usepackage{multirow}
\usepackage{color}
\usepackage{textcomp}
\usepackage{url}
\usepackage{tikz}
\usetikzlibrary{positioning,arrows.meta,fit,backgrounds,shapes.geometric}

\usepackage{algpseudocode}
\usepackage{framed}

\def\br{{\bm r}}
\def\hbr{\hat{\bm r}}

\def\bt{{\bm t}}

\def\vii{{\bm v}}

\def\ace{\varphi}

\def\br{{\bm r}}

\def\Nat{{N_{\text{at}}}}

\newcommand {\braket}[2]{ \langle #1 | #2  \rangle }
\newcommand {\braHket}[3]{ \langle #1 | #2 | #3 \rangle }

\begin{document}

\title{Efficient Machine Learning Interatomic Potentials}

\author{Yury Lysogorskiy}
\affiliation{%
 ICAMS, Ruhr-Universit\"at Bochum, Bochum, Germany
}%
\author{Anton Bochkarev}
\affiliation{%
 ICAMS, Ruhr-Universit\"at Bochum, Bochum, Germany
}%
\author{Ralf Drautz}
\email{ralf.drautz@rub.de}
\affiliation{%
 ICAMS, Ruhr-Universit\"at Bochum, Bochum, Germany
}%

\date{\today}%

\begin{abstract}
Machine learning interatomic potentials (MLIPs) reach an accuracy close to quantum mechanical reference calculations at a small fraction of the cost. Efficiency is the purpose of MLIPs, yet strategies for efficient MLIPs have not been reviewed systematically. This review takes efficiency as its guideline. To compare MLIPs on a common footing, we represent them in a complete basis expansion, the Atomic Cluster Expansion (ACE) and its graph generalization (GRACE). The expansion is canonical, built from products of single-particle functions. Classical interatomic potentials, such as the embedded-atom method, and many modern MLIPs, including equivariant transformers, can be represented within this expansion. Much of the apparent diversity of MLIP architectures therefore reflects how non-linearity, layer coupling and optimization are handled rather than which interactions the models can represent. We show how the cost of an MLIP becomes linear in system size and, at fixed rank of the tensor decomposition, in model complexity, and we discuss the role of machine learning, which provides non-linear parameterizations and optimization. Because efficiency is meaningful only at a given accuracy, we also compare accuracy and cost. On the Matbench Discovery leaderboard, models with three to five layers approach the accuracy of deeper models, and one basis, truncated along the Pareto front of error and speed, spans more than two orders of magnitude in cost. Ultimately, training data determine the quality of any MLIP.
\end{abstract}

\maketitle

\section{Introduction}

Machine learning interatomic potentials (MLIPs) enable simulations with accuracy close to quantum mechanical reference calculations at a computational expense that is several orders of magnitude smaller. In the past two decades many different approaches were demonstrated~\cite{manzhosUsingNeuralNetworks2006,behlerGeneralizedNeuralNetworkRepresentation2007,braamsPermutationallyInvariantPotential2009, bartokGaussianApproximationPotentials2010,thompsonSpectralNeighborAnalysis2015,shapeevMomentTensorPotentials2016,chmielaMachineLearningAccurate2017,smithANI1ExtensibleNeural2017,schuttSchNetDeepLearning2018,zhangDeepPotentialMolecular2018,faberAlchemicalStructuralDistribution2018,thomasTensorFieldNetworks2018,lubbersHierarchicalModelingMolecular2018,drautzAtomicClusterExpansion2019,NEURIPS2019_03573b32,unkePhysNetNeuralNetwork2019a,gasteigerDirectionalMessagePassing2020,satorrasEquivariantGraphNeural2021,gasteigerGemNetUniversalDirectional2021,schuttEquivariantMessagePassing2021, bochkarevMultilayerAtomicCluster2022,nigamUnifiedTheoryAtomcentered2022,chenUniversalGraphDeep2022,haghighatlariNewtonNetNewtonianMessage2022,batznerE3equivariantGraphNeural2022,dussonAtomicClusterExpansion2022, NEURIPS2022_4a36c3c5,musaelianLearningLocalEquivariant2023,dengCHGNetPretrainedUniversal2023,liaoEquiformerV2ImprovedEquivariant2023,pozdnyakovSmoothExactRotational2023,chengCartesianAtomicCluster2024,bochkarevGraphAtomicCluster2024,batatiaDesignSpaceE3equivariant2025,mazitovPETMADLightweightUniversal2025a}, making it hard to keep up with this very fast moving field. An increasing number of reviews of machine learning interatomic potentials are becoming available~\cite{jacobsPracticalGuideMachine2025,deringerMachineLearningInteratomic2019,eyertMachinelearnedInteratomicPotentials2023,unkeMachineLearningForce2021,behlerFourGenerationsHighdimensional2021,musilPhysicsInspiredStructural2021,kocerNeuralNetworkPotentials2022,thiemannIntroductionMachineLearning2024}. Efficiency is the purpose of surrogate models, yet strategies for efficient MLIPs have not been reviewed systematically.

MLIPs differ fundamentally from machine learning in other domains. While typical machine learning establishes models through the stochastic analysis of data for which the underlying laws are unknown or non-existent, MLIPs act as surrogate models for the fully known, yet computationally expensive, many-electron Schrödinger equation or its density functional theory (DFT) approximations.

Because MLIPs are rooted in quantum mechanics, physical and chemical intuition can inform their design. Therefore, many MLIPs build on physical, chemical and mathematical considerations and employ machine learning technology mainly for implementation and training. In contrast, data-driven MLIPs adopt architectures that were developed in other areas of machine learning. The two approaches are, however, less different than they appear. In fact, some of the machine learning technology that has been introduced recently is closely related to established, physically derived models of atomic interactions. For example, the transformer attention mechanism~\cite{vaswaniAttentionAllYou2017} applied to atomic interactions is closely analogous to the realization that atomic interactions are non-linear functions of electron and atomic density, which started the systematic development and derivation of many-body interatomic potentials~\cite{guptaLatticeRelaxationMetal1981,norskovEffectivemediumTheoryChemical1980,dawEmbeddedatomMethodDerivation1984,finnisSimpleEmpiricalBody1984, tersoffNewEmpiricalModel1986}. This may also help to explain why machine learning driven and physics and chemistry derived MLIPs often achieve comparable accuracy.

This review is concerned with the efficiency of MLIPs. Efficiency has several aspects, among them computational cost, accuracy, system-size scaling, data efficiency, robustness and physical interpretability. As these aspects cannot all be evaluated unambiguously and their weight often depends on the application, we focus on the mathematical and architectural foundations essential for designing efficient MLIPs.

As we will show, the requirements of efficiency leave only a narrow design space. Naive representations hit exponential walls in the number of atoms, neighbors, basis functions, body order, layers and chemical species, and only few paths lead around them. Many developments, made independently by different groups, therefore follow from necessity rather than choice, and this is why most efficient MLIPs can be understood from a single model strategy.

This strategy builds on a complete basis expansion, which can in principle represent any interaction within its range, Sec.~\ref{sec:complete}. Once atomic interactions are expressed in a complete basis, machine learning takes on two well-defined roles. It supplies efficient non-linear functions that let the basis expansion converge with fewer terms, and it provides optimizers capable of training the millions of parameters of foundation models.

We structure our review of MLIP efficiency as follows:

{\it {\bf Navigation} --} We present a short overview of the relationships between MLIPs (Sec.~\ref{sec:overview}), identifying the architectural features shared by most of them, before turning to details.

{\it {\bf Completeness} --} A complete representation (Sec.~\ref{sec:complete}) can in principle be converged to arbitrary accuracy and can represent any MLIP with the same interaction range, exactly or as a limit. Achieved in a canonical way, it lets us understand most MLIPs from a few basic considerations and makes efficiency arguments transferable across architectures. Magnetism and charge transfer are treated on the same footing as the atomic positions, Sec.~\ref{sec:extendability}.

{\it {\bf Scaling} --} The cost of naive representations grows steeply with system size and model complexity. Sec.~\ref{sec:scaling} shows how to reduce it to linear or constant scaling. Linear scaling with model complexity is essential for efficiency and a cornerstone of modern MLIP design.

{\it {\bf Symmetry} --} Atomic interactions are invariant under translation, rotation, and permutation of identical atoms (Sec.~\ref{sec:symmetry}). Although symmetry can be built into the expansion exactly, some approaches instead learn it from data, and others favor Cartesian over spherical coordinates.

{\it {\bf Physical and chemical foundation} --} As MLIPs are surrogates for known quantum-mechanical laws, they can be understood from the physics and chemistry of bonding. We show how the graph cluster basis functions emerge from a quantum-mechanical description of interacting atoms (Sec.~\ref{sec:chem}).

{\it {\bf Further numerical aspects} --} More technical, application-dependent aspects, force learning versus exact gradients, numerical precision, memory access and platform-specific implementation, are discussed in Sec.~\ref{sec:numerics}.

 {\it {\bf Machine learning in MLIPs} --} Section~\ref{sec:MLinMLIPs} addresses the role of machine learning, which we argue is of limited relevance for the representation of interactions but critical for their efficient non-linear parameterization and training.

 {\it {\bf Examples} --} In Sec.~\ref{sec:Relations} we briefly discuss selected MLIPs in view of a complete basis expansion, together with the Matbench Discovery leaderboard and accuracy versus computational cost.

{\it {\bf Summary and outlook} --} We conclude in Sec.~\ref{sec:summary}.

Readers who look for an overview of MLIPs rather than for the construction of the expansions may focus on Sec.~\ref{sec:overview}, the summary boxes at the beginning of the sections and the discussion of individual models in Sec.~\ref{sec:Relations}. Sections \ref{sec:complete}--\ref{sec:symmetry} develop the basis expansion, its scaling and its symmetry in detail and address readers who want to understand MLIPs in depth or build their own.

Several aspects that are essential to the practical success of MLIPs are not covered in this review. Much of the recent progress in the field has been driven by the growing availability of large community datasets~\cite{jainCommentaryMaterialsProject2013,ramakrishnanQuantumChemistryStructures2014,chanussotOpenCatalyst20202021,schreinerTransition1xDatasetBuilding2022,tranOpenCatalyst20222023,eastmanSPICEDatasetDruglike2023,schmidtMachineLearningAssistedDetermination2023,dengCHGNetPretrainedUniversal2023,sriramOpenDAC20232024,ganschaQCMLDatasetQuantum2025,kunerMPALOER2SCANDataset2025,kaplanFoundationalPotentialEnergy2025,mazitovPETMADLightweightUniversal2025a,levineOpenMolecules2025,barroso-luqueOpenMaterials20242024,gharakhanyanOpenMolecularCrystals2026}, together with systematic strategies for generating training data, such as SMAX~\cite{bochkarevExploringExtremesAtomic2026} and ASSYST~\cite{poulAutomatedGenerationStructure2025}. We do not attempt to review these developments. Nor do we address how to ensure completeness of configurational sampling and active learning, or how MLIPs are deployed across the many simulation tasks they enable. Models for charge transfer and magnetism are likewise not reviewed in detail.

\section{Navigating Machine Learning Interatomic Potentials} \label{sec:overview}

\begin{framed}
{\it{Summary:}} Most MLIPs can be viewed from a single complete basis expansion. Local MLIPs build on star graphs (ACE), semi-local message-passing MLIPs on tree graphs (GRACE). Models differ mainly in how they terminate the expansion with non-linear functions, how they couple layers and how they are optimized. The requirements of efficiency leave only a narrow design space, which is why a single strategy covers most models.
\end{framed}

Our starting point for the overview is the Atomic Cluster Expansion (ACE)~\cite{drautzAtomicClusterExpansion2019, drautzAtomicClusterExpansion2020,dussonAtomicClusterExpansion2022}. Because its basis is complete, ACE can represent classical potentials and many local MLIPs. Its generalization to graph cluster basis functions, the Graph Atomic Cluster Expansion (GRACE)~\cite{bochkarevGraphAtomicCluster2024}, can in addition represent semi-local message-passing graph neural networks. Here local means that an atom interacts directly only with atoms inside its cutoff sphere. Semi-local MLIPs relay interactions further through intermediate atoms, as in message passing, so that $L$ layers extend the interaction range to the atoms connected to the central atom by chains of at most $L$ bonds, each shorter than the cutoff $r_c$, i.e. to at most $L$ times the cutoff. The relay through intermediate atoms comes at the price of increased communication in large, parallel molecular dynamics simulations~\cite{parkScalableParallelAlgorithm2024}. ACE and GRACE together give a unified view of most MLIPs. In this view we deliberately ignore model-specific details, in order to isolate the architectural ingredients that are necessary for MLIP efficiency.

Throughout, when we state that a given MLIP can be obtained from, or is contained in, ACE or GRACE, we mean this in the sense of representability, i.e. its functional form lies within the complete basis, exactly for polynomial forms and as a limit for general non-linear ingredients, Sec.~\ref{sec:Relations}. The limit is convergence in the norm defined by the orthonormality of the single-bond and single-atom basis functions, Sec.~\ref{sec:completebasis}, and holds for any number of neighbors within the cutoff. Rigorous convergence results under additional assumptions are given in Refs.~\cite{dussonAtomicClusterExpansion2022,bachmayrPolynomialApproximationSymmetric2024}. The statement concerns the space of models and says nothing about the historical development of a particular method, as many methods were proposed independently. The organization also does not depend on ACE and GRACE as particular methods. Because complete bases are related by linear transformations, any complete basis organizes the MLIPs in the same way, Sec.~\ref{sec:completebasis}.

In practice only few other complete constructions exist, and all of them are limited to star graphs. The Moment Tensor Potential (MTP)~\cite{shapeevMomentTensorPotentials2016}, permutation-invariant polynomials~\cite{braamsPermutationallyInvariantPotential2009} in their atomic body-ordered form~\cite{vanderoordRegularisedAtomicBodyordered2020} and the NICE hierarchy~\cite{nigamRecursiveEvaluationIterative2020} are related to ACE by exactly such basis transformations~\cite{dussonAtomicClusterExpansion2022,nigamCompletenessAtomicStructure2024}. Complete global representations, such as the inverse pairwise distances used in GDML~\cite{chmielaMachineLearningAccurate2017,unkeMachineLearningForce2021}, do not decompose into atomic contributions and are limited to small systems. For point clouds, universality, i.e. the ability to approximate any invariant or equivariant function of the atomic positions, was proven for equivariant and for scalar-based networks~\cite{dymUniversalityRotationEquivariant2021,villarScalarsAreUniversal2021}, and atom-centered and message-passing representations were related in Refs.~\cite{nigamUnifiedTheoryAtomcentered2022,batatiaDesignSpaceE3equivariant2025}, but we are not aware of an alternative explicit complete basis beyond star graphs. What distinguishes ACE and GRACE is an explicit basis, which GRACE extends to general graphs, and which can be truncated systematically in body order, angular resolution and number of layers.

\begin{figure}[ht!]
\includegraphics[width=\columnwidth]{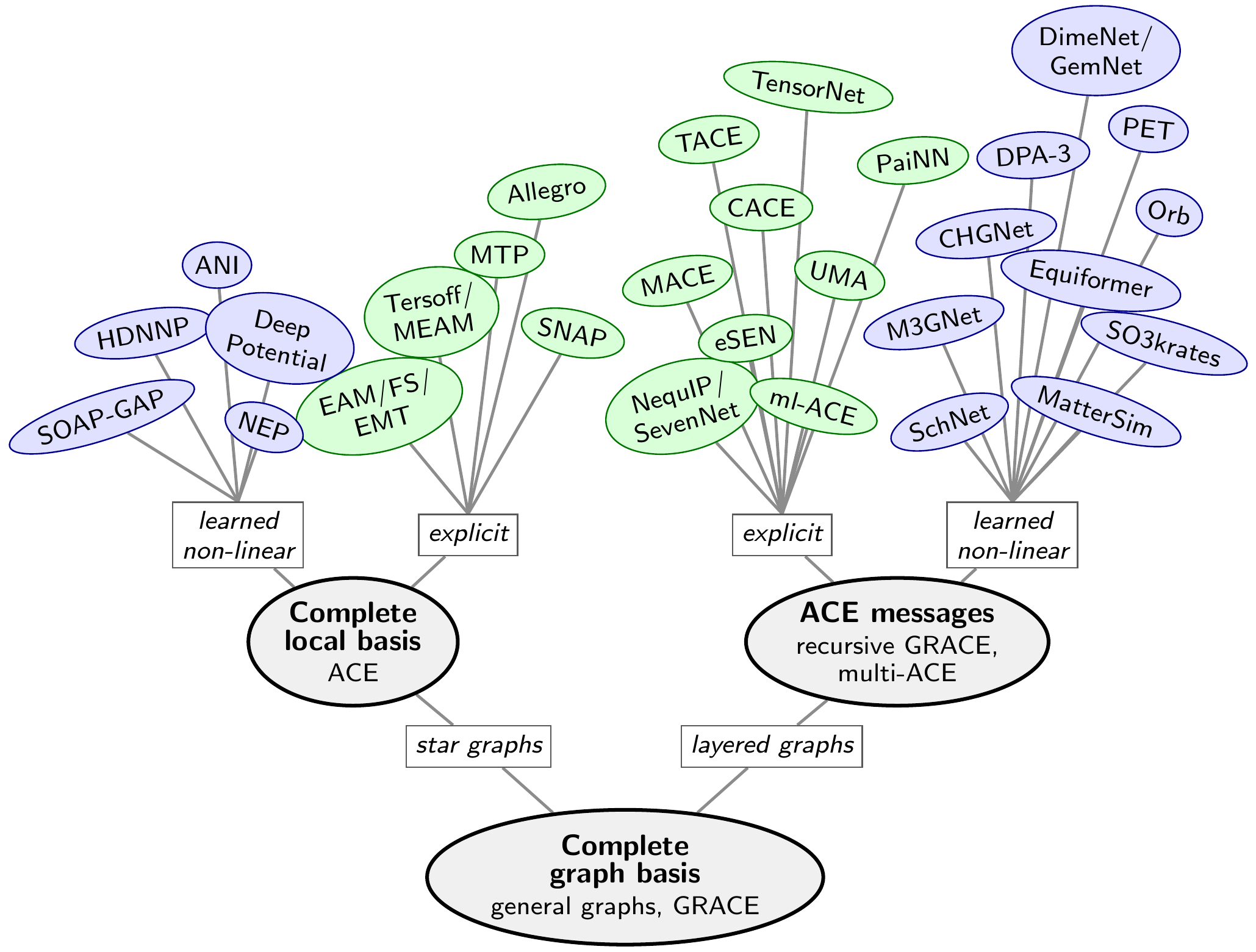}
        \caption{MLIPs viewed from the perspective of a complete basis expansion. Restricted to star graphs, the complete graph basis (GRACE) gives the complete local basis (ACE). Layered graphs, evaluated recursively, give ACE messages (multi-ACE). Green marks models that generate the many-body order by explicit products, blue models that generate it by learned updates of the features, Sec.~\ref{sec:overview} and Table~\ref{tab:mapping}. The placement indicates representability and does not imply historical derivation. The models are discussed in Sec.~\ref{sec:Relations}.}
    \label{fig:ACEfamily}
\end{figure}

Fig.~\ref{fig:ACEfamily} summarizes this view, with the complete graph basis at the bottom, local models on the left and semi-local message-passing models on the right. Models in green generate the many-body order by explicit products of basis functions or equivariant features and use non-linear functions only of distances, as fixed embeddings, as scalar gates or in the readout. In models in blue, learned networks, kernels, updates by multilayer perceptrons (MLPs) or attention act on the features and generate the many-body order, Table~\ref{tab:mapping}. In both cases non-linear functions are employed, and green and blue distinguish only how the many-body order is generated.

As displayed on the left side of Fig.~\ref{fig:ACEfamily}, ACE is obtained from GRACE by limiting the tree graphs to star graphs. In a star graph the central atom connects directly to each neighbor, while a tree graph continues through further generations of edges, see Sec.~\ref{sec:complete}. ACE provides a complete basis in which many classical potentials and local MLIPs can be represented. For example, the Smooth Overlap of Atomic Positions (SOAP)~\cite{bartokRepresentingChemicalEnvironments2013} expands the neighbor density in radial functions and spherical harmonics, the construction on which many of today's MLIPs build, for example the polynomial invariants of Seko et al.~\cite{sekoGrouptheoreticalHighorderRotational2019}. SOAP and the closely related Spectral Neighbor Analysis Potential (SNAP)~\cite{thompsonSpectralNeighborAnalysis2015} can be represented by ACE, see~\cite{drautzAtomicClusterExpansion2019,drautzAtomicClusterExpansion2020,dussonAtomicClusterExpansion2022}. MTP~\cite{shapeevMomentTensorPotentials2016}, an atom-centered complete, in fact overcomplete, basis in Cartesian coordinates, can be represented by ACE~\cite{drautzAtomicClusterExpansion2020}. Employing Cartesian coordinates here means that spherical harmonics are not used explicitly, although the spherical harmonics themselves are computed from Cartesian coordinates, Sec.~\ref{sec:cartesian}. Classical potentials fit the same picture. The embedded-atom method, for example, corresponds to a two-body ACE inserted into a non-linear embedding function, Sec.~\ref{sec:nlopt}.

Many further local MLIPs, shown in blue on the left of Fig.~\ref{fig:ACEfamily}, take as input descriptors that are transformed to atomic energies with the help of machine learning methods such as neural networks or kernels. The high-dimensional neural network potential (HDNNP)~\cite{behlerGeneralizedNeuralNetworkRepresentation2007} with its Atom Centered Symmetry Functions (ACSF)~\cite{behlerAtomcenteredSymmetryFunctions2011} established this construction, and many MLIPs since then built on the same construction~\cite{bartokGaussianApproximationPotentials2010,smithANI1ExtensibleNeural2017,zhangDeepPotentialMolecular2018,fanNeuroevolutionMachineLearning2021,vandermauseActiveLearningReactive2022}. These descriptors can be re-expanded in ACE and in many cases are ACE basis functions. In contrast to empirically chosen descriptors, the ACE basis provides higher body order terms at low computational cost. With such higher body order terms, only linear or mildly non-linear functions were required for accurate and efficient local MLIPs for the systems studied in Refs.~\cite{lysogorskiyPerformantImplementationAtomic2021, bochkarevEfficientParametrizationAtomic2022,kovacsLinearAtomicCluster2021,wittACEpotentialsjlJuliaImplementation2023}, so that neural networks are optional.

As displayed on the right side of Fig.~\ref{fig:ACEfamily}, for semi-local interactions, which are relayed through intermediate atoms, GRACE provides a complete representation based on graph cluster basis functions. Recursive evaluation of the GRACE basis functions, layer by layer from the outermost part of the graph inward, reproduces the structure of graph neural networks (GNNs)~\cite{scarselliGraphNeuralNetwork2009}. The starting point of the right side is recursively evaluated GRACE together with multi-ACE~\cite{batatiaDesignSpaceE3equivariant2025}, which inserts equivariant ACE messages into a GNN architecture. Equivariant quantities rotate with the structure and in this way carry angular information between layers, see Sec.~\ref{sec:symmetry}. For example, multi-ACE, which predates GRACE, may today be understood as recursively evaluated GRACE with additional freedom in coupling the layers of the GNN. Multi-ACE was obtained by merging ACE and NequIP~\cite{batznerE3equivariantGraphNeural2022} and contains NequIP, multilayer ACE (ml-ACE)~\cite{bochkarevMultilayerAtomicCluster2022} and related MLIPs~\cite{NEURIPS2019_03573b32} as special cases. CACE is a variant of multi-ACE formulated in Cartesian coordinates~\cite{chengCartesianAtomicCluster2024}.

The backbone of a GNN can therefore be viewed as a numerically efficient evaluation of a semi-local basis expansion, onto which the network adds learned non-linear updates. By the same reasoning, transformer architectures~\cite{vaswaniAttentionAllYou2017,joshiTransformersAreGraph2025} applied to atomic interactions map onto a semi-local expansion, Sec.~\ref{sec:nlrep}. Recursive evaluation further relates many current MLIPs to GRACE~\cite{haghighatlariNewtonNetNewtonianMessage2022,batznerE3equivariantGraphNeural2022, NEURIPS2022_4a36c3c5,liaoEquiformerV2ImprovedEquivariant2023,pozdnyakovSmoothExactRotational2023,chengCartesianAtomicCluster2024,mazitovPETMADLightweightUniversal2025a,fuLearningSmoothExpressive2025, rhodesOrbv3AtomisticSimulation2025, kimDataefficientMultifidelityTraining2024, zhangGraphNeuralNetwork2025, yinAlphaNetScalingLocalframebased2025, kokerTrainingFoundationModel2025, yanMaterialsFoundationModel2025}. While the representation of GNNs within GRACE rests on the completeness of the GRACE basis functions, most GNNs employ learned layer couplings and non-linear transformations, shown in blue on the right of Fig.~\ref{fig:ACEfamily}, in analogy to the local MLIPs in blue on the left.

The correspondence between GRACE and other MLIPs goes beyond the functional form. Recursive evaluation reproduces message passing, and tensor decomposition, by which one large coefficient tensor is written as a product of small ones, reproduces the mixing of feature channels between layers, so that also the architecture of many MLIPs maps onto GRACE. Representability does not, however, make models equivalent in practice, as models differ in training data, truncation, non-linear terminations and implementation. Numerical efficiency, too, is defined only at a given accuracy. The same complete basis, truncated differently, yields models that range continuously from fast to accurate over more than two orders of magnitude in computational cost, Fig.~\ref{fig:warp}. Table~\ref{tab:mapping} in Sec.~\ref{sec:Relations} summarizes how selected MLIPs map onto this picture.

Model architecture and coverage of the periodic table are independent choices. Universal or foundation MLIPs parameterize the interactions of a large part of the periodic table in a single model. The currently most accurate foundation MLIPs build on GNNs~\cite{batatiaFoundationModelAtomistic2023,woodUMAFamilyUniversal2025,lysogorskiyGraphAtomicCluster2026,riebesellMatbenchDiscovery2025}, but foundation models are not limited to GNNs, as shown by the local, single-layer models NEP89~\cite{liangNEP89UniversalNeuroevolution2026}, GRACE-1L~\cite{lysogorskiyGraphAtomicCluster2026} and Allegro~\cite{tanHighperformanceTrainingInference2025,Leaderboard}. Fast models for specific applications can also be obtained by distillation, i.e. by training a simpler architecture on the predictions of a foundation model~\cite{gardnerDistillationAtomisticFoundation2025}.

\section{Completeness} \label{sec:complete}

\begin{framed}
{\it{Summary:}} A complete graph connects an atom to all atoms that it interacts with. A complete basis on the graph can represent any functional form of the atomic interaction and with it, in principle, the functional form of any other MLIP with the same interaction range. Non-linear functions of the linear basis expansion converge faster. The atomic cluster expansion (ACE) and its generalization to general graphs (GRACE) provide canonical, complete expansions, and any other complete representation is related to them by a basis transformation. The error is observed to decrease as a power law with the number of basis functions, the number of training data and the number of training steps. They are straightforward to extend to further degrees of freedom, such as charge transfer or magnetism, and to represent vectorial or tensorial atomic properties.
\end{framed}

\subsection{Complete graphs}

\begin{figure}[ht!]
    \includegraphics[width=0.23\textwidth]{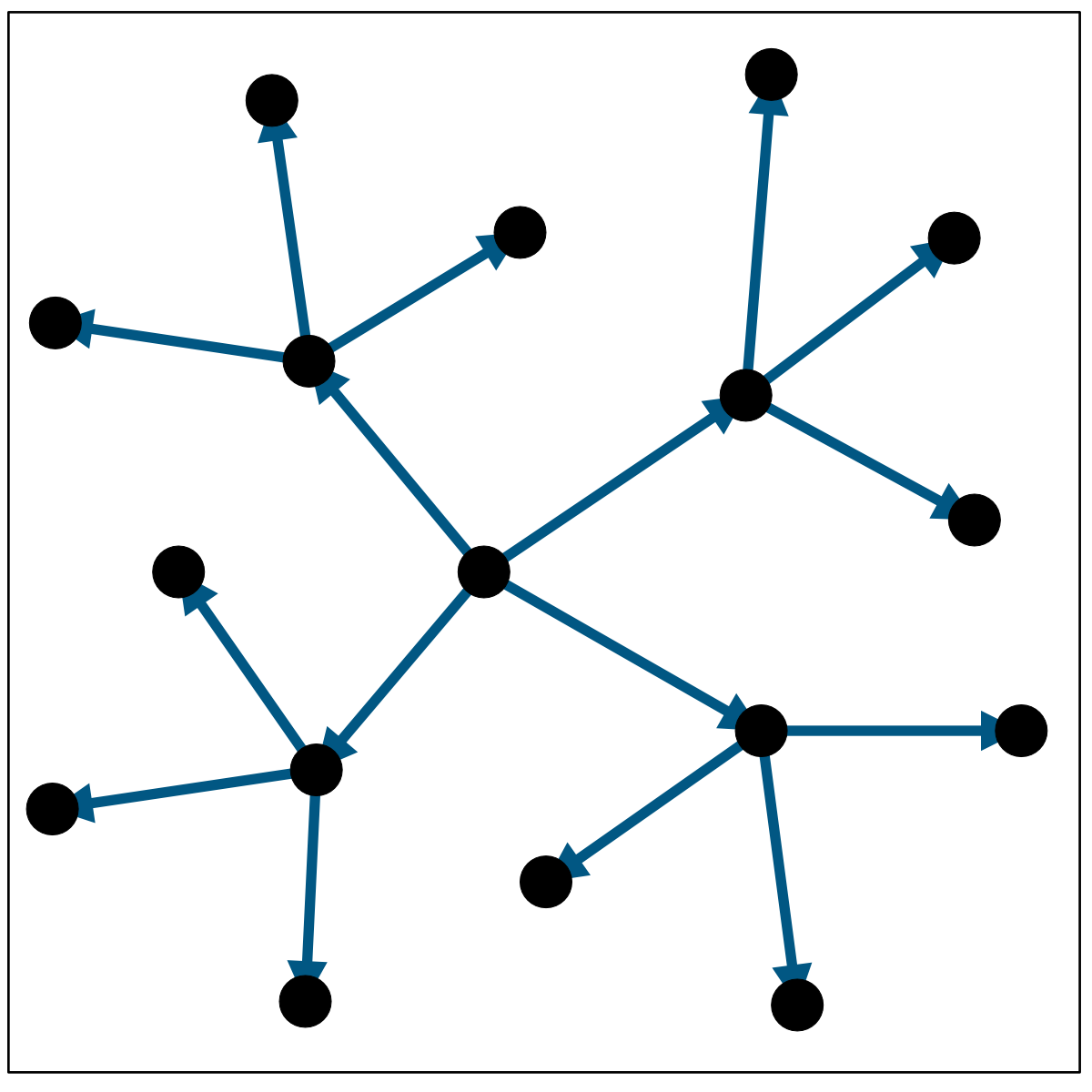}
    \includegraphics[width=0.23\textwidth]{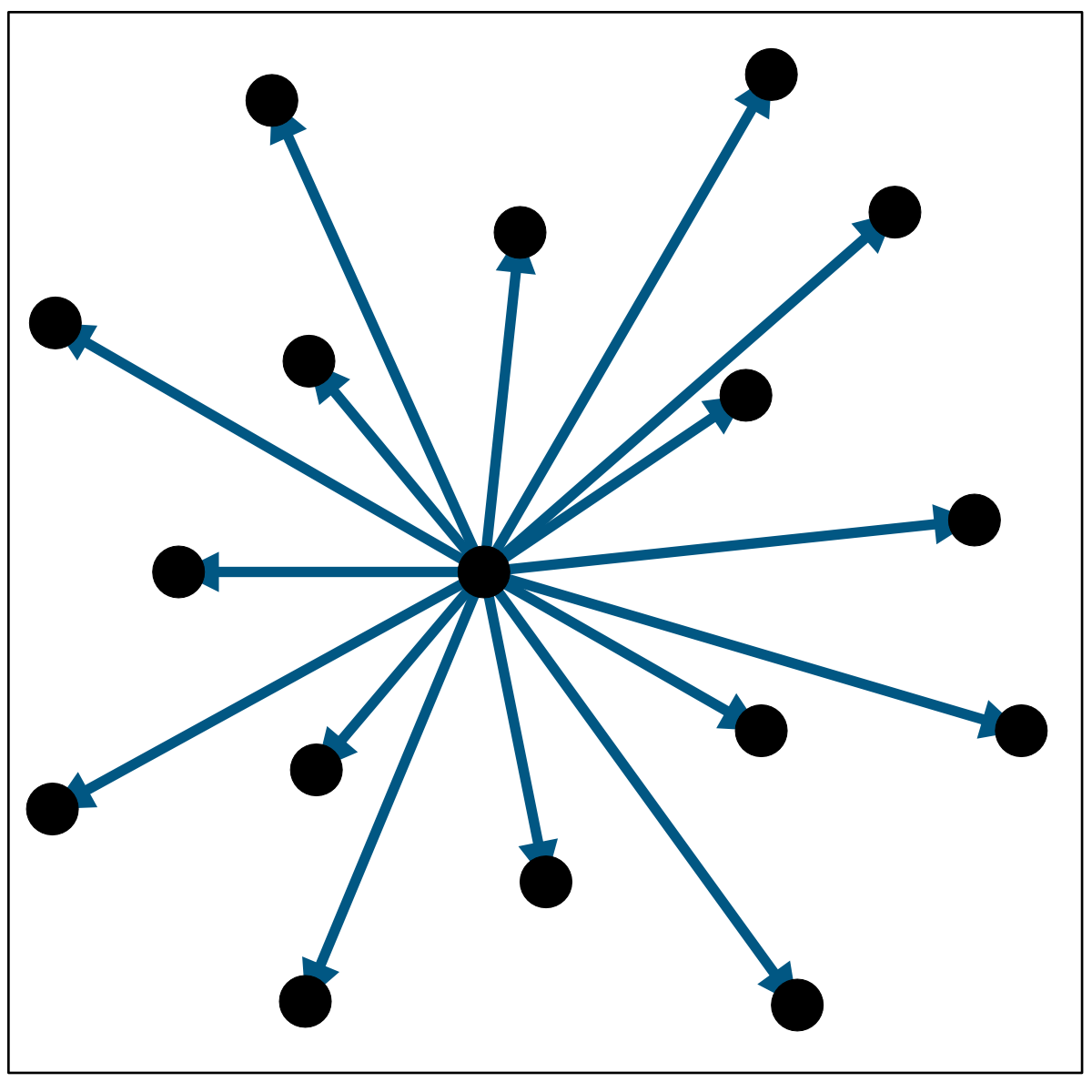}	
    \caption{Complete tree graph (left) and complete star graph (right). The nodes are atoms, the root node is the central node and the arrows are directed edges.}
    \label{fig:completegraphs}
\end{figure}

Completeness starts with geometrical considerations. We take atoms as the nodes of a graph and the vectors that connect atoms as its edges. We are interested in computing the energy of an atom, say of atom $i$, and we refer to atom $i$ as the root node of the graph. A graph is complete if every atom that atom $i$ interacts with can be reached from the root node by traversing edges of the graph. For example, a complete graph must contain as nodes all atoms within the cutoff distance of atom $i$. A direct way of ensuring a complete graph is to take all vectors $\br_j - \br_i$ that reach from atom $i$ to all other atoms $j$ within cutoff distance. The resulting graph is a star, as illustrated in the right hand panel of Fig.~\ref{fig:completegraphs}. A complete graph can also be built in several steps, for example, by first connecting the root node to a selected number of neighboring nodes and continuing from these to further nodes. The left hand panel of Fig.~\ref{fig:completegraphs} gives an example.

Not all graphs are admissible. If a node can be reached from the root node along different paths, the graph contains geometrically redundant information, and we exclude such graphs. Two simple rules are sufficient to construct all complete and admissible graphs. 
\begin{enumerate}
\item A graph has one root node that only has outgoing edges. The root node corresponds to the atom for which the energy is computed.
\item Every other node has exactly one incoming and any number of outgoing edges.
\end{enumerate}
It follows that all admissible graphs are acyclic, i.e. they are trees with $N-1$ edges for $N$ nodes. The exclusion of cycles is taken up again in two places. The density trick of Sec.~\ref{sec:neighbors} includes self-interactions, in which the same atom appears more than once in a term of the expansion, and in the recursive evaluation of Sec.~\ref{sec:scalelayer} also paths that return to an atom already visited. The moments of the electronic density of states in Sec.~\ref{sec:chem} are closed walks that reduce to tree graphs. In addition, a cutoff distance for the interaction between pairs of atoms limits the length of every edge in a graph to the cutoff distance.

\begin{figure}[ht!]
    \includegraphics[width=0.46\textwidth]{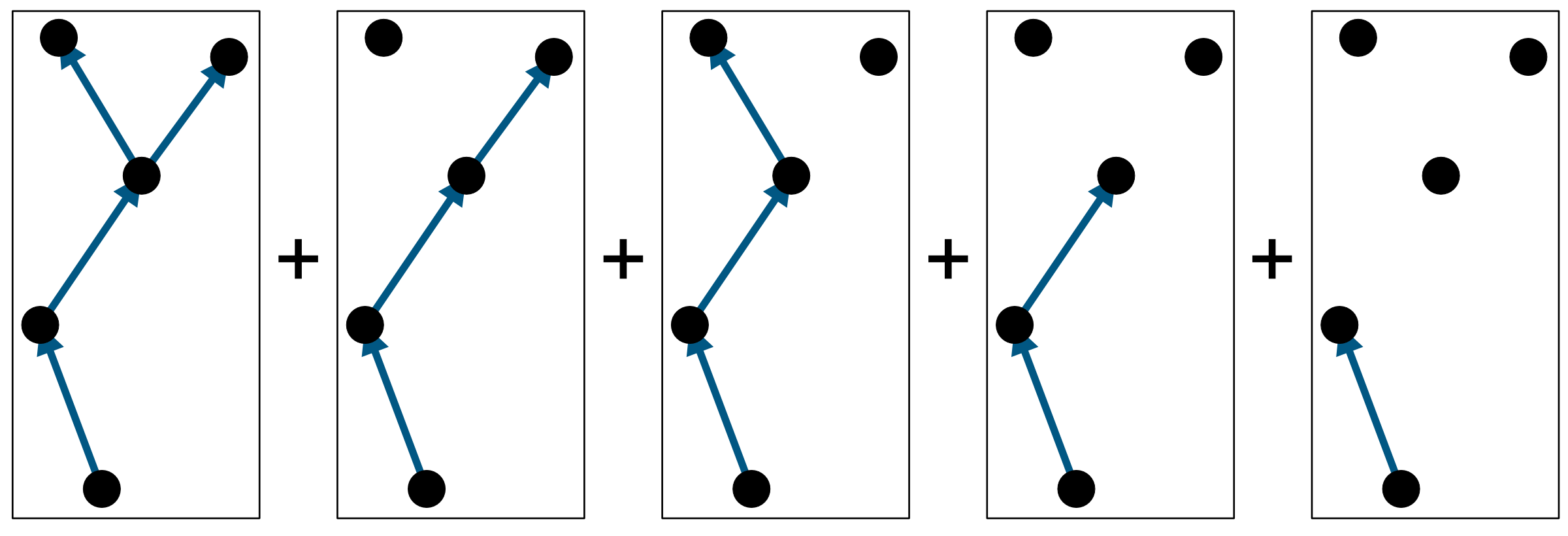}
    \caption{Graphs contributing to a complete expansion for five atoms. The complete graph (left) is a (1233) graph and the other graphs are its admissible subgraphs, obtained by picking florets from the complete graph. The root node is at the bottom and the arrows are directed edges.}
    \label{fig:completebasis}
\end{figure}

Graphs are further characterized by layers. The layer of a node is the number of edges that are traversed from the root node to the node, and the number of layers of a graph is the maximum layer of any node. The graphs in Fig.~\ref{fig:completegraphs} have two layers (left) and one layer (right). Layers also provide a simple notation for graphs. For example, (1233) denotes a graph with one node on the first layer, one node on the second layer and two nodes on the third layer, see Fig.~\ref{fig:completebasis}. With the node layers listed branch by branch, the notation extends to more complex graphs~\cite{bochkarevGraphAtomicCluster2024}. For modeling atomic interactions one often restricts oneself to graphs with the same number of outgoing edges on each layer, $N_0$ from the root node, $N_1$ from every node on layer 1, $N_2$ from every node on layer 2 and so on, which we call dandelions~\cite{bochkarevGraphAtomicCluster2024}.

The graph rules only lay out the structure of a graph and give no guidance on how to choose the complete graph for an actual parameterization. In practice one chooses the complete graph as compact as possible and as large as necessary. The graph determines the region of atoms that may contribute to the energy of the root atom, and this region should reflect the physical interaction range. Message-passing models with many layers reach far beyond the physical interaction range, as with $L$ layers the effective range grows to $L$ times the cutoff, Sec.~\ref{sec:overview}. Models on compact graphs approach the accuracy of deeper models on current benchmarks, Sec.~\ref{sec:Relations}, which indicates that many layers are not a physical necessity for these benchmarks, and effective interaction ranges far beyond the physical interaction range should be viewed with caution.

A complete expansion contains contributions from the complete graph and from all its admissible subgraphs. The subgraphs are obtained by picking florets, where a floret is a terminal node without outgoing edges together with its single incoming edge. Picking a floret removes one node and one edge from the graph, and every node of the remaining subgraph can still be reached from the root node, i.e. the subgraphs are trees, too. All subgraphs are generated from the complete graph by picking florets in all possible ways. Fig.~\ref{fig:completebasis} gives an example of a complete graph and its subgraphs.

\begin{figure}[ht!]
    \includegraphics[width=0.45\textwidth]{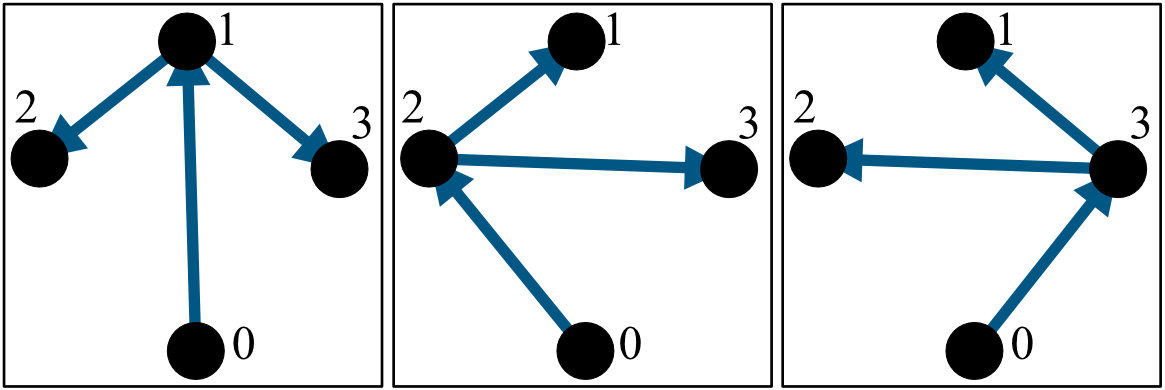}
    \caption{Symmetrization of a (122) graph for invariance under relabeling of atoms. The three graphs are obtained by exchanging the labels of the nodes 1, 2 and 3, and all three contribute to the expansion.}
    \label{fig:symmetrization}
\end{figure}

For atomistic simulation one has to label all atoms as $1, 2, \dots, \Nat$. These labels are arbitrary and relabeling the atoms must not change the results of a simulation. 
Invariance with respect to labeling requires graph symmetrization, i.e. all graphs that can be obtained by relabeling the nodes must be taken into account. For example, a graph from root node 0 with an outgoing edge to node 1 and two outgoing edges from node 1 to nodes 2 and 3, $0 \to 1 \to (2,3)$, needs to be symmetrized as $0 \to 1 \to (2,3)$, $0 \to 2 \to (1,3)$, $0 \to 3 \to (1,2)$. Note that star graphs, for example $0 \to (1,2,3)$, are automatically symmetric. Fig.~\ref{fig:symmetrization} illustrates this symmetrization. Symmetrization over root nodes is achieved by computing the energy as a sum over atomic energies. The recursive evaluation of tree graphs in Sec.~\ref{sec:twostep} makes use of the symmetry of star graphs, as it combines the florets on each layer into star graphs.

\subsection{Complete basis expansion} \label{sec:completebasis}

Once a complete graph has been chosen, a complete set of basis functions is constructed by attaching a basis function to each edge and to each node of the complete graph. To the edge from atom $i$ to atom $j$ we attach a single-bond basis function,
\begin{equation}
 \phi_{i v}(j) =  \phi_{iv}(\br_j - \br_i) \,, \label{eq:varphi}
\end{equation}
which is often chosen as the product of a radial function and a spherical harmonic,
\begin{equation}
 \phi_{i v}(j) =   R_{nl}(|\br_j - \br_i|) Y_{lm}(\theta, \phi) \,,  \label{eq:varphiSH}
\end{equation}
with the combined index $v = (n,l,m)$ and the polar angles $\theta, \phi$ of the direction of $\br_j - \br_i$, see Sec.~\ref{sec:symmetry}. 

We ask these single-bond basis functions to be orthonormal and complete on each atom,
\begin{align}
\int \phi_{iv}^*(\br - \br_i) \phi_{iv'}(\br- \br_i) \, d\br &= \delta_{v v'} \,, \\
\sum_{v}  \phi_{iv}^*(\br- \br_i) \phi_{iv}(\br'- \br_i) &= \delta(\br - \br') \,.
\end{align}
Basis functions on different atoms are in general not orthogonal, cf. the overlap matrix elements in Sec.~\ref{sec:chem}. We further ask that the first single-bond basis function is constant, $\phi_{i0}(\br) = 1$, where its normalization is absorbed in the expansion coefficients.

Analogously, single-atom basis functions $\chi_{\kappa}(\mu_i, q_i, \pmb{m}_i)$ are attached to each node $i$. These basis functions characterize the state of the atoms, as described by chemical species $\mu_i$, charge $q_i$, magnetic moment $\pmb{m}_i$, or other variables of interest. We request orthonormality and completeness of the single-atom basis functions with respect to their variables. In analogy to the single-bond basis functions, the first single-atom basis function is constant, $\chi_{0} = 1$.

Next we build graph cluster basis functions as the product of the single-bond basis functions on all edges and the single-atom basis functions on all nodes of the graph,
\begin{equation}
\Phi_{\vii}^{(\bt)} ( \pmb{\sigma})= \underbrace{\chi_{\kappa_0}(\mu_i)}_{\substack{\mathrm{root}\\ \mathrm{node}\ i}}
\underbrace{\phi_{iv_{1}}(j)  \chi_{\kappa_1}(\mu_j)}_{\substack{\mathrm{floret}\\ {i \to j}}} 
\underbrace{\phi_{jv_2}(k)   \chi_{\kappa_2}(\mu_k)}_{\substack{\mathrm{floret}\\ {j \to k}}}
 \dots \,, \label{eq:Phi}
\end{equation} 
where $\bt$ labels the graph topology, $\vii = (v_1, v_2, \dots, \kappa_0, \kappa_1, \kappa_2, \dots)$ the basis functions on the edges and on the nodes and $\pmb{\sigma} = (\mu_i, \br_j - \br_i, \mu_j, \br_k - \br_j, \dots)$ the node states, here limited to the chemical species, and the edge vectors. The graph topology is labeled by the layer notation of the previous subsection. As every node except the root node has exactly one incoming edge, edge and node basis functions are conveniently combined into floret basis functions, as illustrated in Eq.(\ref{eq:Phi}). On the edges of a graph only basis functions with $v > 0$ are taken, while the constant basis functions $v = 0$ and $\kappa = 0$ serve to relate subgraphs to the complete graph.

A basis function $v=0$ on the edge of a floret shows no variation with distance or direction of the edge and in this way removes the edge from the graph, producing a subgraph. This can be repeated, as $v=0$ on a floret of the subgraph produces a subgraph of the subgraph, so that all subgraphs are obtained by decorating florets with $v=0$. If the edge of a floret carries $v=0$, then the node of the floret must carry $\kappa=0$. Otherwise an atom would contribute to the energy while its position could be changed without changing the energy, which is unphysical. A basis function $v=0$ on an edge inside a graph is excluded for the same reason, as the corresponding bond could be elongated without changing the energy, and it would cut the graph into fragments.

Orthogonality and completeness are formulated on the complete graph. To this end a subgraph is patched with florets that carry $v=0$ and $\kappa=0$ until its topology matches the complete graph. All graphs then take the same product form, and orthogonality follows from the orthonormality of the single-bond and single-atom basis functions. Completeness follows from the sum over all basis function indices, where the sum over graphs includes the patched subgraphs.

The orthogonality and completeness of the graph cluster basis functions then read
\begin{align}
&\int \left( \Phi_{\vii}^{(\bt)}( \pmb{\sigma})  \right)^* \Phi_{\vii'}^{(\bt')}( \pmb{\sigma}) \,d \pmb{\sigma}  = \delta_{\bt \bt'} \delta_{\vii \vii'} \,, \\
&\sum_{\bt \vii} \left( \Phi_{\vii}^{(\bt)} ( \pmb{\sigma}) \right)^* \Phi_{\vii}^{(\bt)} ( \pmb{\sigma}') = \delta( \pmb{\sigma} - \pmb{\sigma}' ) \,, \label{eq:complete}
\end{align}
with $\bt$ and $\bt'$ the complete graph or subgraphs of the complete graph. 

\begin{figure}[ht!]
    \includegraphics[width=0.5\textwidth]{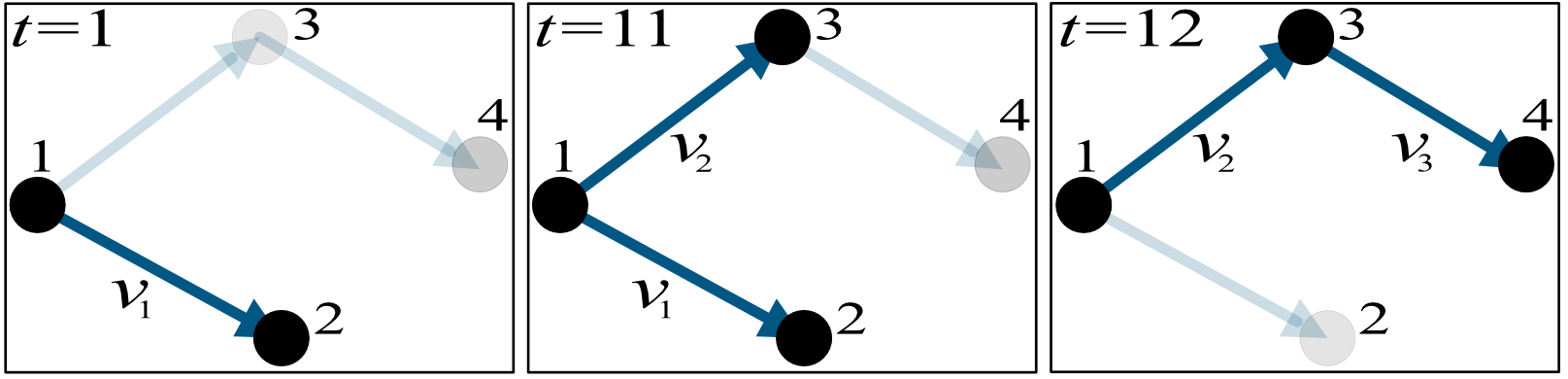}
    \caption{The (1), (11) and (12) subgraphs of a (112) graph, labeled by their topology $t$, with the indices $v$ of the basis functions on the edges. Edges that carry the constant basis function $\phi_{i0}(j)=1$ are drawn in faint color. }
    \label{fig:orthonormal}
\end{figure}

We illustrate the orthogonality of the graph cluster basis functions for a (112) graph. For simplicity all atoms are of the same chemical species and carry no other degrees of freedom, such that the node basis functions are $\chi = 1$. We take the (112) graph as the complete graph and analyze the orthogonality of three subgraphs in Fig.~\ref{fig:orthonormal}. The basis functions that we attach to these three graphs are given as 
\begin{align}
&\Phi_{v_{1}}^{(1)}  = \phi_{1v_{1}}(2) \phi_{10}(3) \phi_{3 0}(4) = \phi_{1v_{1}}(2) \nonumber \,, \\
&\Phi_{v_{1} v_{2}}^{(11)}  = \phi_{1v_{1}}(2) \phi_{1v_{2}}(3) \phi_{3 0}(4) = \phi_{1v_{1}}(2)\phi_{1v_{2}}(3) \nonumber \,,\\
&\Phi_{v_{2}v_{3}}^{(12)}  = \phi_{10}(2) \phi_{1v_{2}}(3) \phi_{3 v_{3}}(4) = \phi_{1v_{2}}(3)\phi_{3v_{3}}(4)\,,\nonumber
\end{align}
i.e. the edges drawn in faint color carry the constant basis function $\phi_{i0}(\br) = 1$, which removes the corresponding florets from the graph.

The (1) and the (11) graph are orthogonal as
\begin{align}
&\int \left( \Phi_{v_{1}}^{(1)} \right)^*  \Phi_{v_{1} v_{2}}^{(11)} \, d\br \nonumber \\
&=  \int  \phi^*_{1v_{1}}(2)   \phi_{1v_{1}}(2) \int \phi_{10}^*(3)  \phi_{1v_{2}}(3)  \nonumber \\
& \times \int \phi_{3 0}^*(4)  \phi_{3 0}(4) \, d \br_2 d \br_3 d \br_4 =0 \,.
\end{align}
The same holds for the (11) and the (12) graph,
\begin{align}
&\int \left( \Phi_{v_{1} v_{2}}^{(11)} \right)^*  \Phi_{v_{2} v_{3}}^{(12)} \, d\br \nonumber \\
&=  \int  \phi^*_{1v_{1}}(2)   \phi_{10}(2) \int \phi_{1v_{2}}^*(3)  \phi_{1v_{2}}(3)  \nonumber \\
& \times \int \phi_{3 0}^*(4)  \phi_{3 v_{3}}(4) \, d \br_2 d \br_3 d \br_4 =0 \,.
\end{align}
In both cases the integral vanishes because on at least one edge a constant basis function meets a basis function with $v > 0$, and the two are orthogonal.

The energy or any other quantity of atom $i$ can now be expanded in terms of the graph cluster basis functions, 
\begin{equation}
E_i = \sum_{\bt \vii} c_{\vii}^{(\bt)} \Phi_{i \vii}^{(\bt)}  =  \pmb{c} \cdot \pmb{\Phi}_i \,,  \label{eq:linear}
\end{equation}
with expansion coefficients $c_{\vii}^{(t)} $. This expansion is the graph atomic cluster expansion (GRACE)~\cite{bochkarevGraphAtomicCluster2024}, and its restriction to star graphs, the atomic cluster expansion (ACE)~\cite{drautzAtomicClusterExpansion2019,dussonAtomicClusterExpansion2022}, was introduced first. Atomic charges, magnetic moments and other node properties enter through the single-atom basis functions~\cite{drautzAtomicClusterExpansion2020}, Sec.~\ref{sec:extendability}.

The expansion Eq.(\ref{eq:linear}) is invariant under basis transformations. If a new set of basis functions is defined as linear combinations of the current basis functions, $\pmb{\Phi}_i' = \pmb{C}  \pmb{\Phi}_i$, with transformation matrix $\pmb{C}$, then the transformed expansion coefficients $\pmb{c}' = \pmb{c} \pmb{C}^{-1} $ leave the result of the expansion unchanged. Any complete basis is related to the graph cluster basis by such a transformation. Among the complete bases the graph cluster basis is canonical, as it is built directly from products of single-particle basis functions.

Before the complete basis Eq.(\ref{eq:linear}) was introduced, atomic environments were characterized by descriptors. Well-known descriptors are the ACSF~\cite{behlerAtomcenteredSymmetryFunctions2011}, SOAP~\cite{bartokRepresentingChemicalEnvironments2013} and the closely related SNAP~\cite{thompsonSpectralNeighborAnalysis2015}. These descriptors can all be represented by ACE, i.e. by the basis functions Eq.(\ref{eq:Phi})~\cite{drautzAtomicClusterExpansion2019,drautzAtomicClusterExpansion2020,dussonAtomicClusterExpansion2022}. The same holds for MTP~\cite{shapeevMomentTensorPotentials2016}, which provides an overcomplete basis in Cartesian coordinates, Sec.~\ref{sec:cartesian}. Descriptors truncated at low body order are incomplete. For example, there exist distinct atomic environments with identical three-body descriptors, and no model built on these descriptors can distinguish them~\cite{pozdnyakovIncompletenessAtomicStructure2020,pozdnyakovIncompletenessGraphNeural2022,nigamCompletenessAtomicStructure2024}.

\subsection{Non-linear expansions and optimal basis} \label{sec:nlopt}

In practice only a finite number of graph cluster basis functions is evaluated, and for numerical efficiency one computes as few as possible. The expansion Eq.(\ref{eq:linear}) is therefore truncated and the residual higher-order contributions are approximated by non-linear functions of the available basis functions,
\begin{equation}
E_i  = \pmb{c} \cdot \pmb{\Phi}_i +  F(\pmb{\Phi}_i) \,,  \label{eq:terminate}
\end{equation}
with a non-linear function $F(\pmb{\Phi}_i)$ of the basis functions. The function $F(\pmb{\Phi}_i)$ can be chosen from physical and chemical considerations or by machine learning. Physically and chemically motivated functions $F(\pmb{\Phi}_i)$ should be scale-invariant, universal and invariant under basis transformation~\cite{bochkarevEfficientParametrizationAtomic2022}. Scale invariance means that $F(\pmb{\Phi}_i)$ has no intrinsic energy scale, so that the representation does not depend on the energy unit, and universality means that $F(\pmb{\Phi}_i)$ does not depend on a particular chemistry. 

To ensure invariance of non-linear expansions under the basis transformations of Sec.~\ref{sec:completebasis}, we apply non-linear functions only to basis expansions that are themselves invariant under basis transformation. We term these expansions atomic properties,
\begin{equation}
\ace_i =  \pmb{c} \cdot \pmb{\Phi}_i =  \pmb{c}' \cdot \pmb{\Phi}_i'  \,.
\end{equation}
By taking different sets of expansion coefficients $\pmb{c}^{(n)}$, with $n = 1, \dots, n_{\max}$ and corresponding atomic properties  $\ace_i^{(n)} =  \pmb{c}^{(n)} \cdot \pmb{\Phi}_i$, we write the energy Eq.(\ref{eq:terminate}) as
\begin{equation}
E_i = \mathcal{F}(\pmb{\ace}_i) \,, \label{eq:F}
\end{equation}
with $\pmb{\ace}_i = (\ace_i^{(1)}, \ace_i^{(2)}, \dots, \ace_i^{(n_{\text{max}})})$. This representation is invariant under basis transformation and relates the expansion to classical interatomic potentials and to local MLIPs. 

Many classical interatomic potentials~\cite{norskovEffectivemediumTheoryChemical1980,puskaAtomsEmbeddedElectron1981,guptaLatticeRelaxationMetal1981,finnisSimpleEmpiricalBody1984,dawEmbeddedatomMethodDerivation1984,stillingerComputerSimulationLocal1985,tersoffNewEmpiricalModel1986,biswasNewClassicalModels1987,baskesModifiedEmbeddedatomPotentials1992} can be cast in the form
\begin{equation}
E_i = \ace_i^{(1)} +  F( \ace_i^{(2)})  \,. \label{eq:EFS}
\end{equation}
In the second-moment approximation~\cite{ducastelleModulesElastiquesMetaux1970} and the Finnis-Sinclair potential~\cite{finnisSimpleEmpiricalBody1984} the non-linear function is given by $F(\ace_i) = \sqrt{\ace_i}$ and $\ace_i^{(2)}$ is interpreted as the second moment of the density of states or as a density, respectively. For ACE we demonstrated numerically that the non-linear term improves convergence considerably~\cite{bochkarevEfficientParametrizationAtomic2022}, and taking several atomic properties as input improved the representation further~\cite{erhardModellingAtomicNanoscale2024}. 

For MLIPs the physically and chemically motivated form is replaced by a neural network or other machine learning algorithm for $\mathcal{F}$ in Eq.(\ref{eq:F}), for which the atomic properties  $\pmb{\ace}_i$ serve as input. In fact, the Behler-Parrinello HDNNP~\cite{behlerGeneralizedNeuralNetworkRepresentation2007} takes exactly this form, with the limitation that the ACSF~\cite{behlerAtomcenteredSymmetryFunctions2011} are empirical two- and three-body functions on a star graph rather than a complete basis expansion. This analysis extends to related approaches based on neural networks~\cite{smithANI1ExtensibleNeural2017,zhangDeepPotentialMolecular2018} or other machine learning methods, such as Gaussian process regression~\cite{bartokGaussianApproximationPotentials2010} built on the SOAP descriptor~\cite{bartokRepresentingChemicalEnvironments2013}, which can all be cast in the form Eq.(\ref{eq:F}). On the Cu and Si benchmark of Ref.~\cite{zuoPerformanceCostAssessment2020}, star graph ACE in the mildly non-linear form of Eq.(\ref{eq:EFS}) reached lower errors at lower cost than local MLIPs with neural network or kernel non-linearities~\cite{lysogorskiyPerformantImplementationAtomic2021}. Recent GNN and transformer architectures can likewise be cast as non-linear functions of a linear basis expansion, subject to the constraint that rotational symmetry imposes on non-linear functions, as made explicit for transformer attention in Sec.~\ref{sec:nlrep}.

In addition to the non-linear termination, an optimal choice of basis functions reduces the number of basis functions required for a given accuracy. To this end the single-bond basis functions Eq.(\ref{eq:varphi}) are optimized during training
~\cite{bochkarevEfficientParametrizationAtomic2022}. Basis optimization is closely related to chemical embedding, Sec.~\ref{sec:scale chem}, where a weighted sum over chemical species defines effective single-bond basis functions. Optimized basis functions are well established in electronic structure theory, including screened bond integrals that depend on the local atomic environment~\cite{nguyen-manhAnalyticEnvironmentDependentTightBinding2000, drautzAnalyticBondorderPotentials2007}, and optimized radial functions are also used in MLIPs~\cite{goscinskiOptimalRadialBasis2021}.

\subsection{Convergence}

For a large class of models, in particular linear expansions of the form Eq.(\ref{eq:linear}), the error is observed to converge as a power law,
\begin{equation}
\Delta \propto N^{-m} \,.
\end{equation}
Here $\Delta$ is a measure of the error of the model, for example, the loss function $ \mathcal{L}$ (Sec.~\ref{sec:opt}) or the error in energy or forces. The power law holds for different choices of $N$, the number of model parameters or basis functions, the number of reference data for training, or the number of training steps, i.e. iterations or epochs of the optimizer. Power laws of this form are derived in learning theory as minimax rates for estimating functions of many variables~\cite{Stone1982} and as spectral rates for regression with kernels or linear bases~\cite{Caponnetto2007,Canatar2021}, and they are observed empirically as the scaling laws of deep learning~\cite{Hestness2017,Kaplan2020}, including deep models for chemistry~\cite{freyNeuralScalingDeep2023}.

The exponent $m$ depends on model, training data and optimizer. For the root-mean-square error and convergence with the number of training data in the presence of noise, $m = 1/2$ is the parametric rate at which noise averages out~\cite{Canatar2021}. A value of $m = 1/2$ is also expected for the convergence of the energy error with the number of model parameters if the contributions of the basis functions are independent~\cite{bochkarevEfficientParametrizationAtomic2022}. Exponents well below $1/2$ indicate that part of the interaction remains unresolved and that model, data or optimizer can be improved, while learning curves that flatten indicate that the error is limited by the model itself or by noise in the reference data. As the exponent $m$ is readily obtained during training, it provides an estimate of how many more parameters, reference data or optimizer iterations are required for a desired accuracy.

\subsection{Extensions: chemistry, charge, magnetism, vectors and tensors} \label{sec:extendability}

The basis expansion Eq.(\ref{eq:linear}) incorporates further degrees of freedom, such as charge or magnetic moments, by extending only the single-bond and single-atom basis functions. For example, for atomic magnetic moments $\pmb{m}_i$ the single-atom basis functions are extended to $\chi_{\kappa}(\mu_j, \pmb{m}_j)$~\cite{drautzAtomicClusterExpansion2020,rinaldiNoncollinearMagneticAtomic2024} and the floret basis function becomes
\begin{equation}
 \phi_{iv \kappa}(\br_j - \br_i, \pmb{m}_j)  =  \phi_{iv}(\br_j - \br_i) \chi_{\kappa}(\mu_j, \pmb{m}_j) \,. \label{eq:varphim}
\end{equation}
Orthonormality and completeness of the resulting graph cluster basis functions follow as in Sec.~\ref{sec:completebasis}~\cite{drautzAtomicClusterExpansion2020,bochkarevGraphAtomicCluster2024}.

The expansion also represents vectorial or tensorial atomic properties, for which the basis expansion Eq.(\ref{eq:linear}) is carried out for each entry of the vector or tensor. Equivariant vectorial or tensorial expansions, which rotate with the structure, follow from the irreducible representations of the rotation group~\cite{drautzAtomicClusterExpansion2020,grisafiSymmetryadaptedMachineLearning2018}, Sec.~\ref{sec:symmetry}. Non-linear vectorial or tensorial functions are obtained from a series expansion of the non-linear function.

Charge transfer is included in charge-constrained ACE~\cite{rinaldiChargeconstrainedAtomicCluster2025}, fourth-generation HDNNP~\cite{koFourthgenerationHighdimensionalNeural2021}, PhysNet~\cite{unkePhysNetNeuralNetwork2019a}, SpookyNet~\cite{unkeSpookyNetLearningForce2021} and AIMNet2~\cite{anstineAIMNet2NeuralNetwork2023}, and magnetism in magnetic ACE~\cite{rinaldiNoncollinearMagneticAtomic2024}, spin-dependent HDNNP~\cite{eckhoffHighdimensionalNeuralNetwork2021} and magnetic MTP~\cite{novikovMagneticMomentTensor2022}. Long-range electrostatics can be added by latent Ewald summation~\cite{chengLatentEwaldSummation2025}, which augments short-range MLIPs such as MACE, CACE or UMA with latent charges~\cite{kimUniversalAugmentationFramework2025}.

Because of its generality, the linear atomic cluster expansion Eq.(\ref{eq:linear}) has also been applied outside MLIPs, to jet tagging in particle physics, to Hamiltonian matrices and electronic wave functions and to coarse-grained molecular dynamics~\cite{munozBoostInvariantPolynomials2022,zhangEquivariantAnalyticalMapping2022,drautzAtomicClusterExpansion2022,zhouMultilevelMethodManyElectron2024,wangManyBodyCoarseGrainedMolecular2025}.

\section{Scaling} \label{sec:scaling}

\begin{framed}
{\it{Summary:}} Naive implementations incur a computational cost that grows as a power law with system size and up to exponentially with model complexity. However, the cost can be reduced to linear with system size and to linear or even constant with model complexity, which enables large and accurate simulations. We demonstrate linear computational cost with the number of atoms, the number of neighbors, the number of basis functions, the number of graph layers and the body order, the latter three at fixed rank of the tensor decomposition, and constant cost with the number of chemical species, as summarized in Table~\ref{tab:scaling}. For each of the reductions in this section we know of essentially one solution, and an efficient model has to adopt all of them, leaving only a narrow design space for efficient models.
\end{framed}

Here we use the term scaling as in physics, i.e. for the change of cost or accuracy with system size or model complexity. In machine learning, scaling also refers to the rescaling of features, which we call normalization and discuss in Sec.~\ref{sec:MLinMLIPs}.

Table~\ref{tab:scaling} summarizes the scaling discussed in the following subsections. In the table, $N$ is the count of the property in the first column and $M$ the exponent of the naive evaluation, except in the rows for body order and graph layers, where the naive cost is exponential in the row variable and $N$ is the number of neighbors. For atoms the naive quadratic cost arises from interactions between all pairs. For neighbors and basis functions the naive cost is a power law with the number of edges $M$ of a graph as exponent, i.e. the body order minus one, Secs.~\ref{sec:neighbors} and \ref{sec:scaleorder}, while for body order and graph layers the efficient cost is linear in the row variable. With tensor decomposition across body orders the cost becomes linear in the maximum body order, Sec.~\ref{sec:scaleorder}. The number of chemical species enters formally linearly but at negligible cost, Sec.~\ref{sec:scale chem}, hence constant cost in practice.

\begin{table}[h!]
\begin{tabular}{|c|c|c|}
\hline
Property & Naive & Efficient \\
\hline \hline
Atoms & $N^2$ & $N$ \\
Neighbors & $N^M$ & $N$ \\
Basis functions & $N^M$ & $N$ \\
Body order & $N^M$ & $N$ \\
Graph layers & $N^M$ & $N$ \\
Chemical species & $N^M$ & $1$ \\
\hline 
\end{tabular}
\caption{Scaling of the computational cost of naive and efficient evaluation with the relevant count $N$ and exponent $M$, see text for their definition in each row. \label{tab:scaling}}
\end{table}

\subsection{Scaling with number of atoms \label{sec:nat}}

\begin{figure}[ht!]
    \includegraphics[width=0.5\textwidth]{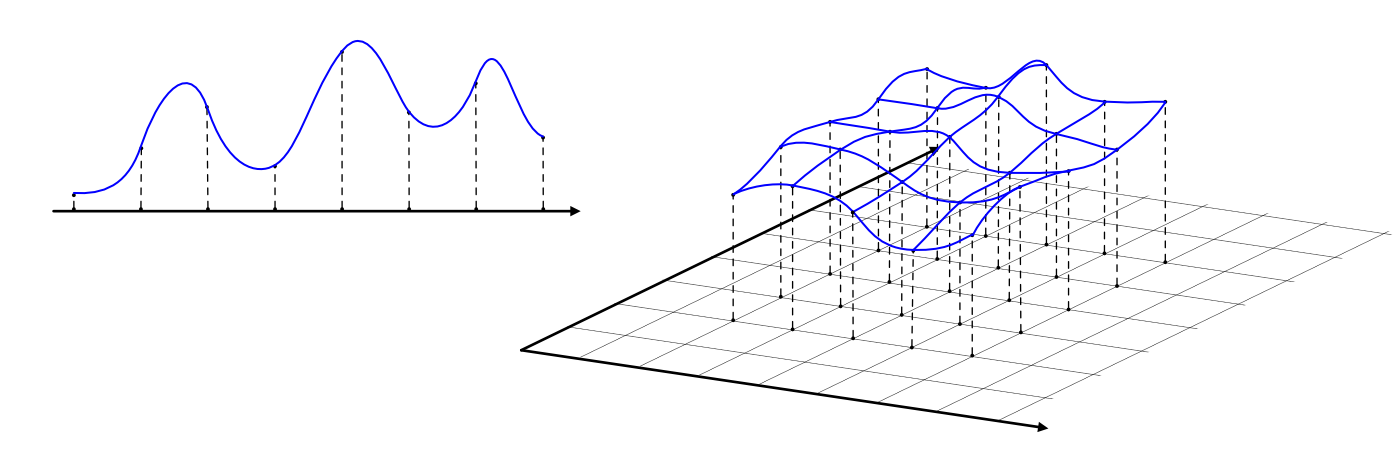}
    \caption{Naive scaling with the number of atoms. If one dimension requires $M$ mesh points, then two dimensions require $M^2$ and $N$ dimensions $M^N$.}
    \label{fig:scalingnumberatoms}
\end{figure}

Parameterizing the energy as a function of the $3N$ atomic coordinates, $E(\br_1, \br_2, \dots, \br_N)$, is a hard task. Without further assumptions, one could attempt to discretize the domain of atomic positions and to interpolate between mesh points.
For $N$ atoms and $M$ mesh points in each dimension, the effort for representing the energy $E$ grows as $M^{3N}$, in analogy to the representation of many-electron wavefunctions~\cite{kohnNobelLectureElectronic1999}. A direct discretization of atomic interactions is therefore impossible, as illustrated in Fig.~\ref{fig:scalingnumberatoms}.

\begin{figure}[ht!]
    \includegraphics[width=0.42\textwidth]{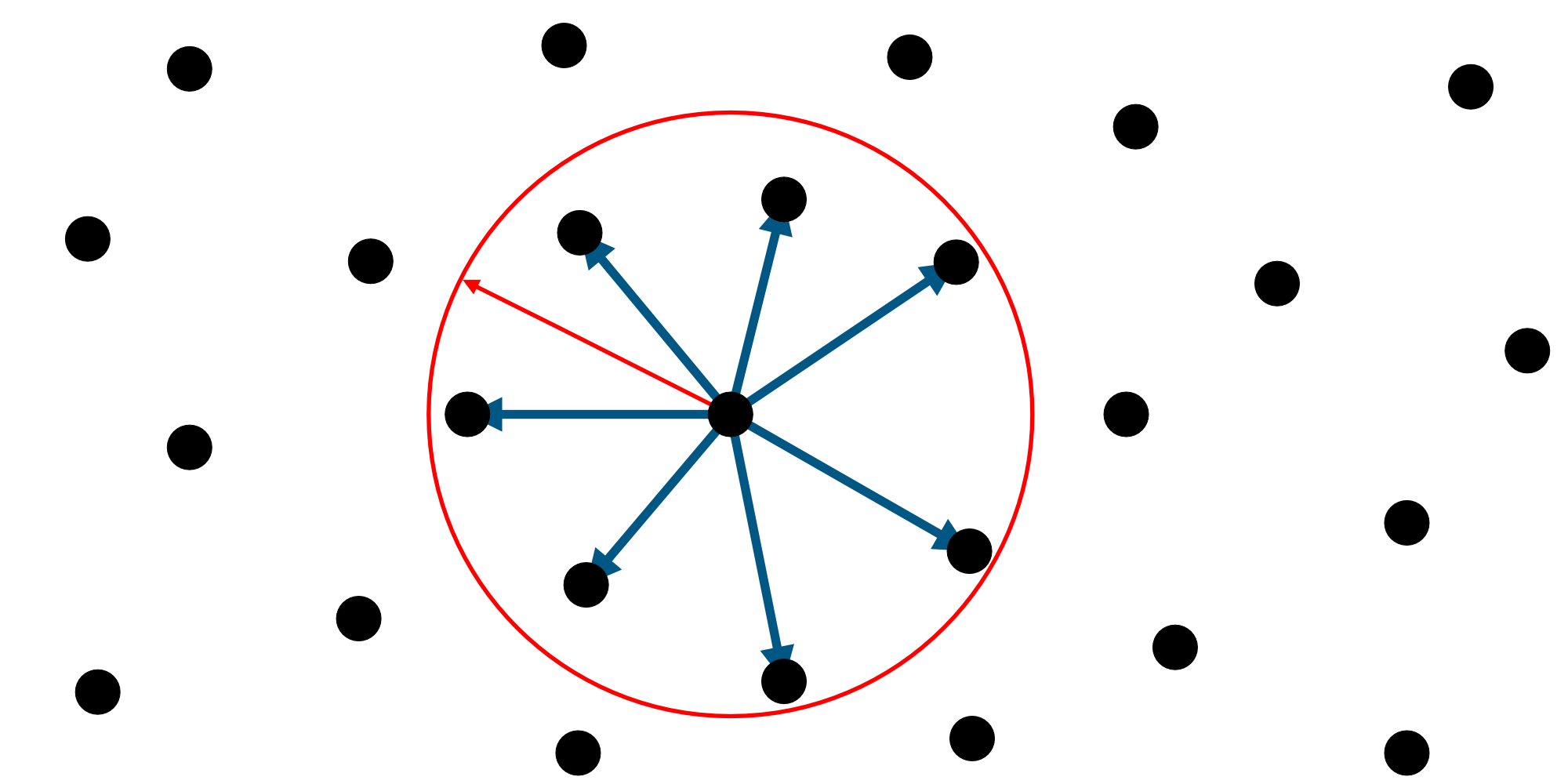}
    \caption{An atom interacts directly only with atoms within the cutoff distance. The length of an edge in a graph is therefore bounded by the cutoff distance.}
    \label{fig:cutoff}
\end{figure}

Even for a parameterized interaction, evaluating direct interactions between all pairs of atoms incurs a cost that is quadratic in the number of atoms, cf. Table~\ref{tab:scaling}. Instead, linear scaling with the number of atoms is achieved by introducing a cutoff distance beyond which there is no direct interaction between pairs of atoms, see Fig.~\ref{fig:cutoff}. To this end the energy is decomposed into atomic contributions,
\begin{equation}
E = \sum_{i=1}^{N} E_i \,.
\end{equation}
As the energy of every atom $i$ only depends on the atoms within the cutoff distance, the effort for computing $E$ is linear with the number of atoms $N$.
The decomposition of the energy into atomic contributions has been used for interatomic potentials for more than forty years~\cite{finnisSimpleEmpiricalBody1984,dawEmbeddedatomMethodDerivation1984,tersoffNewEmpiricalModel1986,baskesModifiedEmbeddedatomPotentials1992,drautzValencedependentAnalyticBondorder2006}. Early MLIPs did not decompose the energy into atomic contributions, which limited them to low-dimensional potential energy surfaces of small molecules or of molecules at frozen surfaces~\cite{lorenzRepresentingHighdimensionalPotentialenergy2004a,manzhosUsingNeuralNetworks2006,braamsPermutationallyInvariantPotential2009}. MLIPs became applicable to the simulation of solids and liquids only when Behler and Parrinello adopted the decomposition of the energy into atomic contributions~\cite{behlerGeneralizedNeuralNetworkRepresentation2007}.

\subsection{Scaling with number of neighbors} \label{sec:neighbors}

The energy of an atom can be decomposed further into a body-ordered many-atom expansion as~\cite{drautzGeneralRelationsManybody2004,drautzAtomicClusterExpansion2019,thomasBodyorderedApproximationsAtomic2022}
\begin{align}
E_i = & V_0 + V_1(\br_i) + \sum_{j} V_2(\br_i,\br_j) + \sum_{j_1 j_2} V_3(\br_i,\br_{j_1},\br_{j_2}) \nonumber \\
&+  \sum_{j_1 j_2 j_3} V_4(\br_i,\br_{j_1},\br_{j_2},\br_{j_3}) + ...  \,.
\end{align}
This expansion is of limited use in practice because of its poor scaling. In a material where each atom has $N$ neighbors, the evaluation of the terms $V_{M+1}$, which connect atom $i$ by $M$ edges to $M$ neighbors, requires $N^M$ summations, so that high body orders cannot be evaluated efficiently.

\begin{figure}[ht!]
    \includegraphics[width=0.5\textwidth]{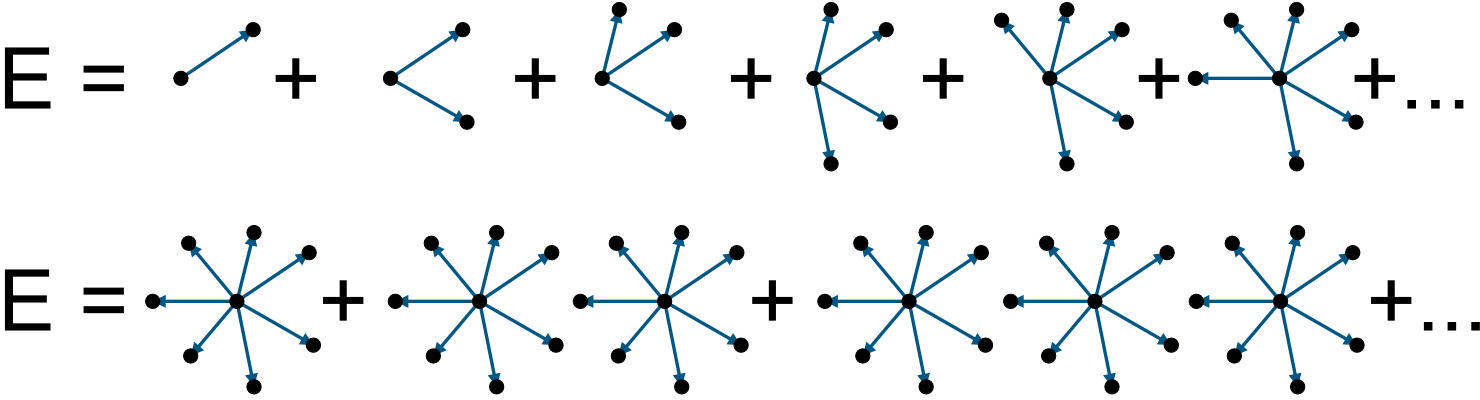}
    \caption{Top: $N^M$ scaling of the $M$-fold summation over $N$ neighbors. Bottom: linear scaling with $N$ for products of the atomic base.}
    \label{fig:densitytrick}
\end{figure}

The basis expansion Eq.(\ref{eq:linear}) overcomes the poor scaling. We illustrate this for a four-body star graph expansion and then generalize to arbitrary body order and, in Sec.~\ref{sec:scalelayer}, to graphs with several layers. The four-body contribution to the energy is expanded as
\begin{align}
E_i^{(4)} &= \sum_{j_1 j_2 j_3} V_4(\br_i,\br_{j_1},\br_{j_2},\br_{j_3}) \nonumber \\ 
&=  \sum_{j_1 j_2 j_3 } \sum_{ v_1  v_2 v_3}  c_{ v_1  v_2 v_3} \phi_{i v_1}(j_1) \phi_{i v_2}(j_2) \phi_{i v_3}(j_3)   \nonumber \\ 
&  = \sum_{ v_1  v_2 v_3} c_{ v_1  v_2 v_3} A_{iv_1} A_{iv_2} A_{iv_3} \,, \label{eq:E4}
\end{align}
where we introduced the atomic base~\cite{drautzAtomicClusterExpansion2019}
\begin{equation}
A_{iv} = \sum_j  \phi_{i v}(j) \,. \label{eq:abase}
\end{equation}
The cost for evaluating the atomic base is linear with the number of neighbors $N$ and therefore the cost for evaluating $E_i^{(4)}$ is also linear with $N$. The rearrangement in Eq.(\ref{eq:E4}) is sometimes referred to as the density trick and extends to arbitrary body order, 
\begin{equation}
\underbrace{\sum_{j_1 \dots j_M } c_{ v_1  \dots v_M} \prod_{k=1}^M \phi_{i v_k}(j_k)}_{N^M} = \underbrace{ c_{ v_1  \dots v_M}   \prod_{k=1}^M A_{iv_k}}_{N}   \,,
\end{equation}
which means that the energy can be evaluated with an effort that is linear in the number of neighbors, irrespective of the maximum body order of the expansion. Fig.~\ref{fig:densitytrick} illustrates the rearranged summation. Several descriptors and MLIPs employ the density trick~\cite{drautzAtomicClusterExpansion2019,shapeevMomentTensorPotentials2016,bartokRepresentingChemicalEnvironments2013,willattAtomdensityRepresentationsMachine2019}. 
In contrast, the ACSF in the Behler-Parrinello neural network potentials and related methods~\cite{behlerAtomcenteredSymmetryFunctions2011,smithANI1ExtensibleNeural2017} are computed without the density trick, and these methods are therefore limited to descriptors of low body order.

The density trick includes self-interactions, in which the same atom appears several times in a many-body term, for example $V_4(\br_i,\br_{j},\br_{j},\br_{k})$. This is not a problem, as the self-interactions are terms of lower body order that are included implicitly~\cite{drautzAtomicClusterExpansion2019}. The term $V_4(\br_i,\br_{j},\br_{j},\br_{k})$, for example, is a three-body contribution as it involves only the three atoms $i$, $j$ and $k$. The self-interactions can be removed if the many-atom expansion is to be reconstructed~\cite{hoAtomicClusterExpansion2024}. 

\subsection{Scaling with number of graph layers \label{sec:scalelayer}}

The linear scaling with the number of neighbors transfers to graphs with several layers. We illustrate this for the $\pmb{t} = (1233)$ graph in Fig.~\ref{fig:completebasis}, which is evaluated as
\begin{equation} 
A_{i v v_{2} v_{3} v_{4} }^{(1233)} =  \sum_{j_{1} j_{2}  j_{3}   j_{4}  } \phi_{i v}(j_{1}) \phi_{j_{1} v_{2}}(j_{2})   \phi_{j_{2} v_{3}}(j_{3})  \phi_{j_{2} v_{4}}(j_{4})  \,.
\end{equation}
The evaluation of the graph involves four nested sums over neighbors and for $N$ neighbors incurs a cost proportional to $N^4$, and $N^M$ for a graph with $M$ edges.

To avoid the power-law scaling, we evaluate the graph recursively, starting with the summations over the florets. The sums over the two florets on atom $j_2$ give the atomic base, Eq.(\ref{eq:abase}), which is evaluated for all required indices in a single summation over the $N$ neighbors of $j_2$ and leads to
\begin{equation} 
A_{i v v_{2} v_{3} v_{4} }^{(1233)} = \sum_{j_{1} j_{2}  } \phi_{i v}(j_{1}) \phi_{j_{1} v_{2}}(j_{2})  A_{j_{2} v_{3}}  A_{j_{2} v_{4}} \,.
\end{equation}
The reach of the graph is reduced by one layer, and the floret $\phi_{j_{1} v_{2}}(j_{2})$ is dressed with the atomic base, $\phi_{j_{1} v_{2}}(j_{2})  A_{j_{2} v_{3}}  A_{j_{2} v_{4}}$. Next we sum over the dressed floret,
\begin{equation}
A_{j_1 v_2 v_3 v_4}^{(122)}  = \sum_j  \phi_{j_{1} v_{2}}(j)  A_{j v_{3}}  A_{j v_{4}} \,,
\end{equation}
which requires another $N$ operations.
A further $N$ operations complete the evaluation
\begin{equation} 
A_{i v v_{2} v_{3} v_{4} }^{(1233)} =  \sum_{j_{1} }  \phi_{i v}(j_{1}) A_{j_1 v_2 v_3 v_4}^{(122)}\,,
\end{equation}
such that the graph with three layers is evaluated with $3 N$ operations. This example generalizes to more complex graphs. A graph with $M$ edges requires at most $M \times N$ summations, an upper bound that is reduced by avoiding duplicate computations, as for the two florets above.
Therefore the evaluation of graph basis functions is linear with the number of neighbors and linear with the number of layers of the graph. The same scaling holds for the gradients~\cite{bochkarevGraphAtomicCluster2024}.

As the density trick is applied on every layer, the recursive evaluation also includes self-interactions and redundant paths that return to an atom already visited, for example from atom $j_1$ to atom $j_2$ and back to atom $j_1$. These terms are the graph analogue of the self-interactions of star graphs, Sec.~\ref{sec:neighbors}. In the limit of a complete basis they can be removed and play no formal role. For a finite basis, however, they may affect the representation, as was shown for the self-interactions of ACE~\cite{hoAtomicClusterExpansion2024}.

\subsection{Scaling with body/graph order \label{sec:scaleorder}}

So far we considered the numbers of atoms and neighbors. The number of single-bond basis functions attached to each edge adds a further cost. For the four-atom interaction in Eq.(\ref{eq:E4}), for example, the effort scales as $N^3$ for $N$ single-bond basis functions at each edge, 
\begin{equation}
E_i^{(4)} =  \sum_{ v_1  v_2 v_3} c_{ v_1  v_2 v_3} A_{iv_1} A_{iv_2} A_{iv_3} \,.
\end{equation}
This extends to arbitrary body order and graphs, with $N^M$ operations for a graph with $M$ edges. The poor scaling would limit expansions to few basis functions and therefore to low body order and small graphs. There are three strategies for reducing the cost. First, atomic interactions are hierarchical, i.e. the contributions are expected to decay with increasing body order. At comparable accuracy, small contributions require fewer basis functions than large contributions, so that $N$ can be reduced with increasing body order. Second, optimizing the single-bond basis functions Eq.(\ref{eq:varphi}) lowers the required number of basis functions $N$, Sec.~\ref{sec:nlopt}. Both strategies make it possible to converge a basis expansion systematically up to body order five or six. However, at a given body order the cost of both strategies still scales as $N^M$, with a reduced $N$.

The third strategy represents the expansion coefficients $c_{ v_1  v_2 v_3}$ by sums of products of smaller tensors, for example as
\begin{equation}
c_{ v_1  v_2 v_3} = \sum_{k} \lambda_k c_{ v_1}^{(k)}  c_{ v_2}^{(k)} c_{ v_3}^{(k)} \,, \label{eq:decomp}
\end{equation}
with $k = 1, \dots, K$ channels for the tensor decomposition. The result is a representation
 \begin{equation}
 E_i^{(4)} = \sum_{k} \lambda_k   \tilde{A}_{ik}^3 \,,
\end{equation} 
that requires a linear effort in the number of neighbors and in the number of basis functions $v = 1, \dots, N$, because the atomic base is formed from effective single-bond basis functions in which $v$ is contracted,
\begin{equation}
 \tilde{A}_{ik} =  \sum_j  \tilde{\phi}_{i k}(j) \,, \,\,\,\text{and}\,\,\, \,
  \tilde{\phi}_{i k}(j)  =  \sum_{v} c_{ v}^{(k)} \phi_{i v}(j) \,.  \label{eq:TDA}
\end{equation}

By extending the range of $k$, the tensor decompositions of the individual body orders are grouped into a single decomposition across body orders. Then Eq.(\ref{eq:TDA}) holds for all body orders and the energy of a star-graph ACE can be written as
 \begin{equation}
E_i = \sum_n \sum_{k} \lambda_k^{(n)}    \tilde{A}_{ik}^n \,.
\end{equation} 
This means that the energy can be computed in principle with an effort that is linear with the number of channels $K$ required for tensor decomposition across body orders, linear with the number of neighbors, linear with the number of basis functions and, through the powers of $\tilde{A}_{ik}$, linear with the maximum body order. The dominant cost, the sum over neighbors in Eq.(\ref{eq:TDA}), is independent of the body order. In this form no expansion coefficients are required, and an optimal choice of the effective single-bond basis functions $ \tilde{\phi}_{i k}(j) $ is sufficient for a general parameterization of the energy. This is seen by using unnormalized functions $\tilde{\phi}_{i k}(j) $, such that the weights can be taken as $| \lambda_k^{(n)}| = 1$.

An analogous expression is derived for tree graphs. We illustrate this for the (123) graph,
\begin{align}
E_i^{(4)} &=  \sum_{ v i_1 v_1 i_2 v_2 i_3}   c_{ v  v_1 v_2} \phi_{iv}(i_1) \phi_{i_1v_1}(i_2) \phi_{i_2v_2}(i_3) \nonumber \\
                 &=  \sum_k \lambda_k \sum_{i_1}     \tilde{\phi}_{i k}(i_1)    \sum_{i_2}     \tilde{\phi}_{i_1 k}(i_2)  \sum_{i_3}     \tilde{\phi}_{i_2 k}(i_3)    \,, \label{eq:rec123}
\end{align}
which is evaluated recursively from right to left. This extends to arbitrary graphs, and with the notation of the recursive evaluation in Sec.~\ref{sec:scalelayer} the expansion reads
 \begin{equation}
E_i = \sum_{\pmb{t}} \sum_{k} \lambda_k^{(\pmb{t})}    \tilde{A}_{ik}^{(\pmb{t})} \,, \label{eq:reducedGRACE}
\end{equation}
where $\tilde{A}_{ik}^{(\pmb{t})}$ denotes the recursive evaluation of graph $\pmb{t}$ in terms of the effective basis functions $\tilde{\phi}_{ik}(j)$, cf. Eq.(\ref{eq:rec123}).

There are two caveats. First, the decomposition in Eq.(\ref{eq:decomp}) ignores that MLIPs need to be invariant under rotation and inversion. We will return to this point in Sec.~\ref{sec:symmetry}. Second, the efficiency of the tensor decomposition Eq.(\ref{eq:decomp}) depends on the rank of the decomposition, i.e. the number of terms in the sum over $k$ that is required to represent the tensors. In practice moderate ranks were sufficient for the models of Refs.~\cite{darbyTensorreducedAtomicDensity2023,hodappEquivariantTensorNetwork2024}. GRACE-3L, for example, uses 64 to 128 channels per product, Sec.~\ref{sec:Relations}. The linear scaling with body order and graph layers holds at fixed rank. Sec.~\ref{sec:chem} relates the basis functions to quantum mechanical matrix elements with only few relevant orbitals, which makes a small rank plausible.

\subsection{Scaling with number of chemical species} \label{sec:scale chem}

The indices $v$ of the single-bond basis functions are compound indices that contain all the information necessary to describe the basis functions. In a typical simulation without charge transfer or magnetism they are given by $v = ( \mu n l m)$, where $\mu$ labels the dependence of the basis function on chemical species, $n$ indexes radial functions and $l$ and $m$ specify the angular character. For the following short analysis we keep $v = (nlm)$ in a combined index but separate out the chemical index $\mu$, such that the single-bond basis function index reads $v \mu$ and the four-body energy is written as
\begin{equation}
E_i^{(4)} =  \sum_{ v_1  \mu_1 v_2 \mu_2  v_3 \mu_3 } c_{ v_1  \mu_1 v_2 \mu_2 v_3 \mu_3 } A_{iv_1 \mu_1} A_{iv_2 \mu_2 } A_{iv_3 \mu_3 } \,. 
\end{equation}
To isolate the cost of chemistry, we apply the tensor decomposition to the chemical index only and write
\begin{equation}
c_{ v_1  \mu_1 v_2 \mu_2 v_3 \mu_3} = \sum_k \lambda_{k v_1 v_2 v_3} w_{ v_1 \mu_1}^{(k)}  w_{ v_2 \mu_2}^{(k)} w_{ v_3 \mu_3}^{(k)} \,,
\end{equation}
from which we obtain
 \begin{equation}
 E_i^{(4)} = \sum_{ v_1 v_2 v_3} \sum_{k} \lambda_{k v_1 v_2 v_3}    \tilde{A}_{ik v_1}  \tilde{A}_{ik v_2}  \tilde{A}_{ik v_3}\,,
\end{equation} 
with
 \begin{equation}
   \tilde{A}_{ik v} = \sum_{j}   \tilde{\phi}_{ik v}(j) \,,\,\,\text{and} \,\,\,\,    \tilde{\phi}_{ik v}(j) = \sum_{\mu} w_{ v \mu}^{(k)} \phi_{iv \mu}(j)   \,. \label{eq:Atilde2}
 \end{equation} 
 
Therefore the energy can be written as
 \begin{equation}
E_i = \sum_{n \vii k} \lambda_{k\vii}^{(n)}    \prod_{v = v_1}^{v_n} \tilde{A}_{ikv} \,, \label{eq:energychem}
\end{equation} 
with $\vii = (v_1, v_2, \dots)$. This equation has no explicit dependence on chemical species, as chemistry is embedded in the atomic base in Eq.(\ref{eq:Atilde2}). Formally, the number of chemical species enters only linearly, through the embedding sum over $\mu$ in Eq.(\ref{eq:Atilde2}). Related low-dimensional chemical embeddings were introduced and applied in Refs.~\cite{willattFeatureOptimizationAtomistic2018,darbyTensorreducedAtomicDensity2023,lopanitsynaModelingHighentropyTransition2023}. The embedding is inexpensive compared to the tensor products in Eq.(\ref{eq:energychem}), and the number of chemical species is in any case bounded by the number of elements in the periodic table. In practice the cost of treating chemistry is therefore constant, Table~\ref{tab:scaling}. Foundation models that cover most of the periodic table remain fast~\cite{lysogorskiyGraphAtomicCluster2026,liangNEP89UniversalNeuroevolution2026}. As in Sec.~\ref{sec:scaleorder}, the embedding extends to tree graphs by tensor decomposition across the layers.

\subsection{Combined scaling with body order, graph layers and chemistry in recursive evaluation\label{sec:twostep}}

So far the cost was analyzed for a given graph. However, the expansion Eq.(\ref{eq:linear}) sums over the contributions of all admissible subgraphs of a complete graph. In this section we show how the evaluation of the subgraphs is combined. 

\begin{figure}[ht!]
    \includegraphics[width=0.45\textwidth]{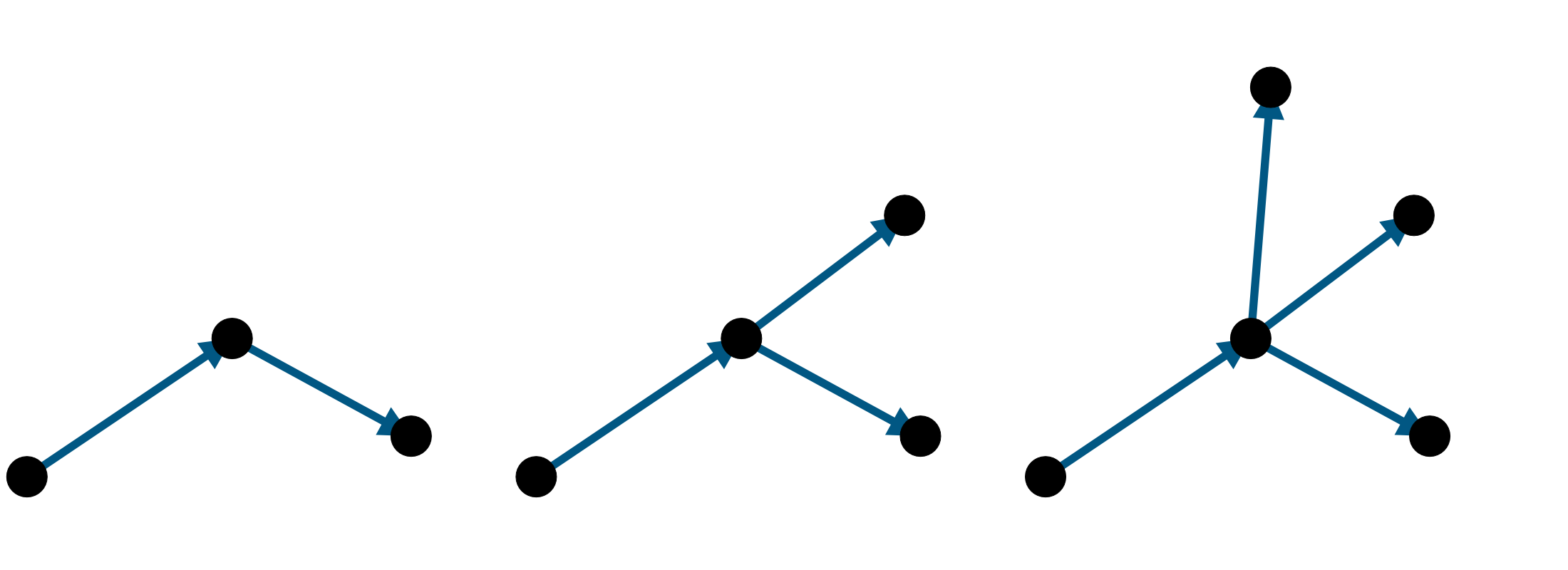}
    \caption{Three graphs that differ only in their florets.}
    \label{fig:combineflorets}
\end{figure}

According to Eq.(\ref{eq:reducedGRACE}), ACE and GRACE are evaluated from effective basis functions $\tilde{\phi}_{i k}(j)$. These effective basis functions differ between graphs and, within a graph, from edge to edge. This means that in principle every graph needs to be evaluated individually. However, as the graphs are subgraphs of a complete graph, they overlap, Fig.~\ref{fig:combineflorets}, and an efficient evaluation combines the identical parts as far as possible.
We illustrate the combined recursive evaluation for the $(1)$, $(12)$, $(123)$ and $(1233)$ graphs in Fig.~\ref{fig:completebasis}. The contributions of the four graphs to the expansion are
\begin{align}
E_i^{(1)} &=   \sum_{ v i_1 }   c_{ v }^{(1)} \phi_{iv}(i_1)  \,, \nonumber \\
E_i^{(12)} &=   \sum_{ v i_1 v_1 i_2 }   c_{ v  v_1 }^{(12)} \phi_{iv}(i_1) \phi_{i_1v_1}(i_2) \,,  \nonumber \\
E_i^{(123)} &=   \sum_{ v i_1 v_1 i_2 v_2 i_3}   c_{ v  v_1 v_2}^{(123)} \phi_{iv}(i_1) \phi_{i_1v_1}(i_2) \phi_{i_2v_2}(i_3) \nonumber \\
E_i^{(1233)} &=   \sum_{ v i_1 v_1 i_2 v_2 i_3 v_3 i_4}   c_{ v  v_1 v_2 v_3}^{(1233)} \phi_{iv}(i_1) \phi_{i_1v_1}(i_2) \phi_{i_2v_2}(i_3) \phi_{i_2v_3}(i_4)\,.  \nonumber 
\end{align} 
Subgraphs of the complete graph are obtained by assigning the constant basis function $\phi_{i0}(j)=1$ to edges, Sec.~\ref{sec:completebasis}. With the $(1233)$ graph as the complete graph, the four contributions read
\begin{align}
E_i^{(1)} &=   \sum_{ v i_1  i_2  i_3 i_4}   c_{ v000}^{(1233)} \phi_{iv}(i_1)   \phi_{i_1 0}(i_2) \phi_{i_2 0}(i_3)  \phi_{i_2 0}(i_4)\,, \nonumber \\
E_i^{(12)} &=   \sum_{ v i_1 v_1 i_2 i_3 i_4}   c_{ v  v_100 }^{(1233)} \phi_{iv}(i_1) \phi_{i_1v_1}(i_2) \phi_{i_20}(i_3)  \phi_{i_2 0}(i_4)\,,  \nonumber \\
E_i^{(123)} &=   \sum_{ v i_1 v_1 i_2 v_2 i_3 i_4}   c_{ v  v_1 v_20}^{(1233)} \phi_{iv}(i_1) \phi_{i_1v_1}(i_2) \phi_{i_2v_2}(i_3)  \phi_{i_2 0}(i_4) \nonumber \\
E_i^{(1233)} &=   \sum_{ v i_1 v_1 i_2 v_2 i_3 v_3 i_4}   c_{ v  v_1 v_2 v_3}^{(1233)} \phi_{iv}(i_1) \phi_{i_1v_1}(i_2) \phi_{i_2v_2}(i_3) \phi_{i_2v_3}(i_4) \,,  \nonumber 
\end{align} 
with re-normalized expansion coefficients $c_{ v }^{(1)} =  c_{ v000 }^{(1233)}  N^3 $, $c_{ v v_1 }^{(12)} =  c_{ v v_1 00 }^{(1233)}  N^2$ and  $c_{ v v_1 v_2}^{(123)} =  2 c_{ v v_1 v_2 0 }^{(1233)}  N$, where $N$ is the number of neighbors, taken as constant here. The factor 2 accounts for the two equivalent ways of obtaining the $(123)$ graph by picking one of the two outermost florets, with expansion coefficients symmetric in the last two indices, $c_{ v v_1 v_2 v_3 }^{(1233)} = c_{ v v_1 v_3 v_2 }^{(1233)}$. The combined contribution of the four graphs is now written as
\begin{align}
&E_i = E_i^{(1)} + E_i^{(12)}+ E_i^{(123)}  + E_i^{(1233)}\nonumber \\
&=  \sum_{ v i_1 i_2 i_3 i_4}     \sum_{ v_1v_2v_3= 0} c_{ v  v_1 v_2 v_3}^{(1233)} \phi_{iv}(i_1) \phi_{i_1v_1}(i_2) \phi_{i_2v_2}(i_3) \phi_{i_2v_3}(i_4)\,,
\end{align} 
where the sum over $v_1, v_2$ and $ v_3$ includes 0. Terms with internal zeros, such as $v_1 = 0$ together with $v_2 \neq 0$ or $v_3 \neq 0$, correspond to fragmented graphs that are not admissible and are excluded from the summation, Sec.~\ref{sec:completebasis}. A tensor decomposition of the form
\begin{equation}
c_{ v v_1 v_2 v_3}^{(1233)}  = \sum_k \lambda_{k} c_{ kv}^{(0)} c_{k v_1}^{(1)} c_{k v_2 v_3}^{(2)} \,,
\end{equation}
with $c_{k v_2 v_3}^{(2)} = c_{k v_3 v_2}^{(2)}$, leads to
\begin{align}
&E_i = \sum_k \lambda_{k}  \sum_{ v i_1 }  c_{k v}^{(0)} \phi_{iv}(i_1)  \nonumber \\
&\times \bigg( N^3 c_{k0}^{(1)} c_{k00}^{(2)}  +     \sum_{v_1 i_2}  c_{k v_1}^{(1)} \phi_{i_1v_1}(i_2) \nonumber \\
&\times  \Big[ N^2 c_{k00}^{(2)}  + 2 N \sum_{v_2 i_3}     c_{k v_2 0}^{(2)} \phi_{i_2v_2}(i_3)  \nonumber \\
&+ \sum_{v_2 v_3  i_3 i_4}     c_{ kv_2 v_3}^{(2)} \phi_{i_2v_2}(i_3) \phi_{i_2v_3}(i_4) \Big] \bigg)\,,
\end{align}
where the remaining sums over $v_1$, $v_2$ and $v_3$ run over non-zero indices. We now rearrange the evaluation layer by layer. The florets of the outermost, third layer, enclosed by square brackets, are grouped as
\begin{align}
 \varphi_{i_2k}^{(2)} &= N^2 c_{k00}^{(2)}  + 2 N \sum_{v_2 i_3}     c_{k v_2 0}^{(2)} \phi_{i_2v_2}(i_3)  \nonumber \\ 
&+ \sum_{v_2 v_3  i_3 i_4}     c_{k v_2 v_3}^{(2)} \phi_{i_2v_2}(i_3) \phi_{i_2v_3}(i_4)  \nonumber \\ 
&= N^2 c_{k00}^{(2)}  + 2 N \sum_{v_2}   c_{k v_2 0}^{(2)} A_{i_2v_2} +   \sum_{v_2 v_3}     c_{k v_2 v_3}^{(2)}A_{i_2v_2}A_{i_2v_3} \,.
\end{align}
This is a star graph ACE up to body order three, cf. Sec.~\ref{sec:neighbors}. We next define 
\begin{align}
 \varphi_{i_1k}^{(1)} &=  N^3 c_{k0}^{(1)} c_{k00}^{(2)}  +     \sum_{v_1 i_2}  c_{ kv_1}^{(1)} \phi_{i_1v_1}(i_2)  \varphi_{i_2k}^{(2)} \nonumber \\
&=  N^3 c_{k0}^{(1)} c_{k00}^{(2)}  +     \sum_{v_1 }  c_{k v_1}^{(1)} A_{i_1kv_1}^{(2)}  \,,
\end{align}
with the layer-dependent atomic base
\begin{equation}
A_{i_1kv_1}^{(2)}  =  \sum_{i_2} \phi_{i_1v_1}(i_2)  \varphi_{i_2k}^{(2)} \,.
\end{equation}
The third layer of the $(1233)$ graph carries two florets, and $ \varphi_{i_2k}^{(2)}$ is therefore a three-body ACE. The remaining $(12)$ graph carries one floret, and $ \varphi_{i_1k}^{(1)} $ is therefore a two-body ACE. The last floret is the one of the $(1)$ graph,
\begin{equation}
 \varphi_{ik}^{(0)} = \sum_{ v }  c_{k v}^{(0)} A_{ikv}^{(1)} \,,
\end{equation}
where $ A_{ikv}^{(1)} $ is defined in analogy to $A_{i_1kv_1}^{(2)}$. Finally, we obtain
\begin{equation}
E_i = \sum_k \lambda_{k}   \varphi_{ik}^{(0)} \,.
\end{equation}
This analysis extends to general tree graphs. An efficient evaluation combines the layer-wise tensor decomposition of the expansion coefficients with the layer-wise evaluation, Secs.~\ref{sec:scalelayer} and \ref{sec:scaleorder}. All subgraphs that differ only in the florets on the outermost layer are collected, Fig.~\ref{fig:combineflorets}, and the florets are summed together with the corresponding factors of the tensor-decomposed expansion coefficients. This leads to a two-step recursion. The recursion is initialized by computing the atomic base on the outermost layer $t_{max}$ according to Eqs.(\ref{eq:abase}) and (\ref{eq:Atilde2}), $A^{(t_{max})}_{j k v} =  A_{j k v}$. First, the florets are combined into an ACE on each layer,
 \begin{align}
\varphi^{(t)}_{jk} &=  c^{(t,0)}_{k }  +  \sum_v c^{(t,1)}_{k v}   A^{(t+1)}_{j k v} + \sum_{v_1 v_2}  c^{(t,2)}_{k v_1 v_2}   A^{(t+1)}_{j k v_1}  A^{(t+1)}_{j k v_2} + \dots  \,,\label{eq:layerACE}
\end{align}
where $t$ is the layer index. A further tensor decomposition could be applied to the expansion coefficients $c^{(t,n)}_{k v_1 v_2 \dots v_n}$, but as chemistry is embedded in the single-bond basis functions and the star graphs on a layer typically have only three or four edges, in practice it is often not required.  

Second, the resulting ACE dresses the next inner layer of the graph, which reduces the number of layers by one. To this end a modified atomic base is computed as
\begin{equation}
A^{(t)}_{i k v} = \sum_{j} \phi^{(t)}_{iv}(j) \varphi^{(t)}_{jk} \,.\label{eq:message}
\end{equation} 
Here the information assembled in $\varphi^{(t)}_{jk}$ is passed from atom $j$ to atom $i$ through the single-bond basis functions $ \phi^{(t)}_{iv}(j)$. Note that $ \varphi^{(t)}_{jk}$ carries only the tensor decomposition index $k$, which is maintained across the layers, and there is no equivariant coupling. Equivariant coupling is needed only for rotational symmetry, Sec.~\ref{sec:symmetry}.

\begin{figure}[ht!]
    \includegraphics[width=0.5\textwidth]{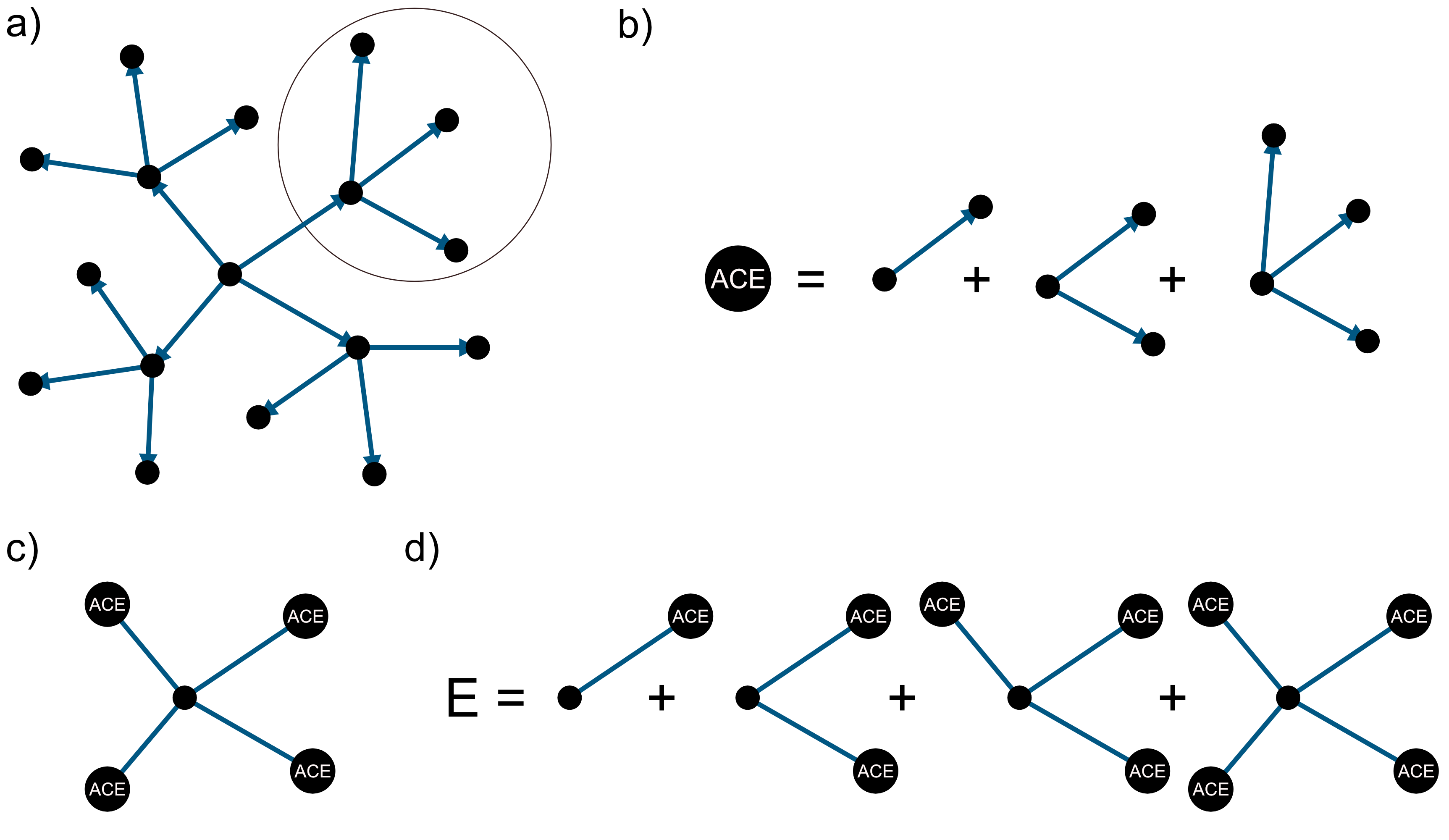}
    \caption{Recursive evaluation of graphs with several layers. a) The subgraphs in the outermost layer are separated. b) An ACE is evaluated for the separated subgraphs. In equivariant models the ACE carries scalar, vectorial and tensorial contributions, Sec.~\ref{sec:equirec}. c) The ACE replaces the subgraphs of the outermost layer. d) A final ACE is evaluated for the reduced graph. In a graph with more layers, ACE is carried out on each layer.}
    \label{fig:ACEmessage}
\end{figure}

The recursion ends with the computation of the energy
\begin{equation}
E_i = \sum_k \lambda_k \varphi^{(0)}_{ik} \,. \label{eq:recfin}
\end{equation}
Fig.~\ref{fig:ACEmessage} illustrates the two-step recursion. With information passed along the edges in Eq.(\ref{eq:message}) and combined on the nodes in Eq.(\ref{eq:layerACE}), the recursive evaluation of GRACE has the structure of a message-passing graph neural network.
In practice the recursion is slightly more complex because of rotational invariance, Sec.~\ref{sec:equirec}.  %

\section{Symmetry \label{sec:symmetry}}

\begin{framed}
{\it{Summary:}} The energy does not change when a structure is translated, rotated or inverted, or when two atoms of the same chemical species are exchanged. These symmetries can be built into an MLIP exactly. A quantity that does not change under rotation is called invariant, and a quantity that rotates with the structure is called equivariant. Equivariant intermediates carry angular information through a model before it is combined into the invariant energy. The canonical tools are spherical harmonics and angular momentum coupling. An equivalent algebra can be formulated in Cartesian coordinates, and some models instead learn rotational symmetry approximately from training data. Building in symmetry is also a matter of efficiency, as a symmetric expansion requires fewer expansion coefficients and no augmented training data, and the exact angular coupling can be contracted efficiently.
\end{framed}

\subsection{Translation and permutation}

MLIPs generally incorporate translational invariance by using difference vectors or distances between atoms. The great majority of MLIPs are also constructed to be invariant under the exchange of two atoms of the same chemical species. This is usually achieved by first decomposing the energy into atomic contributions and then formulating the atomic energy to be invariant under the exchange of two atoms of the same species. The GRACE expansion Eq.(\ref{eq:linear}) is invariant under permutation for the same reason, as the summation over neighbors in the evaluation of the basis functions, Eq.(\ref{eq:abase}), together with the graph symmetrization of Sec.~\ref{sec:complete}, symmetrizes the expansion with respect to the exchange of atoms of the same chemical species.

In contrast, classical force fields such as CHARMM and AMBER are not permutation invariant~\cite{brooksCHARMMProgramMacromolecular1983,weinerAllAtomForce1986}. These force fields assign fixed types to atoms according to their bonding environment, so that two atoms of the same chemical species are in general not equivalent and the potential is not invariant under their exchange. Assigning atom types simplified the parameterization of complex molecules at a time when flexible functional forms for distinguishing bonding environments were not available.

\subsection{Rotation and inversion} \label{sec:rotinv}

Invariance under rotation and inversion needs two ingredients. The first are functions with a known behavior under rotation, the spherical harmonics, and the second is a rule for multiplying them such that products again have a known behavior, the angular momentum coupling. Spherical harmonics are basis functions of the irreducible representations of the group of rotations and inversions in three dimensions O(3). The angular momentum $l$ indexes the irreducible representation and the magnetic quantum number $m$ the basis function within the representation. Under inversion of the argument, $\hat{r} \to -\hat{r}$, spherical harmonics are multiplied by their parity $p = (-1)^l$. Spherical harmonics are rotated with matrices that only mix basis functions within a representation,
\begin{equation}
\hat{R} Y_{lm} = \sum_{m'=-l}^l D_{m m'}^{(l)} Y_{lm'} \,,
\end{equation}
where $\hat{R}$ is a rotation operator and $D^{(l)}$ a Wigner $D$-matrix, with the index order of $D_{m m'}$ transposed relative to Refs.~\onlinecite{Wigner31,varshalovichQuantumTheoryAngular1988}.  The product of two spherical harmonics rotates as
\begin{equation}
\hat{R} \, Y_{l_1 m_1} Y_{l_2 m_2} = \sum_{m'_1=-l_1}^{l_1} \sum_{m'_2=-l_2}^{l_2} D_{m_1 m'_1}^{(l_1)} D_{m_2 m'_2}^{(l_2)} Y_{l_1 m'_1} Y_{l_2 m'_2} \,.
\end{equation}
The product representation is reducible, and angular momentum coupling brings it to block-diagonal form,
\begin{equation}
 D^{(l_1)} \otimes D^{(l_2)} \cong \bigoplus_{l = | l_1 - l_2|}^{l_1 + l_2} D^{(l)} \,, \label{eq:angularcoupling}
\end{equation}
where $\cong$ denotes equivalence of the representations, i.e. equality up to a unitary transformation, which is built from Clebsch-Gordan coefficients. The tensor product on the left hand side has $(2 l_1 +1)^2(2 l_2 +1)^2$ elements, while the direct sum on the right hand side is block-diagonal with $ \sum_{l = | l_1 - l_2|}^{l_1 + l_2} (2l+1)^2$ non-zero matrix elements. For example, for $l_1 = l_2 = 1$ we have $3^2 3^2 = 81$ matrix elements on the left and only $1^2 + 3^2 + 5^2 = 35$ matrix elements that are non-zero on the right. The Clebsch-Gordan coefficients couple products of spherical harmonics to a definite angular character,
\begin{equation}
Y^{(lm)}_{l_1 l_2} = \sum_{m_1=-l_1}^{l_1} \sum_{m_2=-l_2}^{l_2} C^{lm}_{l_1 m_1 l_2 m_2} Y_{l_1 m_1} Y_{l_2 m_2} \,, \label{eq:CG}
\end{equation}
such that $Y^{(lm)}_{l_1 l_2}$ transforms under simultaneous rotation of both arguments in the same way as a single spherical harmonic $Y_{lm}$ and carries the parity of the product, $p = p_1 p_2 = (-1)^{l_1 + l_2}$, in general different from the parity $(-1)^l$ of $Y_{lm}$. The indices $l_1 l_2$ are kept, as different couplings to the same $l$ and $p$ result in general in different, linearly independent functions.

Note that reducing a product of two spherical harmonics to a single spherical harmonic is valid only if both have the same argument. In our case the spherical harmonics $Y_{l_1 m_1}$ and $Y_{l_2 m_2}$ depend on different interatomic direction vectors. Therefore we keep the product and use the coupling only to characterize its transformation properties under simultaneous rotation and inversion.

The Clebsch-Gordan coefficients are unitary and sparse. They are different from zero only if $| l_1 - l_2 | \leq l \leq l_1 + l_2$ and $m = m_1 + m_2$. The coupling is extended to products of more than two spherical harmonics by applying it recursively, which generates generalized Clebsch-Gordan coefficients that couple a product of an arbitrary number of spherical harmonics to a definite angular character. For example, for a product of three spherical harmonics
\begin{align}
Y^{(lm)}_{l_1 l_2 l_3} &= \sum_{m_1=-l_1}^{l_1} \sum_{m_2=-l_2}^{l_2} \sum_{m_3=-l_3}^{l_3} C^{lm}_{l_1 m_1 l_2 m_2 l_3 m_3} \nonumber \\
& \times Y_{l_1 m_1} Y_{l_2 m_2} Y_{l_3 m_3} \,,
\end{align}
with parity $p = (-1)^{l_1 + l_2 + l_3}$, and analogously for higher order products.
Generalized Clebsch-Gordan coefficients are obtained from products of Clebsch-Gordan coefficients, for example as,
\begin{equation}
C^{\tau lm}_{l_1 m_1 l_2 m_2 l_3 m_3} = \sum_{m' m''} C^{lm}_{l_1 m_1 l' m'} C^{l'm'}_{l_2 m_2 l'' m''} C^{l''m''}_{l_3 m_3 0 0} \label{eq:CGprod} \,,
\end{equation}
where the intermediate angular momenta $l', l'', \dots$ form the multiplicity index $\tau$. Sequential coupling generates more basis functions than are required for a complete representation, as it is not adapted to the permutation symmetry of the products it acts on~\cite{yutsisMathematicalApparatusTheory1962,dussonAtomicClusterExpansion2022,drautzAtomicClusterExpansion2020,bochkarevGraphAtomicCluster2024,goffPermutationadaptedCompleteIndependent2024,goffYoungE3TensorProduct2026}. These extra basis functions are, however, tolerated, as the product form facilitates the tensor decomposition of the graph expansion coefficients~\cite{bochkarevGraphAtomicCluster2024}, Eq.(\ref{eq:symC-TD}). 

In general, the generalized Clebsch-Gordan coefficients for coupling $K$ spherical harmonics $l_1, l_2, \dots, l_K$ are written as
\begin{equation}
C^{\tau lm}_{l_1 m_1 l_2 m_2 \dots l_K m_K} \label{eq:CGgen} \,,
\end{equation}
where the multiplicity index $\tau$ again labels the different ways to achieve the same coupling $l$. 

\subsection{Representation of expansion coefficients}

The energy is invariant under rotation and inversion. For the representation of the energy therefore only basis functions that couple to angular momentum $l=0$ and even parity, i.e. $\sum_i l_i$ even, are relevant. Coupling to $l=0$ alone is not sufficient. For example, three $l_i = 1$ contributions can couple to a pseudoscalar, $\pmb{v}_1 \cdot (\pmb{v}_2 \times \pmb{v}_3)$, the determinant of three vectors in three dimensions, which is invariant under rotation but changes sign under inversion. For a vectorial quantity one requires instead $l=1$ with odd parity, and for a symmetric traceless second-rank tensor $l=2$ with even parity.

We next apply the coupling to the linear combination of atomic orbitals (LCAO)-type single-bond basis functions Eq.(\ref{eq:varphiSH}). We multiply the basis expansion with generalized Clebsch-Gordan coefficients, which is a linear transformation of the product basis and characterizes the product basis functions by their resulting angular momentum. 
For a four-body star with three edges as in Eq.(\ref{eq:E4}) the product of basis functions is then written as,
\begin{align}
B_{i n_1 l_1 n_2 l_2 n_3 l_3}^{(\tau lm)} &= \sum_{m_1 m_2 m_3} C^{\tau lm}_{l_1 m_1 l_2 m_2 l_3 m_3} \nonumber \\
& \times  A_{i n_1 l_1 m_1}  A_{i n_2 l_2 m_2}  A_{i n_3 l_3 m_3} \,.
\end{align}
The invariant form of the four-body energy Eq.(\ref{eq:E4}) is obtained by keeping only the basis functions that couple to $l=0$ with even parity, while all other products do not contribute~\cite{drautzAtomicClusterExpansion2019,dussonAtomicClusterExpansion2022},
\begin{equation}
E_i^{(4)} = \sum_{\tau n_1 l_1 n_2 l_2 n_3 l_3}   c_{\tau n_1 l_1 n_2 l_2 n_3 l_3} B_{i n_1 l_1 n_2 l_2 n_3 l_3}^{(\tau 00)} \,,
\end{equation}
with expansion coefficients $ c_{\tau n_1 l_1 n_2 l_2 n_3 l_3}$. By comparing to the representation Eq.(\ref{eq:E4}) that does not make any assumptions on rotational symmetry, one sees that a rotationally invariant representation of $E_i^{(4)}$  is achieved by employing expansion coefficients of the form
\begin{equation}
c_{n_1 l_1 m_1 n_2 l_2 m_2 n_3 l_3 m_3} =  \sum_\tau c_{\tau n_1 l_1 n_2 l_2 n_3 l_3} C^{\tau00}_{l_1 m_1 l_2 m_2 l_3 m_3} \,.
\end{equation}
This analysis extends to arbitrary body and graph order. This means that in rotationally invariant models expansion coefficients $c_{n_1 l_1 m_1 n_2 l_2 m_2 n_3 \dots}$ take the form
\begin{equation}
c_{n_1 l_1 m_1 n_2 l_2 m_2 \dots} =  \sum_\tau c_{\tau n_1 l_1 n_2 l_2 \dots} C^{\tau00}_{l_1 m_1 l_2 m_2 \dots} \,,\label{eq:symC}
\end{equation}
where the expansion coefficients $c_{\tau n_1 l_1 n_2 l_2 \dots} $ are independent training parameters and the coupling is restricted to even parity, $\sum_i l_i$ even.

To facilitate the recursive evaluation of the basis functions, we write the generalized Clebsch-Gordan coefficients as a product of Clebsch-Gordan coefficients, Eq.(\ref{eq:CGprod}), which admits more basis functions than a minimal complete set, Sec.~\ref{sec:rotinv}. If the expansion coefficients $c_{\tau n_1 l_1 n_2 l_2 \dots}$ are in addition tensor decomposed as in Eq.(\ref{eq:decomp}), the resulting expansion coefficients can be distributed over layers,
\begin{align}
c^{(t_1 t_2 t_3)}_{n_1 l_1 m_1 n_2 l_2 m_2 n_3 l_3 m_3} &= \sum_{k}  \lambda_{k} c^{(k,t_1)}_{n_1 l_1}  \sum_{l' m'}  C^{00}_{l_1 m_1 l' m'} \nonumber \\ \times  c^{(k,t_2)}_{n_2 l_2}  \sum_{ l'' m''}  & C^{l'm'}_{l_2 m_2 l'' m''}   c^{(k,t_3)}_{n_3 l_3}  C^{l''m''}_{l_3 m_3 0 0} \,, \label{eq:symC-TD}
\end{align}
where $t_1, t_2, t_3$ are layer indices that describe the graph topology. For the three edges shown, the intermediate sums collapse to a single invariant and no multiplicity is lost. For higher orders the intermediate angular momenta are retained in the channel index, $k \to (k, l', l'', \dots)$, which absorbs $\tau$ into $k$. In addition, slots with equal angular momentum require different factors $c^{(k,t)}$, as couplings that are antisymmetric under their exchange vanish for identical factors~\cite{darbyTensorreducedAtomicDensity2023}. The products and sums are further reordered to prepare for the recursive evaluation. 

\subsection{Equivariant recursive evaluation \label{sec:equirec}}

The recursive evaluation of GRACE in Sec.~\ref{sec:twostep} can now be formulated in equivariant form. To this end we expose the angular indices $lm$, which determine the behavior under rotation. Then Eq.(\ref{eq:layerACE}) becomes
 \begin{align}
&\varphi^{(t)}_{jknlm} =  c^{(t,0)}_{knlm}  +  \sum_{n_1 l_1 m_1}  c^{(t,1)}_{k nlm n_1 l_1 m_1}   A^{(t+1)}_{j k n_1 l_1 m_1} \nonumber \\
&+ \sum_{n_1 l_1 m_1 n_2 l_2 m_2}  c^{(t,2)}_{k nlm n_1 l_1 m_1  n_2 l_2 m_2}   A^{(t+1)}_{j k n_1 l_1 m_1}  A^{(t+1)}_{j k n_2 l_2 m_2} + \dots  \,.\label{eq:layerACEsym}
\end{align}
The radial and chemical indices $n, n_1, n_2, \dots$ could be absorbed into the basis functions as discussed in Secs.~\ref{sec:scaleorder} and \ref{sec:scale chem}, and are kept here for clarity. The expansion coefficients channel the input angular momenta $l_1 l_2 \dots$ to the target output $lm$, in analogy to Eq.(\ref{eq:symC}) with the coupling to $00$ replaced by the coupling to $lm$. Equivariance further requires that the constant term is limited to the invariant channel, $c^{(t,0)}_{knlm} \propto \delta_{l0} \delta_{m0}$, and that the linear term reduces to $c^{(t,1)}_{knlm\, n_1 l_1 m_1} \propto \delta_{l l_1} \delta_{m m_1}$.

The recursion is initialized by computing the atomic base on the outermost layer $t_{max}$, $A^{(t_{max})}_{j k nlm} =  A_{j k nlm}$. As before, the resulting ACE is used to dress the next inner layer in the graph, which reduces the number of layers in the graph by one. This is achieved by computing a modified atomic base as,
\begin{align}
A^{(t)}_{i k nlm} &= \sum_{n_1 l_1 m_1 n_2 l_2 m_2}  \sum_{j} C^{(t)}_{knlm n_1 l_1 m_1 n_2 l_2 m_2} \nonumber \\
& \times \phi^{(t)}_{i n_1 l_1 m_1}(j) \varphi^{(t)}_{jk n_2 l_2 m_2} \,.\label{eq:messagesym}
\end{align} 
Different from Eq.(\ref{eq:message}), the angular characters $l_1$ and $l_2$ are contracted into a single index $l$, so that the atomic base keeps a definite angular character across the layers. The tensor $C^{(t)}$ can be taken as a Clebsch-Gordan coefficient, in which case the contraction is a unitary transformation if all admissible output channels $l$ are kept. However, the coupling may also be used to combine radial contributions, if these are not embedded in the radial basis functions, as the radial indices of $C^{(t)}$ in Eq.(\ref{eq:messagesym}) indicate.

The recursion ends with the computation of the energy
\begin{equation}
E_i = \sum_{kn} \lambda_{kn} \varphi^{(0)}_{ikn00} \,, \label{eq:recfinsym}
\end{equation}
where only the invariant contributions to $\varphi^{(0)}$, with $l=m=0$ and even parity, are kept. For vectorial and tensorial quantities, the contributions with $l=1$ and odd parity or with $l=2$ and even parity are kept instead.

\subsection{Scaling of angular coupling} \label{sec:angscaling}

The direct evaluation of the Clebsch-Gordan coupling, Eq.(\ref{eq:CG}), requires of the order of $l_{\max}^5$ operations, counting the non-zero coupling coefficients. This cost has been regarded as a computational bottleneck and has motivated giving up exact rotational symmetry~\cite{rhodesOrbv3AtomisticSimulation2025} or employing Cartesian coordinates~\cite{simeonTensorNetCartesianTensor2023}.

However, the Clebsch-Gordan coupling of Eq.(\ref{eq:CG}) can be evaluated exactly at a much lower cost \cite{bochkarevFastContractedClebschGordan2026,xieAsymptoticallyFastClebschGordan2026}, and restricted couplings allow further speed-ups \cite{passaroReducingSO3Convolutions2023,luoEnablingEfficientEquivariant2024}. We illustrate this for the parity-even elements of the Clebsch-Gordan tensor, which can be represented by the Gaunt integral over three spherical harmonics,
\begin{align}
&\int Y_{l_1 m_1}(\Omega) Y_{l_2 m_2}(\Omega) Y_{l m}^*(\Omega) \, d\Omega \nonumber \\
&= \sqrt{\frac{(2l_1+1)(2l_2+1)}{4\pi(2l+1)}} C_{l_1 0 l_2 0}^{l 0}  C_{l_1 m_1 l_2 m_2}^{l m} \,.
\end{align}
The parity-odd contribution is represented analogously using vector spherical harmonics \cite{xieAsymptoticallyFastClebschGordan2026} or Poisson brackets on the unit sphere \cite{bochkarevFastContractedClebschGordan2026,heyraudIntegralFormulasVector2026}. The integral over the sphere is replaced by a discrete sum, for example over a tensor-product grid, which exploits the separation of variables in the spherical harmonics
\begin{align}
& \int Y_{l_1 m_1} Y_{l_2 m_2} Y_{l m}^* \, d\Omega \nonumber \\
&= \sum_{u=1}^{U} w_u p_{u l_1 m_1} p_{u l_2 m_2} p_{u l m} \left[ \frac{2\pi}{V} \sum_{v=1}^{V} q_{v m_1} q_{v m_2} q_{v m}^* \right] \,,
\end{align}
with
\begin{equation}
p_{ulm} = \tilde{P}_l^m(\cos\theta_u) \,\,\,\, \mathrm{and} \,\,\,\, q_{vm} = e^{im\phi_v} \,,
\end{equation}
where $\tilde{P}_l^m$ are the normalized associated Legendre polynomials, $Y_{lm} = \tilde{P}_l^m(\cos\theta)\, e^{im\phi}$, and the grid consists of $U$ Gauss-Legendre quadrature points in $\cos\theta_u$ with weights $w_u$ and $V$ uniformly spaced points for the azimuthal angle $\phi_v = 2\pi v / V$. For an exact integration, one requires $U \ge (3 l_{\max} + 1)/2$ and $V \ge 3 l_{\max} + 1$. The total number of grid points is $U \times V = \mathcal{O}(l_{\max}^2)$.

The grid sum replaces the sparse Clebsch-Gordan contraction. Combined with the tensor decomposition of the expansion coefficients, the exact equivariant coupling is then evaluated at a cost of $l_{\max}^3$ at fixed decomposition rank, compared to $l_{\max}^5$ for the direct contraction~\cite{bochkarevFastContractedClebschGordan2026}. Cubic cost was also reached with Gaunt products and $SO(2)$ convolutions~\cite{luoEnablingEfficientEquivariant2024,passaroReducingSO3Convolutions2023}. In benchmarks the wall-clock time grows only as $l_{\max}^2$ up to the values of $l_{\max}$ used in practice, because for message passing the evaluation is limited by memory access to the $\mathcal{O}(l_{\max}^2)$ grid tensors rather than by arithmetic, while the on-site contraction approaches the $l_{\max}^3$ scaling only at large $l_{\max}$~\cite{bochkarevFastContractedClebschGordan2026}. The cost of the exact equivariant coupling is therefore comparable to the cost of the other dense operations in an MLIP, and there is no evidence that the cost of the Clebsch-Gordan coupling requires giving up exact rotational symmetry or spherical harmonics. This argument does not affect other reasons for not enforcing symmetry, such as unconstrained non-linearities and ease of training, Secs.~\ref{sec:learnsym} and \ref{sec:MLinMLIPs}.

\subsection{Cartesian basis} \label{sec:cartesian}

Rotational symmetry can also be built in without spherical harmonics. The moment tensor potential~\cite{shapeevMomentTensorPotentials2016} introduced a complete, in fact overcomplete, set of basis functions in Cartesian coordinates in the form
\begin{equation}
f_1(r) \hbr,\,\,\,  f_2(r) \hbr \otimes \hbr, \,\,\,  f_3(r) \hbr \otimes \hbr \otimes \hbr, \,\, \dots \,. \label{eq:carttensor}
\end{equation}
This representation is linked directly to the LCAO-type basis functions Eq.(\ref{eq:varphiSH}), as a unit vector can be expressed as
\begin{equation}
\hbr_n = \sum_m a_{nm} Y_{1m} \,,
\end{equation}
with a $3\times3$ matrix $a$~\cite{drautzAtomicClusterExpansion2020}. The Cartesian tensors Eq.(\ref{eq:carttensor}) therefore form a reducible basis of the rotation group. The unit vector $\hbr$ transforms as $D^{(1)}$, and the general tensor products reduce as $D^{(1)} \otimes D^{(1)} \cong  D^{(0)} \oplus  D^{(1)} \oplus  D^{(2)}$ and $D^{(1)} \otimes D^{(1)} \otimes D^{(1)} \cong D^{(0)} \oplus  3   D^{(1)} \oplus  2 D^{(2)} \oplus  D^{(3)}$. The moment tensors are symmetric, so that $\hbr \otimes \hbr$ carries only $D^{(0)} \oplus D^{(2)}$ and $\hbr \otimes \hbr \otimes \hbr$ only $D^{(1)} \oplus D^{(3)}$, and as $\hbr \cdot \hbr = 1$ the lower components reduce to lower-rank moments. The Cartesian basis may appear an attractive alternative to spherical harmonics. However, the spherical harmonics are themselves polynomials in the Cartesian components of $\hbr$ with efficient recursive relations~\cite{bigiFastEvaluationSpherical2023}, and a symmetric rank-$l$ Cartesian tensor has $(l+1)(l+2)/2$ independent components, and a general one $3^l$, compared to $2l+1$ for the spherical harmonics. Spherical expansions are therefore more compact at higher ranks, unless the Cartesian tensors are reduced to their irreducible, traceless parts.

The Cartesian formulations of ACE in CACE and TACE were published recently~\cite{chengCartesianAtomicCluster2024,xuSpectralSpatialTensor2025} and reach high efficiency. The cost of the Clebsch-Gordan coupling, a motivation for Cartesian formulations, can be reduced to the cost of dense operations, Sec.~\ref{sec:angscaling}. Spherical and Cartesian coordinates both enable efficient implementations, and implementation details are more important than the seemingly fundamental difference between the two.

\subsection{Learning rotational symmetry from data} \label{sec:learnsym}

Instead of building rotational invariance and invariance under inversion into the model, one can also start from the non-symmetrized representation of Sec.~\ref{sec:twostep} and symmetrize the data. To this end the training dataset is augmented. Copies of the reference structures are rotated and inverted at random and added to the training data with identical energies and correspondingly rotated forces. Exact symmetrization by augmentation is achieved with a quadrature rule with a degree at least equal to the polynomial degree of the model, and the number of quadrature points grows at least with the cube of this degree, while random augmentation converges only as the inverse square root of the number of copies~\cite{klintebackHighCostData2026}. This multiplies the training data and with it the cost of training.

Not enforcing symmetry explicitly removes constraints from the numerical optimization, which may ease training. On the other hand, a linear expansion without symmetry requires more coefficients, as the model must learn contributions that symmetry would otherwise fix. The symmetric form Eq.(\ref{eq:symC}) requires only the coefficients $c_{\tau n_1 l_1 n_2 l_2 \dots}$ in place of all components $c_{n_1 l_1 m_1 n_2 l_2 m_2 \dots}$. For example, four single-bond basis functions with $l_1 = l_2 = l_3 = l_4 = 2$ have $5^4 = 625$ components, but only five rotationally invariant couplings. For non-linear architectures the count is less direct. Orb-v3, for example, has fewer parameters than several equivariant models on the Matbench Discovery leaderboard~\cite{rhodesOrbv3AtomisticSimulation2025,Leaderboard}.

Completeness of the graph cluster basis does not rest on symmetrization, as the non-symmetrized representation of Sec.~\ref{sec:twostep} is complete. Models without built-in rotational invariance therefore remain representable within GRACE.

\section{Physical and chemical foundation} \label{sec:chem}

\begin{framed}
{\it{Summary:}} The basis functions of ACE and GRACE are closely related to matrix elements that appear in quantum mechanical calculations with atom-centered orbitals. Overlap and Hamiltonian matrix elements take the form of single-bond basis functions, and the moments of the electronic density of states unfold into closed walks that reduce to tree graphs. Through the moments, the bond energy becomes an explicit function of graph cluster basis functions. The basis is adapted to the physics of bond formation, and the small number of independent two-center integrals indicates why moderate numbers of basis functions suffice.
\end{framed}

To relate the graph cluster basis to quantum mechanical matrix elements, we start from the expansion of single-electron wave functions in a linear combination of atomic orbitals, 
\begin{equation}
\Psi( \br) = \sum_{inlm} c_{inlm}  \psi_{inlm} (\br) \,,
\end{equation}
with
\begin{equation}
\psi_{inlm} (\br) = R_{nl}(| \br - \br_i|) Y_{lm}(\theta, \phi) \,.
\end{equation}
This expansion is commonly used for single-electron wave functions in density functional theory and also for the construction of many-electron wave functions. 

\subsection{Overlap}

The overlap matrix elements give the most direct connection to the single-bond basis functions of ACE, as they depend on one interatomic vector only. The overlap matrix
\begin{equation}
\braket{i  n_1 l_1 m_1}{j n_2 l_2 m_2} = \int d\br \, \psi^*_{i n_1 l_1 m_1}( \br)\psi_{j n_2 l_2 m_2}( \br)
\end{equation}
transforms under rotation of the structure, $\br_j - \br_i \to \hat{R} (\br_j - \br_i)$, as
\begin{align}
&\hat{R}: \braket{i n_1 l_1 m_1}{j n_2 l_2 m_2} \nonumber \\
&= \sum_{m'_1 m'_2} D^{(l_1)*}_{m_1 m'_1} D^{(l_2)}_{m_2 m'_2} \braket{i n_1 l_1 m'_1}{j n_2 l_2 m'_2} \,,
\end{align}
where the complex conjugation originates from the bra state.
The product representation is reduced using Eq.(\ref{eq:angularcoupling}), and the overlap matrix element becomes a linear combination of functions of the form of atomic orbitals, which in the terminology of ACE are the single-bond basis functions $\phi_{inlm}(j)$ of Eq.(\ref{eq:varphiSH}),
\begin{equation}
 \braket{i n_1 l_1 m_1}{j n_2 l_2 m_2} = \sum_{nlm} C_{n_1 n_2 l_1 l_2 m_1 m_2}^{nlm} \phi_{inlm}(j) \,, \label{eq:ACEforOverlap}
\end{equation}
where the coefficients $C_{n_1 n_2 l_1 l_2 m_1 m_2}^{nlm}$ combine Clebsch-Gordan coefficients with radial expansion coefficients. The angular sum is finite, restricted by the selection rules $|l_1 - l_2| \leq l \leq l_1 + l_2$, $m = m_2 - m_1$ and even $l + l_1 + l_2$, while the sum over $n$ is an expansion in the complete radial basis. In this sense the single-bond basis functions of ACE are the basis in which overlap matrix elements are expanded. Related derivations are given in Refs.~\cite{bochkarevGraphAtomicCluster2024,Sharma79}, and equivariant representations of multi-center matrix elements were developed in Refs.~\cite{nigamEquivariantRepresentationsMolecular2022,zhangEquivariantAnalyticalMapping2022}.

The selection rules leave only a few couplings. For $l_1 = l_2 = 1$, for example, only $l = 0, 2$ contribute, i.e. two independent radial functions. The number of allowed $l$, $\min(l_1, l_2) + 1$, equals the number of independent two-center integrals in the Slater-Koster construction~\cite{Slater54}. For a single chemical species this gives no more than ten independent integrals up to $d$ orbitals, and for two different species integrals such as $sp$ and $ps$ count separately. A two-center matrix element therefore involves few independent radial functions, which indicates why moderate numbers of basis functions suffice.

\subsection{Further matrix elements}

The expansion extends to other operators that are invariant under simultaneous translation, rotation and inversion of all particle coordinates. The kinetic energy matrix elements, in Hartree atomic units, depend on $\br_j - \br_i$ only and are expanded in the same single-bond basis functions
\begin{equation}
 \braHket{i n_1 l_1 m_1}{-\frac{1}{2} \nabla^2}{j n_2 l_2 m_2} = \sum_{nlm} T_{n_1 n_2 l_1 l_2 m_1 m_2}^{nlm} \phi_{inlm}(j) \,.
\end{equation}
The Coulomb interaction with an ionic core of charge $Z_k$ at position $\br_k$ adds a third center and leads to matrix elements on a (11) star graph,
\begin{align}
& \braHket{i n_1 l_1 m_1}{ \frac{-Z_k}{|\br - \br_k|} }{j n_2 l_2 m_2} =\nonumber \\
& \sum_{n_3 l_3 m_3 n_4 l_4 m_4} U_{n_1 n_2 l_1 l_2 m_1 m_2}^{n_3 l_3 m_3 n_4 l_4 m_4} \phi_{in_3l_3m_3}(j)\phi_{in_4l_4m_4}(k)  \,.
\end{align}
Two-particle four-center integrals are expanded on tree graphs, for example,
\begin{align}
& \braHket{i n_1l_1 m_1, i' n'_1 l'_1 m'_1}{ \frac{1}{|\br - \br'|} }{j n_2 l_2 m_2, j' n'_2 l'_2 m'_2} =\nonumber \\
& \sum_{n_3 l_3 m_3 n_4 l_4 m_4 n_5 l_5 m_5} V_{n_1 n_1' n_2 n_2' l_1  l_1' l_2 l_2' m_1 m_1' m_2 m_2'}^{n_3 l_3 m_3 n_4 l_4 m_4  n_5 l_5 m_5} \nonumber \\
&\times \phi_{in_3l_3m_3}(i') \phi_{in_4l_4m_4}(j) \phi_{jn_5l_5m_5}(j')  \,,
\end{align}
where $\br$ and $\br'$ are the coordinates of the two electrons. Symmetrization over equivalent centers proceeds as discussed in Sec.~\ref{sec:complete}. Fig.~\ref{fig:ACEphysics} a)-c) shows the graphs for the three types of matrix elements.

The efficient evaluation of these three- and four-center integrals is a classical topic of quantum chemistry. Gaussian basis functions were introduced for exactly this purpose~\cite{boysElectronicWaveFunctions1950} and recursion relations enable their fast evaluation~\cite{mcmurchieOneTwoelectronIntegrals1978,obaraEfficientRecursiveComputation1986}. Quantum chemistry further established that the four-center integral tensor is of low numerical rank, which is exploited in density fitting~\cite{whittenCoulombicPotentialEnergy1973,vahtrasIntegralApproximationsLCAOSCF1993}, Cholesky decomposition~\cite{beebeSimplificationsGenerationTransformation1977} and tensor hypercontraction~\cite{hohensteinTensorHypercontractionDensity2012}. The integral tensors of quantum chemistry therefore compress well, which makes efficient tensor decompositions of the expansion coefficients plausible, cf. Sec.~\ref{sec:scaleorder}.

\subsection{Moments}

The moments theorem of Cyrot-Lackmann~\cite{cyrot-lackmannElectronicStructureLiquid1967} relates the local density of states, and through it the energy, to the Hamiltonian matrix elements and is the starting point of the bond-order potentials~\cite{pettiforNewManybodyPotential1989,horsfieldBondorderPotentialsTheory1996,drautzValencedependentAnalyticBondorder2006,drautzValencedependentAnalyticBondorder2011}. In an orthonormal basis, the moments of the local density of states are sums over closed walks, i.e. hopping paths that return to the starting orbital. With the Hamiltonian matrix elements abbreviated as $H_{i n l m \, j n' l' m'} = \braHket{i n l m}{\hat{H}}{j n' l' m'}$, the moments read
\begin{align}
\mu_{inlm}^{(N)} &= \braHket{i n l m}{\hat{H}^N}{i n l m}  \nonumber \\
&=  \sum H_{i n l m \, i_1 n_1 l_1 m_1} H_{i_1 n_1 l_1 m_1 \, i_2 n_2 l_2 m_2} \nonumber \\
& \quad \dots \times H_{i_{N-1} n_{N-1} l_{N-1} m_{N-1} \, i n l m} \,,
\end{align}
where the sum extends over all intermediate states $i_1 n_1 l_1 m_1, \dots, i_{N-1} n_{N-1} l_{N-1} m_{N-1}$.

The moments are polynomials in the Hamiltonian matrix elements. These have the same symmetry as the overlap integrals and in two-center approximation are represented in the form of Eq.(\ref{eq:ACEforOverlap}). The three- and four-center contributions are graph basis functions themselves, so that the representation does not rest on the two-center approximation. The atomic orbitals of the LCAO expansion are not orthogonal, but L\"owdin orthogonalization~\cite{lowdinNonOrthogonalityProblem1950} restores orthonormality and preserves the transformation properties of the matrix elements. Orthogonalization is not a two-center operation and introduces environment dependence in the matrix elements. This dependence is a function of the interatomic vectors and is again representable in the graph basis, in analogy to screened, environment-dependent matrix elements~\cite{tangEnvironmentdependentTightbindingPotential1996,nguyen-manhAnalyticEnvironmentDependentTightBinding2000,drautzAnalyticBondorderPotentials2007}.

\begin{figure}[ht!]
    \includegraphics[width=0.5\textwidth]{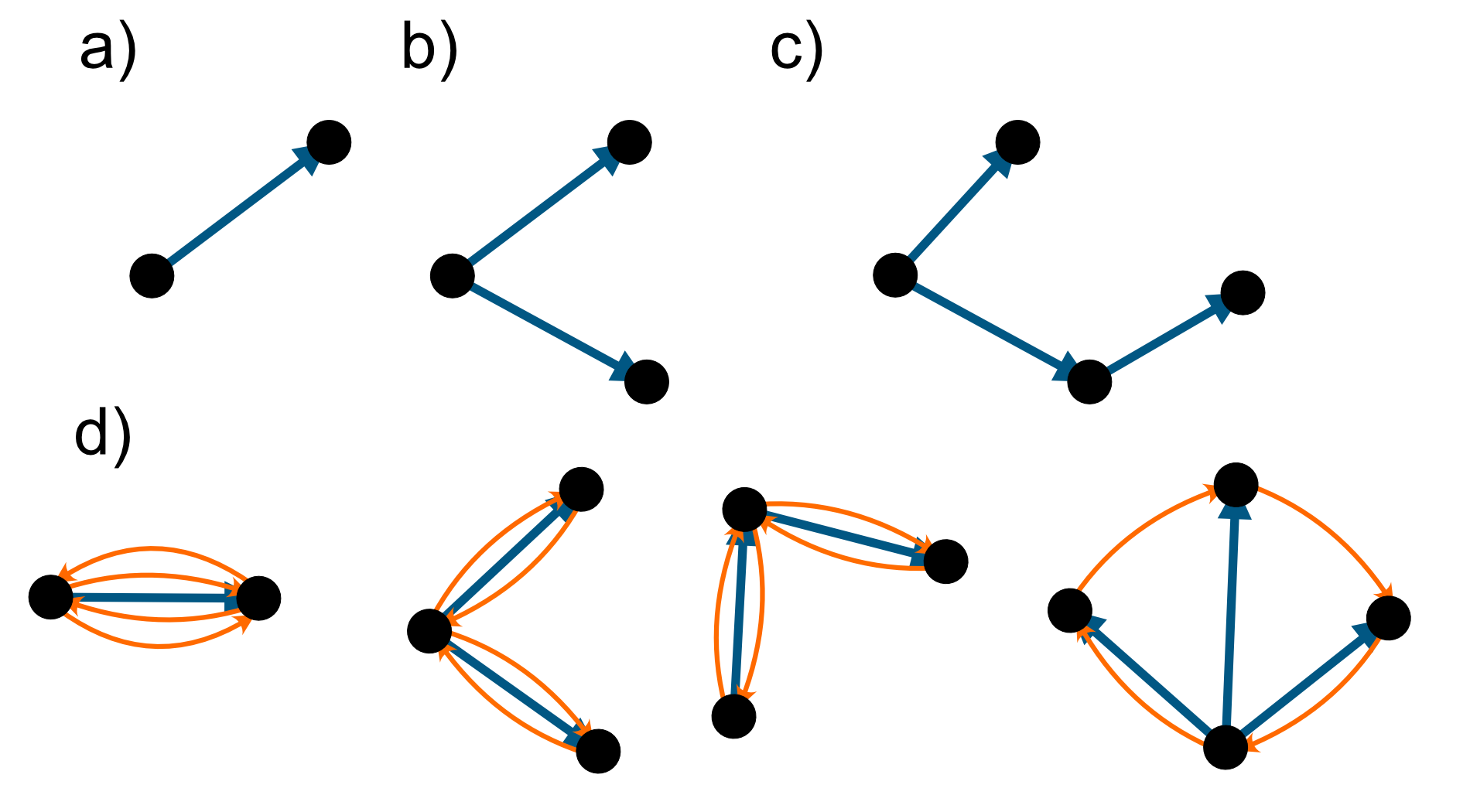}
    \caption{Graphs for a) two-center matrix elements, b) three-center one-particle matrix elements, c) two-particle four-center matrix elements. d) The closed walks contributing to the fourth moment and their representation by tree graphs, the two-atom walk, the (11) star, the (12) path and the four-atom ring on a (111) star, cf. the $s$-orbital example in the text. In d) orange arrows are the hops of the walks and blue arrows the edges of the tree graphs.}
    \label{fig:ACEphysics}
\end{figure}

The closed walks return to atom $i$, while the graph basis functions of Sec.~\ref{sec:complete} are trees. A closed walk that visits $\Nat$ atoms involves, however, only $\Nat - 1$ independent difference vectors, as many as a tree with $\Nat$ nodes has edges. A matrix element between atoms $j$ and $k$ away from the root depends on $\br_k - \br_j = (\br_k - \br_i) - (\br_j - \br_i)$ and can be expanded in products of single-bond basis functions on the edges to the root, while revisited atoms appear as the self-interactions of Sec.~\ref{sec:neighbors}. In this way closed walks are represented by tree basis functions, as illustrated in Fig.~\ref{fig:ACEphysics} d) for the fourth moment, where the four-atom ring reduces to a (111) star. The restriction to acyclic graphs in Sec.~\ref{sec:complete} is therefore no formal limitation for the moments.

The reduction also makes the required cutoff explicit. The single-bond basis on the edges to the root must reach every atom visited by the walk, and a closed walk of length $N$ extends up to $N/2$ hops from the root. The moments of order $N$ therefore require a cutoff of $N/2$ times the range of the matrix elements. This quantifies the guideline of Sec.~\ref{sec:complete}, as large as necessary, since higher moments resolve finer features of the density of states and in turn require a larger cutoff.

We illustrate the reduction for $s$ orbitals with one orbital per atom, where we suppress the indices $l = m = 0$ and the radial index, $H_{ij}$, and set the on-site levels to zero, $H_{ii} = 0$. The matrix element depends only on the distance, $H_{ij} = H(r_{ij})$ with $r_{ij} = |\br_j - \br_i|$, and is the single two-center integral for $l_1 = l_2 = 0$. The fourth moment collects the closed walks of length four,
\begin{align}
\mu_i^{(4)} &= \sum_{j_1 j_2 j_3} H_{i j_1} H_{j_1 j_2} H_{j_2 j_3} H_{j_3 i} \nonumber \\
&= \sum_{j_1} H_{i j_1}^4 + \sum_{j_1 \neq j_2} H_{i j_1}^2 H_{i j_2}^2 + \sum_{j_1,\, j_2 \neq i} H_{i j_1}^2 H_{j_1 j_2}^2 \nonumber \\
& \quad + {\sum_{j_1 j_2 j_3}}' H_{i j_1} H_{j_1 j_2} H_{j_2 j_3} H_{j_3 i} \,,
\end{align}
the two-atom walk, the (11) star, the (12) path and the four-atom ring of Fig.~\ref{fig:ACEphysics}~d), where the primed sum runs over pairwise distinct atoms $i, j_1, j_2, j_3$. As $H_{ij}$ depends only on distance, every term can be expanded in the graph cluster basis. The first two terms combine into $\big( \sum_{j_1} H_{i j_1}^2 \big)^2$, a power of the atomic base, Eq.(\ref{eq:abase}), with the two-atom walk as the self-interaction $j_1 = j_2$ of Sec.~\ref{sec:neighbors}. The (12) path is a tree basis function that is evaluated recursively as in Sec.~\ref{sec:scalelayer}.

The hops $j_1 \to j_2$ and $j_2 \to j_3$ of the four-atom ring connect atoms away from the root and are expanded in single-bond functions of the edges to the root. The $s$-orbital matrix element carries no angular character of its own, and as $|\br_{j_2} - \br_{j_1}|$ depends only on the lengths of the two edges and the angle between them, the expansion is a simple Legendre series,
\begin{equation}
H(|\br_{j_2} - \br_{j_1}|) = \sum_{l} H_l(r_{i j_1}, r_{i j_2}) P_l(\hbr_{i j_1} \cdot \hbr_{i j_2}) \,,
\end{equation}
with Legendre polynomials $P_l$, unit vectors $\hbr_{i j_1}$ along $\br_{j_1} - \br_i$ and coefficient functions
\begin{equation}
H_l(u, v) = \frac{2l+1}{2} \int_{-1}^{1} H\Big(\sqrt{u^2 + v^2 - 2 u v x}\Big) P_l(x) \, dx \,.
\end{equation}
Expanding the coefficient functions in the radial basis, $H_l(u,v) = \sum_{n n'} H^{(l)}_{n n'} R_{nl}(u) R_{n'l}(v)$, and using the addition theorem, $P_l(\hbr \cdot \hbr') = \frac{4\pi}{2l+1} \sum_{m} Y^*_{lm}(\hbr) Y_{lm}(\hbr')$, the fourth moment factorizes into neighbor sums over the three edges,
\begin{align}
\mu_i^{(4)} - \big(\mu_i^{(2)}\big)^2 &= \sum_{\substack{l_1 l_2 m_1 m_2 \\ n_1 n_1' n_2 n_2'}} \frac{(4\pi)^2 \, H^{(l_1)}_{n_1 n_1'} H^{(l_2)}_{n_2 n_2'}}{(2 l_1 + 1)(2 l_2 + 1)} \nonumber \\
&\times \sum_{j_1} H(r_{i j_1}) R_{n_1 l_1}(r_{i j_1}) Y^*_{l_1 m_1}(\hbr_{i j_1}) \nonumber \\
&\times \sum_{j_2} R_{n_1' l_1}(r_{i j_2}) R_{n_2 l_2}(r_{i j_2}) Y_{l_1 m_1}(\hbr_{i j_2}) Y^*_{l_2 m_2}(\hbr_{i j_2}) \nonumber \\
&\times \sum_{j_3} R_{n_2' l_2}(r_{i j_3}) H(r_{i j_3}) Y_{l_2 m_2}(\hbr_{i j_3}) \,.
\end{align}
The right hand side is an expansion on the (111) star alone. The subtracted term $\big(\mu_i^{(2)}\big)^2$ collects the walks through the root, $j_2 = i$, i.e. the two-atom walk and the (11) star, and as the square of an atomic-base entry it lies in the graph basis itself. With unrestricted sums the star expansion contains the (12) path as the coincidence $j_1 = j_3$, a self-interaction in the sense of Sec.~\ref{sec:neighbors} that in the graph basis without self-interactions belongs to the (12) tree, while terms with $j_1 = j_2$ or $j_2 = j_3$ vanish with $H(0) = 0$. Products of radial functions and of spherical harmonics on the same edge are re-expanded in the single-bond basis, using $Y^*_{lm} = (-1)^m Y_{l\,-m}$ and the Gaunt coupling of Sec.~\ref{sec:angscaling}, so that each neighbor sum becomes an entry of the atomic base, Eq.(\ref{eq:abase}). All contributions to the fourth moment are thereby invariant contractions of atomic-base entries, $\sum c_{v_1 v_2 v_3} A_{i v_1} A_{i v_2} A_{i v_3}$, i.e. explicit ACE basis functions of Eq.(\ref{eq:E4}). A single matrix element thus generates two-, three- and four-body contributions to the moments and through them to the energy. For example, the second moment, $\mu_i^{(2)} = \sum_{j} H_{ij}^2$, is the density of the second-moment approximation of Sec.~\ref{sec:nlopt}.

The moments determine the local density of states, $n_{inlm}(E) = \braHket{i n l m}{\delta(E - \hat{H})}{i n l m}$ with $\mu^{(N)}_{inlm} = \int E^N n_{inlm}(E) \, dE$ and $n_i(E) = \sum_{nlm} n_{inlm}(E)$, and through it the bond energy,
\begin{equation}
E_{\text{bond},i} = \int^{E_F} (E - \epsilon_i) \, n_i(E) \, dE \,,
\end{equation}
the attractive part of the atomic energy in the tight-binding bond model~\cite{suttonTightbindingBondModel1988}, with the on-site level $\epsilon_i$ and the Fermi energy $E_F$. For an atom-resolved energy the Fermi energy is replaced by local charge neutrality~\cite{drautzValencedependentAnalyticBondorder2006,drautzValencedependentAnalyticBondorder2011}. The bond energy is thereby an explicit function of the moments, and through them of closed walks and of graph cluster basis functions. The remaining repulsive contributions are pair potentials or functions of a density and are likewise representable in the graph basis. In this way the graph expansion of the energy is obtained from electronic structure theory. The coefficient tensor of the (111) star in the worked example is the product $H^{(l_1)}_{n_1 n_1'} H^{(l_2)}_{n_2 n_2'}$ of one matrix per hop, which makes plausible that trained expansion coefficients compress equally well and supports the tensor decompositions of Sec.~\ref{sec:scaleorder}.

\section{Further numerical aspects} \label{sec:numerics}

\begin{framed}
{\it{Summary:}} Next to architecture and truncation, the implementation determines the efficiency. Single precision arithmetic is faster on GPUs and sufficient if the computation is organized such that numbers of comparable magnitude are operated on. On current hardware the cost is often set by memory access rather than by arithmetic, which favors dense operations and explains part of the adoption of attention. Platform-specific code evaluates the same GRACE model about five times faster than the TensorFlow implementation. Learning forces directly instead of differentiating the energy is two to three times faster, but the learned forces are not conservative, so that the choice depends on the application.
\end{framed}

\subsection{Single versus double precision}

The numerical simulation of physical and chemical systems has traditionally relied on double-precision (FP64) arithmetic, with about 16 significant decimal digits. While single-precision (FP32) arithmetic, with about 7 significant decimal digits, is often considered insufficiently accurate for these simulations, modern GPU architectures execute single and mixed precision arithmetic substantially faster than double precision~\cite{abdelfattahSurveyNumericalLinear2021,micikeviciusMixedPrecisionTraining2018}, and mixed and single precision schemes have been used successfully in molecular dynamics~\cite{legrandSPFPSpeedCompromise2013,fanGPUMDPackageConstructing2022}.
Single precision suffices when the computation combines numbers of similar magnitude and small intermediate values do not matter. The GPU times in Fig.~\ref{fig:warp}, for example, were obtained in single precision. The full speed-up from double to single precision is obtained only if the evaluation is limited by arithmetic rather than by memory access, which on current hardware often sets the cost, Sec.~\ref{sec:memaccess}.

\subsection{Memory access} \label{sec:memaccess}

GPUs reach their nominal throughput only for operations with high arithmetic intensity~\cite{williamsRooflineInsightfulVisual2009}, with many floating point operations per byte loaded from memory, as in dense matrix multiplications. Sparse operations, in contrast, load scattered data for few operations. The direct evaluation of Clebsch-Gordan couplings is therefore slow beyond the operation count of Sec.~\ref{sec:angscaling}, as the sparse contractions~\cite{geigerE3nnEuclideanNeural2022} are hard to organize in GPU memory~\cite{bharadwajEfficientSparseKernel2025}. Efficient implementations recast the coupling into dense matrix multiplications~\cite{bochkarevFastContractedClebschGordan2026}, generate specialized sparse or fused kernels~\cite{bharadwajEfficientSparseKernel2025,leeFlashTPFusedSparsityaware2025,cuEquivariance2026} or evaluate the coupling on structured grids, where the cost is set by memory bandwidth~\cite{bochkarevFastContractedClebschGordan2026}.

Transformer models consist almost entirely of dense products and match current hardware particularly well. Memory-aware implementations such as FlashAttention organize the computation to minimize data movement in the GPU memory hierarchy~\cite{daoFlashAttentionFastMemoryefficient2022}. This match to the hardware explains part of the adoption of attention in MLIPs, as attention adds no interactions that the complete basis does not represent, Sec.~\ref{sec:MLinMLIPs}. The advantage is, however, not fundamental. Equivariant GRACE models nonetheless reached the Pareto front of accuracy versus computational cost at the time of Ref.~\onlinecite{lysogorskiyGraphAtomicCluster2026}, so that built-in symmetry is no obstacle to efficiency.

\subsection{Platform-specific implementation} \label{sec:runtime}

The same model can be evaluated at very different speed, depending on its implementation. MLIPs are usually trained and evaluated in general machine learning frameworks such as TensorFlow~\cite{abadiTensorFlowLargeScaleMachine2016}, PyTorch~\cite{paszkePyTorchImperativeStyle2019} or JAX~\cite{jax2018github}, with compilers that generate generic code for the computational graph. Dedicated implementations exist for several models, for example GPUMD for the neuroevolution potential (NEP)~\cite{fanGPUMDPackageConstructing2022}, DeePMD-kit and Allegro, with which simulations of a hundred million atoms and more were demonstrated~\cite{jiaPushingLimitMolecular2020,musaelianScalingLeadingAccuracy2023}.

The grace-runtime engine~\cite{graceRuntime2026} separates training from deployment. GRACE models with one to three layers and ACE models are exported in a versioned format that is independent of the training framework and are evaluated without a machine learning framework, by code written directly in C++ for CPUs and with Kokkos~\cite{trottKokkos3Programming2022} for NVIDIA and AMD GPUs. The code is specific to the platform, with hardware-optimized code for each architecture, such as NVIDIA H200 or A100 GPUs, AMD GPUs or different CPUs. Compared with the TensorFlow implementation compiled with XLA, GRACE models are evaluated about five times faster on GPUs and CPUs (our benchmarks, unpublished), and for GRACE-1L and GRACE-2L simulations of tens of millions of atoms fit on a single GPU. The model and its accuracy are unchanged, and the gain is due to the implementation alone. Hardware-specific code used to be too laborious to write and maintain for most MLIPs. AI-assisted programming, with which the grace-runtime code was written, makes it practical.

\subsection{Forces from energy gradients versus force learning} \label{sec:forcelearning}

Automatic differentiation computes the forces as gradients of the energy, and the parameter derivatives for training, efficiently~\cite{baydinAutomaticDifferentiationMachine2018}. Instead of differentiating the energy one can also learn the forces on the atoms directly, as a vectorial quantity expanded on each atom, Sec.~\ref{sec:equirec}. 
As the gradient evaluation is about two to three times more expensive than the evaluation of learned forces~\cite{pmlr-v267-bigi25a}, several recent models learn the forces directly~\cite{quImportanceBeingScalable2024,rhodesOrbv3AtomisticSimulation2025}. Empirically, directly force-learned models sometimes report lower errors on energy and force benchmarks than force-conservative models~\cite{quImportanceBeingScalable2024,fuLearningSmoothExpressive2025}. A plausible explanation is the additional flexibility of fitting forces independently, as in a conservative model the forces are fixed exactly by $\pmb{F}_i = -\nabla_i E$, whereas a non-conservative model fits energies and forces separately.

Whether a speed-up by a factor of two to three outweighs conservative forces depends on the application and its consistency requirements for forces. For example, force learning may be sufficient for some relaxations, although convergence of geometry optimization is not guaranteed~\cite{pmlr-v267-bigi25a}. On the other hand, learned forces are not conservative, so that the work done along a path depends on the path and the energy is not conserved in molecular dynamics, which affects many properties of atomistic systems~\cite{pmlr-v267-bigi25a}.

\section{Machine learning in MLIPs} \label{sec:MLinMLIPs}

\begin{framed}
{\it Summary:} We base our analysis on a complete linear basis expansion, which in principle represents any atomic interaction within its range. Machine learning in MLIPs can then be understood from the basis expansion and plays two roles,
\begin{enumerate}
\item Non-linear functions of atomic properties provide representations of the energy that converge faster and therefore are more accurate at a given number of basis functions than a linear basis expansion. Machine learning provides efficient non-linear functions, on which the accuracy of state-of-the-art MLIPs relies.
\item MLIPs have a large number of trainable parameters. Machine learning provides loss functions, optimizers and autodifferentiation for problems with many parameters and enables efficient training of MLIPs, up to foundation models with millions of parameters.
\end{enumerate}
The representation therefore follows from the complete basis, and machine learning provides the non-linear functions and the technology for training and implementation.
\end{framed}

\subsection{Non-linear representations} \label{sec:nlrep}

GRACE provides a complete linear basis expansion of the local and semi-local atomic environment. However, non-linear functions of the basis expansion converge faster, so that fewer terms are needed for a given accuracy, Sec.~\ref{sec:nlopt}. Rotational symmetry constrains these non-linear functions. General non-linear functions do not commute with rotations, so that in equivariant models they are applied only to invariant scalars or as scalar gates that multiply equivariant features~\cite{weiler3DSteerableCNNs2018,thomasTensorFieldNetworks2018}, while equivariant non-linearity remains available in the form of Clebsch-Gordan coupled products, Sec.~\ref{sec:equirec}.

This constraint motivates architectural choices. Networks with gated non-linearities apply scalar gates to equivariant features, as in eSEN~\cite{fuLearningSmoothExpressive2025}. Attention-based networks weight graph edges with non-linear scalar scores, as in EquiformerV2~\cite{liaoEquiformerV2ImprovedEquivariant2023}. Alternatively, rotational equivariance is not enforced in the architecture. For example, the Point Edge Transformer (PET)~\cite{pozdnyakovSmoothExactRotational2023} does not organize its features into irreducible representations, needs no Clebsch-Gordan coupling of products and applies general non-linear functions to all features. The shift towards attention is also driven by hardware, Sec.~\ref{sec:memaccess}.

Transformer models generate many-atom interactions from a basic layer coupling that is repeated non-linearly and iteratively. The coupling is the attention score, at lowest order a three-body term of the expansion. The softmax makes the coupling non-linear within a layer, and stacking repeats it across layers. To make this explicit, consider attention between the edges of atom $i$ and take queries and keys in the simplest case as the edge vectors themselves, $\pmb{q}_{ij_1} = \br_{j_1} - \br_i$ and $\pmb{k}_{ij_2} = \br_{j_2} - \br_i$, where the bold $\pmb{k}$ denotes the key vector and is not to be confused with the channel index $k$. The attention score is then the scalar product of edge vectors, with the customary scaling of the score by the inverse square root of the key dimension absorbed in the expansion coefficients,
\begin{align}
(\br_{j_1} - \br_i) \cdot (\br_{j_2} - \br_i) &= \frac{4\pi}{3} |\br_{j_1} - \br_i| |\br_{j_2} - \br_i| \nonumber \\
& \times \sum_{m=-1}^{1} Y_{1m}^*(\hbr_{ij_1}) Y_{1m}(\hbr_{ij_2}) \,, \label{eq:attcos}
\end{align}
with unit vectors $\hbr_{ij_1}$ along $\br_{j_1} - \br_i$ and where the right hand side follows from the spherical harmonics addition theorem. The score is the cosine of the bond angle multiplied by the two bond lengths, written as an invariant, $m$-summed product of two single-bond basis functions with $l=1$ and the bond length as radial function, and summation over neighbors yields invariant three-body atomic properties built from the atomic base, Eq.(\ref{eq:abase}). The attention output for the edge from atom $i$ to neighbor $j_1$ collects the values $\pmb{v}_{ij_2}$ with softmax weights,
\begin{equation}
\pmb{y}_{ij_1} = \sum_{j_2} \frac{\exp\left( \pmb{q}_{ij_1} \cdot \pmb{k}_{ij_2} \right)}{\sum_{j_3} \exp\left( \pmb{q}_{ij_1} \cdot \pmb{k}_{ij_3} \right)} \, \pmb{v}_{ij_2} \,, \label{eq:attout}
\end{equation}
where all sums run over the neighbors of atom $i$.

The exponential of the score, expanded in a power series, raises the angular resolution of the three-body terms. Contributions of arbitrary body order arise from the softmax normalization, with the reciprocal expanding into products of neighbor sums, and from stacking layers. The attention weight carries a free edge index and is a non-linear function of edge-resolved basis expansions. In the recursive evaluation, Eq.(\ref{eq:message}), it enters the sum over neighbors in the position of the single-bond basis function $\phi^{(t)}_{iv}(j)$, while the attention values take the position of the messages $\varphi^{(t)}_{jk}$. Summed over the edges of atom $i$, the attention output takes the form of a non-linear function of atomic properties, Eq.(\ref{eq:F}).

Linearly transformed queries and keys, $\pmb{q}_{ij_1} = \pmb{W}_q (\br_{j_1} - \br_i)$ and $\pmb{k}_{ij_2} = \pmb{W}_k (\br_{j_2} - \br_i)$, add products of $l=1$ single-bond basis functions that are no longer invariant, as the score remains invariant only for $\pmb{W}_q^T \pmb{W}_k$ proportional to the unit matrix. More complex queries, keys and values lead to further non-linear many-body terms. This is within the closure of the span of the complete basis, as any permutation-invariant, square-integrable function of the atomic positions within the cutoff is representable by completeness, Eq.(\ref{eq:complete}), and the power series provide the explicit low-order terms. Stacked attention passes these quantities along graph edges and maps onto the recursive, semi-local evaluation of Sec.~\ref{sec:twostep}. In this sense transformer models applied to atomic interactions can be represented within GRACE, and, as the basis is complete before symmetrization, Sec.~\ref{sec:learnsym}, this holds equally for models such as PET that do not enforce rotational invariance.

Completeness therefore determines which interactions can be represented, and the parameters of attention determine how fast the truncated expansion converges. Overparameterization can support smooth interpolation~\cite{belkinReconcilingModernMachine2019}, and we suggest that this is the role of the many parameters of attention in transformer MLIPs.

\subsection{Training and implementation} \label{sec:opt}

Beyond the representation, machine learning contributes the technology for training and for implementation, from loss functions and optimizers to autodifferentiation. The loss function determines which errors the model is trained to avoid. Squared errors are the most common choice for training MLIPs~\cite{behlerGeneralizedNeuralNetworkRepresentation2007,bochkarevEfficientParametrizationAtomic2022},
\begin{equation}
\mathcal{L} = \sum_n w_n \left(y_n - y_n^{(0)} \right)^2 \,,
\end{equation}
where $n$ indexes all training data, $y_n$ are predictions and $y_n^{(0)}$ reference values, such as energies, forces on atoms and stresses. The weights $w_n$ set the relative contribution of the different data, for example of energies against forces. The quadratic dependence means that data points with large errors, i.e. outliers, dominate the loss and its gradient. If a few outliers are acceptable, a loss that grows only linearly for large errors is more appropriate
\begin{equation}
\mathcal{L} = \sum_n w_n f(y_n - y_n^{(0)}) \,,
\end{equation}
with the Huber loss~\cite{huberRobustEstimationLocation1964}
\begin{align}
f(x) = 
\begin{cases} 
\frac{1}{2} x^2 & \,\,\,\, \mathrm{for} \,\,\,\,  |x| \leq \Delta \,,\\
( |x| - \frac{1}{2} \Delta) \Delta & \,\,\,\, \mathrm{for} \,\,\,\,  |x| > \Delta \,.
\end{cases}
\end{align}
Here $\Delta$ is an empirical parameter at which the loss switches from quadratic to linear. The Huber loss has a discontinuous second derivative at $|x| = \Delta$, whereas the pseudo-Huber loss~\cite{charbonnierDeterministicEdgepreserving1997,hartleyMultipleViewGeometry2018} has continuous derivatives of all orders,
\begin{equation}
f(x) = \Delta^2 \left( \sqrt{1 + (x/\Delta)^2} - 1 \right) \,.
\end{equation}

The loss is minimized with respect to thousands or millions of parameters, which is a difficult non-linear optimization. The Adam optimizer~\cite{kingmaAdamMethodStochastic2014} and its variants~\cite{loshchilovDecoupledWeightDecay2019} are the standard choice, up to the foundation models with millions of parameters~\cite{batatiaFoundationModelAtomistic2023,lysogorskiyGraphAtomicCluster2026}. Adam adapts the step size for each parameter individually, but this does not remove the influence of feature scales on conditioning and initialization, so that feature scaling and normalization remain important~\cite{lecunEfficientBackProp2012,ioffeBatchNormalizationAccelerating2015,baLayerNormalization2016}. In an MLIP this concerns, for example, per-atom energy shifts and scales and the magnitudes of messages across layers.

Autodifferentiation frameworks such as TensorFlow~\cite{abadiTensorFlowLargeScaleMachine2016}, PyTorch~\cite{paszkePyTorchImperativeStyle2019} or JAX~\cite{jax2018github} propagate parameter derivatives efficiently. However, across many layers the gradients may become too small for the optimizer, known as the vanishing gradient problem~\cite{bengioLearningLongtermDependencies1994,hochreiterVanishingGradientProblem1998}. Shortcut or residual connections~\cite{heDeepResidualLearning2016} then help to maintain the gradient across the layers.
Related is a readout from every layer, $E_i = \sum_{t=0}^{t_{max}-1} \sum_k \lambda^{(t)}_k \varphi^{(t)}_{ik}$, instead of the last layer only as in Eq.(\ref{eq:recfin})~\cite{NEURIPS2022_4a36c3c5,lysogorskiyGraphAtomicCluster2026}. In a graph expansion with two layers the outer layer then contributes to the energy directly and its coefficients receive a gradient that does not pass through the recursion, which gives them more significance in the optimization.

For these reasons the implementation of an MLIP can appear much more complex than the linear expansion Eq.(\ref{eq:linear}) from which we started. Much of the variety of MLIPs is due to different non-linear representations and optimization strategies.

\section{Examples  \label{sec:Relations}}

\begin{framed}
{\it{Summary:}} Viewed from the complete basis expansion, classical potentials are low-order expansions inside physically motivated non-linear embedding functions, local MLIPs truncate the expansion at higher order and terminate it linearly or with machine learning functions in place of the embedding, and semi-local message-passing MLIPs evaluate multi-layer graph expansions recursively with explicit products or learned updates between the layers. Fig.~\ref{fig:ACEfamily} and Table~\ref{tab:mapping} summarize the mapping. We then compare accuracy, number of layers and number of parameters on the Matbench Discovery leaderboard, and error and GPU time for models trained on the same data.
\end{framed}

We show how to view selected MLIPs from the perspective of a complete basis expansion in GRACE. Table~\ref{tab:mapping} summarizes how the models of this section map onto the complete basis, grouped into classical potentials, local MLIPs on single-layer star graphs and semi-local message-passing MLIPs on multi-layer tree graphs. The last column of Table~\ref{tab:mapping} states the mapping in the sense of representability, i.e. the functional form of a model lies within the basis or its closure, independent of the historical development of the model. Potentials with additional degrees of freedom, such as charge transfer and magnetism, extend the single-particle functions of the expansion, Sec.~\ref{sec:extendability}, where representative models are cited.

Table~\ref{tab:mapping} further lists the symmetry and the dominant non-linear ingredient of each model. Invariant models carry only rotation-invariant internal features, while equivariant models carry features with angular character. In both cases the predicted energy is invariant by construction. For the models marked learned in the symmetry column, in contrast, the invariance is approximate, Sec.~\ref{sec:learnsym}. Among these models, Orb is regularized by energy gradients with respect to rotations, and PET is trained with rotational data augmentation. The many-body column of Table~\ref{tab:mapping} states how the many-body order is generated. In the models marked expl.\ it is generated by explicit products of basis functions or equivariant features, with non-linear functions only of distances, as fixed embeddings, as scalar gates or in the readout. In the models marked learned the features are updated by learned networks, kernels, MLP updates or attention. This distinction sets the colors in Fig.~\ref{fig:ACEfamily}.

The truncation of each model is given by $\nu$, $l_{\max}$ and $L$ in Table~\ref{tab:mapping}. Throughout this section and in Table~\ref{tab:mapping} we call the number of atomic-base factors in a product or message, the order $n$ in Eq.(\ref{eq:layerACE}), the correlation order $\nu$. The body order is then $\nu+1$. For message-passing models $\nu$ refers to the linear part of one layer update, before any non-linear function, i.e. to the number of bonds multiplied in one term. The numbers in Table~\ref{tab:mapping} refer to the parameterization given in parentheses after the model name, or to the Matbench Discovery entry~\cite{Leaderboard} where listed, and their sources are given in Supplementary Table S1.

The truncation determines which many-body terms a model contains and which structures the model can distinguish. For polynomial layer updates, a tree graph with $\nu$ bonds at each node and $L$ layers contains at most $\nu + \nu^2 + \dots + \nu^L$ bonds, so that two layers with $\nu = 3$ give at most 13-body terms before the non-linear readout. Message passing with distance-only messages, $l_{\max} = 0$ as in SchNet, cannot distinguish certain pairs of distinct structures for any cutoff~\cite{pozdnyakovIncompletenessGraphNeural2022,joshiExpressivePowerGeometric2023}. Accuracy and computational cost, however, depend on training data, truncation and implementation and do not follow from the mapping alone. They are compared on the Matbench Discovery leaderboard, Fig.~\ref{fig:matbench}, for models trained on the same data, Fig.~\ref{fig:warp}, and in Ref.~\cite{lysogorskiyGraphAtomicCluster2026}.

\subsection{Accuracy, depth and model size}

\begin{figure}[ht!]
    \includegraphics[width=\columnwidth]{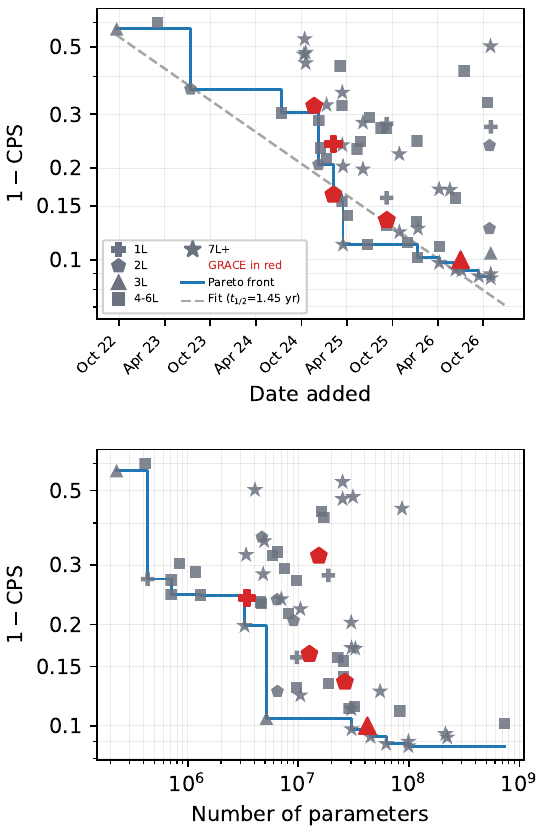}
    \caption{Distance to the perfect combined performance score, $1-\mathrm{CPS}$, Eq.(\ref{eq:cps}), for the 62 fully scored models on the Matbench Discovery leaderboard, accessed 15 November 2026~\cite{Leaderboard,riebesellMatbenchDiscovery2025}. Top: $1-\mathrm{CPS}$ over the date a model was added to the leaderboard, with the Pareto front in blue and an exponential fit to the front. Bottom: $1-\mathrm{CPS}$ over the number of model parameters. Markers indicate the number of message-passing layers. GRACE models are shown in red.}
    \label{fig:matbench}
\end{figure}

Fig.~\ref{fig:matbench} summarizes the Matbench Discovery leaderboard \cite{Leaderboard,riebesellMatbenchDiscovery2025} in terms of the combined performance score (CPS). The CPS is a weighted sum of the discovery metric F1, the thermal conductivity error $\kappa_{\mathrm{SRME}}$ and the relaxation error RMSD,
\begin{align}
\mathrm{CPS} = \; & 0.5 \, \mathrm{F1} + 0.4 \, \max\left(0, 1-\kappa_{\mathrm{SRME}}/2\right) \nonumber \\
 & + 0.1 \, \frac{0.15-\mathrm{RMSD}}{0.15} \,, \label{eq:cps}
\end{align}
where the last term is clamped to the interval $[0,1]$. Fig.~\ref{fig:matbench} shows the distance to the perfect score, $1-\mathrm{CPS}$, for the 62 fully scored models on the leaderboard, accessed 15 November 2026. In the upper panel $1-\mathrm{CPS}$ is plotted over the date a model was added to the leaderboard. An exponential fit to the Pareto front has a half time of 1.45 y. The fit describes the past data and is not a forecast. In the lower panel $1-\mathrm{CPS}$ is plotted over the number of model parameters. The number of message-passing layers of each model, indicated by the markers, was taken from the leaderboard metadata, the model papers or the released parameterizations.

The highest-scoring models, among them DPA-4.1~\cite{liDPA4PushingAccuracycost2026}, the variants of TACE with radial rotary attention~\cite{xuSpectralSpatialTensor2025,xuEdgeClusterExpansion2026} and EquiformerV3~\cite{liaoEquiformerV3ScalingEfficient2026}, have seven or more layers, followed by a three-layer GRACE model, a five-layer PET model~\cite{bigiPushingLimitsUnconstrained2026} and a three-layer DPA-4.1 model, with a CPS 1.4 to 2.0 percent below that of the best model. GRACE-3L extends the GRACE foundation models of Ref.~\onlinecite{lysogorskiyGraphAtomicCluster2026} that contain up to two layers, and is listed on the leaderboard~\cite{Leaderboard}.

GRACE-3L has three layers with a cutoff of 6~\AA{} each. In every layer, a small non-linear network maps ten radial basis functions together with a 64-dimensional chemical embedding onto 64 radial channels. The angular basis and the equivariant messages extend to $l_{\max} = 4$, and products of the atomic base up to fourth order, $\nu = 4$, are formed in tensor-decomposed form with 64 to 128 channels. Invariant features of all three layers enter a small non-linear readout. With 42.1 million parameters for 89 elements, the model was trained on OMat24 and fine-tuned on sAlex~\cite{barroso-luqueOpenMaterials20242024} and MPtrj~\cite{dengCHGNetPretrainedUniversal2023}, as the two-layer models of Ref.~\onlinecite{lysogorskiyGraphAtomicCluster2026}.

The interaction range of GRACE-3L is smaller than that of the models with seven to ten layers by roughly the ratio of layer counts at comparable cutoffs. The benchmark therefore provides no evidence that the large atomic regions covered by models with many layers are required for a CPS within two percent of the best model, in line with the guideline of Sec.~\ref{sec:complete} that the graph should be as compact as possible and as large as necessary. The same holds for the number of parameters. The Pareto front in the lower panel of Fig.~\ref{fig:matbench} reaches $1-\mathrm{CPS} \approx 0.1$ at $5 \times 10^6$ parameters, with the three-layer DPA-4.1 model, and models with up to $7 \times 10^8$ parameters improve on it only marginally. Among the components of the CPS, $\kappa_{\mathrm{SRME}}$, which probes anharmonic phonons, differs most between models, and the leading values of $\kappa_{\mathrm{SRME}} \approx 0.06$--$0.13$ are reached by models with three to ten layers.

However, Matbench Discovery tests bulk inorganic crystals only. It does not probe molecules, interfaces, liquids or long-range interactions, for which the required size of the atomic region may differ, and broader benchmarks address some of these~\cite{pengLAMBenchBenchmarkLarge2026,chiangMLIPArena2025}.

\subsection{Accuracy versus computational cost and model complexity} \label{sec:warp}

Computational efficiency is meaningful only at a given accuracy~\cite{zuoPerformanceCostAssessment2020,leimerothMachinelearningInteratomicPotentials2025}. At comparable model size, fewer layers mean lower computational cost. With two layers the GRACE foundation models reach accuracy close to that of leaderboard models with many more layers, and in Ref.~\onlinecite{lysogorskiyGraphAtomicCluster2026} they lie on the Pareto front of accuracy versus computational cost among the foundation MLIPs compared there.

\begin{figure}[ht!]
    \includegraphics[width=\columnwidth]{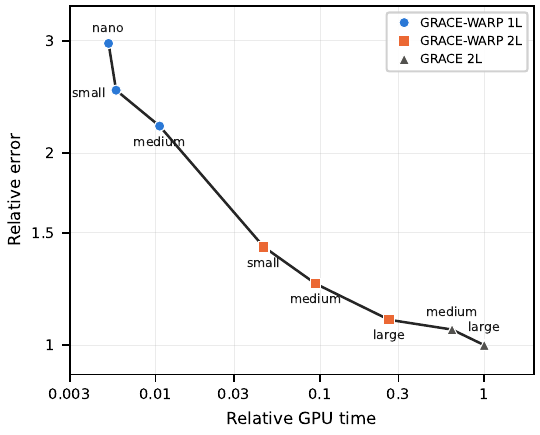}
    \caption{Relative error, Eq.(\ref{eq:warperror}), versus relative GPU time for GRACE-WARP models with one (blue) and two (orange) layers and standard GRACE models with two layers (grey), trained on the same five datasets. Point labels give the model size. Error and GPU time are relative to GRACE-2L (large), see Sec.~\ref{sec:warp}.}
    \label{fig:warp}
\end{figure}

Because the complete basis can be truncated in the number of layers, the number of basis functions and the body order, Sec.~\ref{sec:scaling}, a single basis spans a continuous range from fast to accurate models. Typically, basis settings are chosen ad hoc before a model is trained. For the GRACE-WARP models (Y.L., A.B. and R.D., in preparation), instead, all settings and parameters were optimized to reach a given error at the highest possible speed, so that the models lie along the Pareto front of error and speed, Fig.~\ref{fig:warp}. The foundation model NEP89~\cite{liangNEP89UniversalNeuroevolution2026}, which feeds selected ACE basis functions into a neural network, Sec.~\ref{sec:nep}, and the equivariant DPA4C model~\cite{liUniversalMachinelearningMolecular2026} approach the computational cost of empirical potentials. A comparison with these models on the same datasets is left to a separate paper in preparation.

Fig.~\ref{fig:warp} compares GRACE-WARP models with one and two layers to standard GRACE models with two layers. All models were trained from scratch for 250 epochs on each of the same five DFT datasets, which cover carbon, Al--Li, water, ethanol and a high-entropy alloy, and the settings of the GRACE-WARP models were optimized on these datasets. The relative error combines the energy and force errors over the datasets,
\begin{equation}
\frac{1}{5}\sum_s \left( \frac{1}{2} \frac{\Delta E_s}{\Delta E_s^{\mathrm{ref}}} + \frac{1}{2} \frac{\Delta F_s}{\Delta F_s^{\mathrm{ref}}} \right) , \label{eq:warperror}
\end{equation}
where $\Delta E_s$ and $\Delta F_s$ are the mean absolute energy and force errors on the held-out validation set of dataset $s$ after removal of outlier structures. The reference is the most accurate model, GRACE-2L (large). The GPU time is for LAMMPS with Kokkos in single precision on one NVIDIA L40S and bcc W with 128\,000 atoms. It is given relative to the GPU time of the reference, 714~ms per molecular dynamics step or 5.6~$\mu$s per atom and step.

With increasing model size the error decreases along a single front. The fastest single-layer model has about three times the error of the reference at 0.5\% of the GPU time, or 0.03~$\mu$s per atom and step, and the largest two-layer GRACE-WARP model has an error within 10\% of the error of the reference at about a quarter of the GPU time. For these datasets, the choice between fast and accurate models is therefore a choice of truncation within one basis.

\providecommand{\tcell}[2]{\parbox[t]{#1}{\raggedright #2\strut}}
\begin{table*}[tp]
\caption{Selected MLIPs viewed through the complete ACE/GRACE basis, grouped into classical potentials, local MLIPs and semi-local message-passing MLIPs, see Sec.~\ref{sec:Relations}. In the symmetry column, $O(3)$ and $SO(3)$ denote equivariant features, with $SO(3)$ for equivariance under rotations without parity, Cart.\ denotes Cartesian tensors, and learned denotes symmetry encouraged by data augmentation or gradient regularization. The non-linearity column lists the dominant non-linear ingredient. The many-body column states whether the many-body order is generated by explicit products (expl.) or by learned feature updates (learned). $\nu$ is the correlation order of one layer update, and for local models the maximum order of the basis or descriptor. $l_{\max}$ is the highest angular momentum of the angular basis, of equivariant features or of the angular functions inside invariant features, for Cartesian models the maximum tensor rank, for SNAP $2J_{\max}$ and for HDNNP the exponent $\zeta$ of the angular symmetry functions, which contain Legendre polynomials up to $l = \zeta$. $L \times r_c$ gives the number of layers, i.e. of sums over neighbors carried out in sequence, and the cutoff in \AA, with a separate three-body, angular or interaction cutoff in parentheses, and $L = 1$ for local models. A dash indicates that no angular basis is built in. Low rank denotes restricted tensor-decomposition channels, gated denotes non-linear scalar weighting of edges or features, line graph denotes message passing between bonds, and MoLE a mixture of linear experts. OAM denotes training on OMat24, sAlex and MPtrj, and MP training on the Materials Project. \label{tab:mapping}}
\footnotesize
\setlength{\tabcolsep}{3pt}
\begin{tabular*}{\textwidth}{@{\extracolsep{\fill}}lllcccll}
\hline\hline
Model & Symmetry & Non-linearity & Many-body & $\nu$ & $l_{\max}$ & $L \times r_c$ & Maps onto \\
\hline
\multicolumn{8}{l}{\textit{Classical potentials}}\\
\tcell{3.6cm}{EAM (Cu) / FS / EMT} & \tcell{1.5cm}{invariant} & \tcell{2.5cm}{embedding $F$} & expl. & 1 & 0 & \tcell{2.5cm}{$1 \times 5.5$} & \tcell{3.7cm}{2-body ACE $+$ embedding} \\
\tcell{3.6cm}{Tersoff (Si) / MEAM} & \tcell{1.5cm}{invariant} & \tcell{2.5cm}{bond order / embedding $F$} & expl. & 2 & $\infty$ / 3 & \tcell{2.5cm}{$1 \times 3.2$ / screened} & \tcell{3.7cm}{3-body ACE $+$ non-linear function} \\
\hline
\multicolumn{8}{l}{\textit{Local descriptors and MLIPs}}\\
\tcell{3.6cm}{ACE (PACE, Cu)} & \tcell{1.5cm}{invariant} & \tcell{2.5cm}{linear or FS-like $F$} & expl. & 6 & 6 & \tcell{2.5cm}{$1 \times 7.4$} & \tcell{3.7cm}{local basis, truncated} \\
\tcell{3.6cm}{HDNNP (Si) / ANI (1x)} & \tcell{1.5cm}{invariant} & \tcell{2.5cm}{neural network} & learned & 2 & $\zeta$ & \tcell{2.5cm}{$1 \times 6$ / $1 \times 5.2\,(3.5)$} & \tcell{3.7cm}{2-/3-body ACE $\to$ NN} \\
\tcell{3.6cm}{SOAP--GAP (Si)} & \tcell{1.5cm}{invariant} & \tcell{2.5cm}{kernel} & learned & 2 & 12 & \tcell{2.5cm}{$1 \times 5$} & \tcell{3.7cm}{3-body ACE $\to$ kernel} \\
\tcell{3.6cm}{SNAP / qSNAP (Ta)} & \tcell{1.5cm}{invariant} & \tcell{2.5cm}{linear / quadratic} & expl. & 3 / 6 & 6 / 8 & \tcell{2.5cm}{$1 \times 4.7$ / $1 \times 5.6$} & \tcell{3.7cm}{4-body ACE / its square} \\
\tcell{3.6cm}{MTP (level 16)} & \tcell{1.5cm}{invariant, Cart.} & \tcell{2.5cm}{linear} & expl. & 4 & 6 & \tcell{2.5cm}{$1 \times 5.2$} & \tcell{3.7cm}{ACE (Cartesian)} \\
\tcell{3.6cm}{NEP (NEP89)} & \tcell{1.5cm}{invariant} & \tcell{2.5cm}{neural network} & learned & 4 & 4 & \tcell{2.5cm}{$1 \times 6\,(5)$} & \tcell{3.7cm}{selected ACE $\to$ NN} \\
\tcell{3.6cm}{DeepPot-SE (Cu)} & \tcell{1.5cm}{invariant} & \tcell{2.5cm}{fitting network} & learned & 2 & 1 & \tcell{2.5cm}{$1 \times 6$} & \tcell{3.7cm}{3-body ACE ($l \le 1$) $\to$ NN} \\
\tcell{3.6cm}{Allegro (OAM-L)} & \tcell{1.5cm}{$SO(3)$} & \tcell{2.5cm}{latent MLPs} & expl. & 5 & 3 & \tcell{2.5cm}{$1 \times 7$} & \tcell{3.7cm}{edge-resolved star ACE} \\
\hline
\multicolumn{8}{l}{\textit{Semi-local, message passing}}\\
\tcell{3.6cm}{GRACE (2L-OAM / 3L-OAM-L)} & \tcell{1.5cm}{$O(3)$} & \tcell{2.5cm}{MLP embedding} & expl. & 4 & 4 & \tcell{2.5cm}{$2 \times 6$ / $3 \times 6$} & \tcell{3.7cm}{graph basis, truncated} \\
\tcell{3.6cm}{NequIP (OAM-L) / SevenNet (0)} & \tcell{1.5cm}{$SO(3)$ / $O(3)$} & \tcell{2.5cm}{gate} & expl. & 1 & 3 / 2 & \tcell{2.5cm}{$6 \times 6$ / $5 \times 5$} & \tcell{3.7cm}{recursive GRACE} \\
\tcell{3.6cm}{PaiNN} & \tcell{1.5cm}{$O(3)$} & \tcell{2.5cm}{scalar MLPs, vector gating} & expl. & 1 & 1 & \tcell{2.5cm}{$3 \times 5$} & \tcell{3.7cm}{recursive GRACE ($l \le 1$)} \\
\tcell{3.6cm}{multi-ACE / MACE (MP-0)} & \tcell{1.5cm}{$O(3)$} & \tcell{2.5cm}{radial MLPs, readout} & expl. & 3 & 3 & \tcell{2.5cm}{$2 \times 6$} & \tcell{3.7cm}{recursive GRACE / multi-ACE} \\
\tcell{3.6cm}{SchNet (MP)} & \tcell{1.5cm}{invariant} & \tcell{2.5cm}{filter network} & learned & 1 & 0 & \tcell{2.5cm}{$6 \times 5$} & \tcell{3.7cm}{recursive GRACE (two-body)} \\
\tcell{3.6cm}{ml-ACE (revMD17)} & \tcell{1.5cm}{invariant} & \tcell{2.5cm}{$\tanh$ of scalar updates} & expl. & 3 & 4 & \tcell{2.5cm}{$4 \times 4$} & \tcell{3.7cm}{subset of multi-ACE} \\
\tcell{3.6cm}{M3GNet} & \tcell{1.5cm}{invariant} & \tcell{2.5cm}{gated MLP} & learned & 2 & 2 & \tcell{2.5cm}{$3 \times 5\,(4)$} & \tcell{3.7cm}{low-rank multi-ACE} \\
\tcell{3.6cm}{MatterSim (v1-5M)} & \tcell{1.5cm}{invariant} & \tcell{2.5cm}{gated MLP} & learned & 2 & 3 & \tcell{2.5cm}{$4 \times 5\,(4)$} & \tcell{3.7cm}{low-rank multi-ACE} \\
\tcell{3.6cm}{DimeNet / GemNet-Q} & \tcell{1.5cm}{invariant} & \tcell{2.5cm}{bilinear, MLP} & learned & 2 / 3 & 6 & \tcell{2.5cm}{$6 \times 5$ / $4 \times 5\,(10)$} & \tcell{3.7cm}{recursive GRACE (three-/four-body)} \\
\tcell{3.6cm}{CHGNet (0.3.0)} & \tcell{1.5cm}{invariant} & \tcell{2.5cm}{gated MLP} & learned & 2 & 15 & \tcell{2.5cm}{$4 \times 6\,(3)$} & \tcell{3.7cm}{low-rank multi-ACE $+$ magmom target} \\
\tcell{3.6cm}{CACE (HEA25)} & \tcell{1.5cm}{$O(3)$, Cart.} & \tcell{2.5cm}{products} & expl. & 3 & 3 & \tcell{2.5cm}{$2 \times 4.5$} & \tcell{3.7cm}{GRACE (Cartesian)} \\
\tcell{3.6cm}{TACE (OAM-L)} & \tcell{1.5cm}{$SO(3)$, Cart.} & \tcell{2.5cm}{products} & expl. & 2 & 5 & \tcell{2.5cm}{$5 \times 6$} & \tcell{3.7cm}{ACE/GRACE (Cartesian)} \\
\tcell{3.6cm}{TensorNet (SPICE)} & \tcell{1.5cm}{$O(3)$, Cart.} & \tcell{2.5cm}{matrix products} & expl. & 2 & 2 & \tcell{2.5cm}{$2 \times 10$} & \tcell{3.7cm}{GRACE (Cartesian, low rank)} \\
\tcell{3.6cm}{eSEN (30M-OAM)} & \tcell{1.5cm}{$SO(3)$} & \tcell{2.5cm}{gating} & expl. & 1 & 3 & \tcell{2.5cm}{$10 \times 6$} & \tcell{3.7cm}{gated recursive GRACE} \\
\tcell{3.6cm}{EquiformerV2 (M) / V3 (OAM)} & \tcell{1.5cm}{$SO(3)$} & \tcell{2.5cm}{attention} & learned & 1 & 6 / 4 & \tcell{2.5cm}{$10 \times 12$ / $7 \times 6$} & \tcell{3.7cm}{attention-weighted GRACE} \\
\tcell{3.6cm}{SO3krates} & \tcell{1.5cm}{$O(3)$} & \tcell{2.5cm}{Euclidean attention} & learned & 1 & 3 & \tcell{2.5cm}{$3 \times 5$} & \tcell{3.7cm}{attention-weighted GRACE} \\
\tcell{3.6cm}{UMA (S / M)} & \tcell{1.5cm}{$SO(3)$} & \tcell{2.5cm}{gating, MoLE} & expl. & 1 & 2 / 4 & \tcell{2.5cm}{$4 \times 6$ / $10 \times 6$} & \tcell{3.7cm}{gated recursive GRACE} \\
\tcell{3.6cm}{DPA-3 (3.1-3M)} & \tcell{1.5cm}{invariant} & \tcell{2.5cm}{MLP updates} & learned & 2 & -- & \tcell{2.5cm}{$16 \times 6\,(4)$} & \tcell{3.7cm}{line-graph GRACE approx.} \\
\tcell{3.6cm}{Orb (v3)} & \tcell{1.5cm}{learned} & \tcell{2.5cm}{MLP updates} & learned & 1 & 3 & \tcell{2.5cm}{$5 \times 6$} & \tcell{3.7cm}{non-linear GRACE approx.} \\
\tcell{3.6cm}{PET (OAM-XL)} & \tcell{1.5cm}{learned} & \tcell{2.5cm}{attention} & learned & 1 & -- & \tcell{2.5cm}{$5 \times 10$} & \tcell{3.7cm}{attention-weighted GRACE} \\
\hline\hline
\end{tabular*}
\end{table*}

\subsection{Classical potentials}

Second-moment, effective medium theory (EMT), embedded-atom method (EAM) and Finnis-Sinclair (FS) potentials~\cite{ducastelleModulesElastiquesMetaux1970,norskovEffectivemediumTheoryChemical1980,dawEmbeddedatomMethodDerivation1984,finnisSimpleEmpiricalBody1984} can all be obtained by inserting a lowest-order, two-body ACE into non-linear embedding functions. Potentials with angular terms can also be cast in the form of ACE, as shown for the modified embedded-atom method (MEAM)~\cite{baskesModifiedEmbeddedatomPotentials1992} in Ref.~\onlinecite{drautzAtomicClusterExpansion2019}, and the bond order of the Tersoff potential~\cite{tersoffNewEmpiricalModel1986} follows from the lowest-order bond-order potential~\cite{horsfieldBondorderPotentialsTheory1996}. The non-linear embedding functions are derived from density functional theory or from the lowest-order tight-binding approximation. The second-moment approximation of tight-binding gives the square-root form of the embedding function~\cite{guptaLatticeRelaxationMetal1981,finnisSimpleEmpiricalBody1984}, and the embedding functions derived from density functional theory are likewise concave.

\subsection{Local descriptors and MLIPs}

\subsubsection{HDNNP, ANI, DP and NEP} \label{sec:nep}

The high-dimensional neural network potential (HDNNP) builds on atom-centered symmetry functions (ACSF)~\cite{behlerGeneralizedNeuralNetworkRepresentation2007,behlerAtomcenteredSymmetryFunctions2011}, which are two- and three-body invariants and can be expanded in two-body and three-body ACE. The angular functions that contain the neighbor-neighbor distance require a series expansion of the exponential, analogous to the expansion of the attention score in Sec.~\ref{sec:nlrep}. 
The neural network then acts as the non-linear function $\mathcal{F}$ that takes several invariant ACE properties $\ace_i^{(n)}$ ($n = 1, 2, \dots$) as input. The ANI models~\cite{smithANI1ExtensibleNeural2017} are HDNNP-type potentials for molecules and rest on the same construction with modified symmetry functions.

The Deep Potential (DP)~\cite{zhangDeepPotentialMolecular2018} refers the neighbor coordinates of each atom to a local frame and maps them to the atomic energy with a deep neural network. DeepPot-SE~\cite{zhangEndtoendSymmetryPreserving2018} replaces the frame by an embedding network of radial functions. The resulting invariant descriptor contains distances and angles, corresponding to two- and three-body ACE, and a fitting network acts as $\mathcal{F}$. The DPA models extend this construction to message passing and are discussed with the semi-local models.

The neuroevolution potential (NEP)~\cite{fanNeuroevolutionMachineLearning2021} combines Chebyshev radial functions with three-body angular terms of ACE type, and later versions add four- and five-body ACE terms~\cite{fanGPUMDPackageConstructing2022}. These descriptors serve as input to a neural network acting as $\mathcal{F}$. The name NEP derives from training with a natural-evolution strategy in place of gradient descent.

\subsubsection{SOAP-GAP}

The smooth overlap of atomic positions (SOAP) descriptor~\cite{bartokRepresentingChemicalEnvironments2013} can be cast directly into ACE~\cite{drautzAtomicClusterExpansion2019,willattAtomdensityRepresentationsMachine2019}, and the SOAP power spectrum corresponds to three-body invariants (correlation order two). The Gaussian process~\cite{bartokGaussianApproximationPotentials2010} or any other kernel may then be viewed as the non-linear function $\mathcal{F}$ for the prediction of atomic energies.

\subsubsection{SNAP}

The Spectral Neighbor Analysis Potential (SNAP)~\cite{thompsonSpectralNeighborAnalysis2015} and its quadratic extension (qSNAP)~\cite{woodExtendingAccuracySNAP2018} build on the bispectrum of the atomic density expanded in four-dimensional hyperspherical harmonics~\cite{bartokGaussianApproximationPotentials2010}, corresponding to four-body invariants (correlation order three), and can be written directly in ACE~\cite{drautzAtomicClusterExpansion2019,drautzAtomicClusterExpansion2020,dussonAtomicClusterExpansion2022}.

\subsubsection{MTP}

The moment tensor potential (MTP)~\cite{shapeevMomentTensorPotentials2016} builds on Cartesian moment tensors and can be transformed directly into ACE ~\cite{drautzAtomicClusterExpansion2020,dussonAtomicClusterExpansion2022}.

\subsubsection{Allegro}

Allegro~\cite{musaelianLearningLocalEquivariant2023} builds on iterated tensor products of equivariant representations of the local environment. Although it stacks several tensor-product layers, these raise the body order of a single, strictly local expansion and do not pass information between atomic environments. In the message-passing sense Allegro is therefore a single-layer, local model. The energy is decomposed into edge contributions summed over neighbors, so that Allegro corresponds to an edge-resolved star-graph ACE. Ref.~\onlinecite{musaelianLearningLocalEquivariant2023} compares the structure of Allegro to star-graph ACE in detail and maps the tensor-product depth to the ACE body order.

\subsection{Semi-local, message-passing MLIPs}

Models with layer-wise operations polynomial in the atomic base are represented exactly by a finite GRACE expansion. Non-polynomial ingredients, such as softmax attention or neural-network updates, are represented in the closure of its span, i.e. as limits of convergent expansions in the sense of Sec.~\ref{sec:completebasis}, as shown in Sec.~\ref{sec:nlrep}. The non-linear updates can improve convergence, Secs.~\ref{sec:nlopt} and \ref{sec:nlrep}.

\subsubsection{GRACE}

The complete basis of the graph atomic cluster expansion, Sec.~\ref{sec:completebasis}, is the reference point of this subsection. Its recursive evaluation, Sec.~\ref{sec:twostep}, is implemented in the gracemaker software~\cite{lysogorskiyGraphAtomicCluster2026}.

\subsubsection{NequIP and SevenNet}

NequIP~\cite{batznerE3equivariantGraphNeural2022} builds on tensor field networks~\cite{thomasTensorFieldNetworks2018} and can be represented within GRACE as a model with two-body messages (correlation order one) in the linear part of each layer, so that higher body order arises from stacking layers and from the gate non-linearity. SevenNet~\cite{parkScalableParallelAlgorithm2024} is an implementation of the NequIP architecture for parallel, large-scale molecular dynamics.

\subsubsection{PaiNN}

PaiNN~\cite{schuttEquivariantMessagePassing2021} passes invariant and vectorial features between atoms. The vector features carry $l=1$ angular information. Within GRACE, PaiNN is a recursive evaluation with messages restricted to scalars and vectors.

\subsubsection{Multi-ACE and MACE}

Multi-ACE ~\cite{batatiaDesignSpaceE3equivariant2025} inserts multi-body ACE messages into an equivariant graph neural network. Multi-ACE was obtained by combining ACE with NequIP, and it can equally be represented as a recursive evaluation of GRACE with additional design choices for the layer coupling. MACE~\cite{NEURIPS2022_4a36c3c5} is a realization of multi-ACE with two layers of high body-order messages.

\subsubsection{SchNet}

SchNet~\cite{schuttSchNetDeepLearning2018} is one of the earliest deep message-passing potentials. It employs continuous-filter convolutions with filters that depend only on interatomic distances, so that each message is a rotationally invariant, two-body quantity. Angular information is built up only through successive message-passing layers, and higher body orders also through the non-linear updates within each layer. Within GRACE, SchNet can be represented as a recursive evaluation restricted to invariant, two-body messages, related to ml-ACE, discussed below, but without explicit angular terms on each layer.

\subsubsection{ml-ACE}

ml-ACE~\cite{bochkarevMultilayerAtomicCluster2022} is a multi-layer message-passing potential with full ACE expansions on each layer and scalar, non-linear updates as per-layer terminations, and it represents a subset of multi-ACE and therefore of GRACE. Its layers model semi-local interactions in analogy to the relaxation of the electronic structure.

\subsubsection{M3GNet}

M3GNet~\cite{chenUniversalGraphDeep2022} is restricted to three-body interactions (correlation order two), with the angular dependence expanded in $m=0$ spherical harmonics of the bond angle, and generates higher-order many-body correlations with deep MLPs and recursive message passing. Symmetry is incorporated through invariant features, such as interatomic distances and angles, without equivariant features. Within GRACE, M3GNet is a low-rank, non-linear multi-ACE approximation.

\subsubsection{MatterSim}

MatterSim~\cite{yangMatterSimDeepLearning2024} is a family of foundation models trained across elements, temperatures and pressures, with an invariant M3GNet backbone or a transformer-based Graphormer backbone. Within GRACE, the M3GNet backbone is a low-rank multi-ACE approximation and the Graphormer backbone an attention-weighted recursive evaluation, Sec.~\ref{sec:nlrep}.

\subsubsection{Directional message passing}

DimeNet~\cite{gasteigerDirectionalMessagePassing2020} and GemNet~\cite{gasteigerGemNetUniversalDirectional2021} pass messages between bonds, i.e. on the line graph, and include bond angles. The angular messages are three-body terms, and GemNet-Q adds dihedral angles, i.e. four-body terms. Within GRACE, both are recursive evaluations with invariant three- or four-body messages.

\subsubsection{CHGNet}

CHGNet~\cite{dengCHGNetPretrainedUniversal2023} is an invariant message-passing model trained on energies, forces, stresses and magnetic moments. The magnetic moments are read out from the node features by a linear projection, so that training on them constrains the node features to carry information on the local charge state. Within GRACE, CHGNet is a low-rank multi-ACE with node features supervised by scalar atomic quantities, a variant of the node variables of Sec.~\ref{sec:extendability}.

\subsubsection{CACE}

The Cartesian Atomic Cluster Expansion (CACE) implements ACE in Cartesian coordinates~\cite{chengCartesianAtomicCluster2024}, analogous to MTP. It further includes message passing, typically with at most one message-passing step, and in this way can be represented within GRACE.

\subsubsection{TACE}

The Tensor Atomic Cluster Expansion (TACE) implements ACE and GRACE with Cartesian tensors~\cite{xuSpectralSpatialTensor2025}. The irreducible Cartesian components carry the parity of their tensor rank, so that TACE is in principle equivariant under the full orthogonal group $O(3)$, including inversion, while the published models are implemented with $SO(3)$.

\subsubsection{TensorNet}

TensorNet~\cite{simeonTensorNetCartesianTensor2023} is an $O(3)$-equivariant message-passing model built on Cartesian tensors of rank zero, one and two. The rank-two tensors are decomposed into scalar, vector and symmetric traceless parts. Within GRACE, TensorNet is a Cartesian recursive evaluation with tensors of rank up to two.

\subsubsection{eSEN, EquiformerV2 and EquiformerV3}

eSEN~\cite{fuLearningSmoothExpressive2025} applies scalar gates to equivariant features and provides conservative forces, while EquiformerV2~\cite{liaoEquiformerV2ImprovedEquivariant2023} weights graph edges with scalar attention scores and adds non-linear activations on the sphere. In both cases the non-linear scalar weighting adds many-atom interactions to those generated by stacking the equivariant products, in analogy to the non-linear embeddings of classical potentials. EquiformerV3~\cite{liaoEquiformerV3ScalingEfficient2026} changes the normalization and the activations of EquiformerV2 and reduces its computational cost. Within GRACE, eSEN is a gated recursive evaluation, and EquiformerV2 and V3 are attention-weighted recursive evaluations.

\subsubsection{SO3krates}

SO3krates~\cite{frankEuclideanTransformerFast2024} combines sparse equivariant features with self-attention and separates invariant from equivariant information to avoid tensor products. Within GRACE, SO3krates is a recursive evaluation with attention-weighted, low-rank equivariant messages.

\subsubsection{UMA}

UMA~\cite{woodUMAFamilyUniversal2025} is a family of foundation models trained on several large datasets, built on the eSEN backbone with a mixture of linear experts. The expert weights are combined once per structure from global information, so that a single recursive evaluation with structure-dependent coefficients is carried out. Within GRACE, UMA is a gated recursive evaluation.

\subsubsection{DPA-3}

DPA-3~\cite{zhangGraphNeuralNetwork2025} updates invariant node, edge and angle features along a line-graph series, so that each layer carries three-body terms (correlation order two), and higher-order interactions are generated by non-linear MLP updates and by stacking layers, sixteen in the reference model. Within GRACE, DPA-3 is an invariant line-graph approximation of the complete expansion. Attention-based descriptors were introduced with the predecessors DPA-1~\cite{zhangPretrainingAttentionbasedDeep2024} and DPA-2~\cite{zhangDPA2LargeAtomic2024}.

\subsubsection{Orb}

Orb-v3~\cite{rhodesOrbv3AtomisticSimulation2025} is a non-equivariant message-passing network with MLP updates on a graph-network backbone, with edge features built from radial Bessel functions and spherical harmonics. Rotational symmetry is encouraged by a gradient-based regularization that penalizes the change of the energy under an infinitesimal rotation, Sec.~\ref{sec:learnsym}. Within GRACE, Orb-v3 is a non-linear approximation of the complete expansion without built-in symmetrization. Orb-v3 prioritizes computational speed and is available in conservative variants as well as variants with directly predicted, non-conservative forces, Sec.~\ref{sec:forcelearning}.

\subsubsection{PET}

The Point Edge Transformer (PET)~\cite{pozdnyakovSmoothExactRotational2023} is a transformer on edges that is not constructed to be rotationally equivariant. Symmetry is restored by exact symmetrization or approximately by data augmentation, Sec.~\ref{sec:learnsym}. Its attention acts on edge tokens built from linear maps of the displacement vectors, so that the lowest-order score generalizes the invariant three-body product with $l=1$ of Sec.~\ref{sec:nlrep}. PET can therefore be represented within GRACE as an attention-weighted recursive evaluation. The attention scores act as non-linear weights of the messages, and $\nu = 1$ in Table~\ref{tab:mapping} therefore refers to the values. PET-MAD~\cite{mazitovPETMADLightweightUniversal2025a} and PET-OAM~\cite{bigiPushingLimitsUnconstrained2026} are foundation models built on PET.

\section{Summary and outlook \label{sec:summary}}

Efficiency decides how many atoms and time steps a simulation at near quantum mechanical accuracy can cover. This review therefore took efficiency as its guideline and represented MLIPs in a complete basis expansion, ACE and its graph generalization GRACE, to compare them on a common footing. The expansion is canonical, built from products of single-particle functions on the edges and nodes of a graph. Because complete bases are related by linear transformations, the organization of MLIPs does not depend on ACE and GRACE as particular methods.

Within the expansion, local and semi-local MLIPs differ in their graphs. Local MLIPs, in which an atom interacts directly with the atoms inside its cutoff sphere, are built on star graphs, and the expansion on star graphs is ACE. Semi-local MLIPs relay interactions through intermediate atoms and are built on tree graphs with several layers. For efficiency the tree graphs are evaluated recursively, which reproduces message passing and is closely related to multi-ACE~\cite{batatiaDesignSpaceE3equivariant2025}. Therefore, classical potentials such as the embedded-atom method and local MLIPs can be represented within ACE, and most message-passing MLIPs, such as NequIP and MACE, within recursive GRACE or multi-ACE. Seen from the complete basis, physically motivated and data-driven MLIPs are less different than they appear. Both are represented in the same basis and differ in how they terminate the expansion with non-linear functions, how they couple layers and how they are optimized, which helps to explain why they often reach comparable accuracy when trained on the same data.

The cost of a naive evaluation of the expansion grows steeply with system size and model complexity. Several of the steps that reduce this cost are well established. The decomposition of the energy into atomic contributions has been used for interatomic potentials for more than forty years, and the density trick makes the cost linear in the number of neighbors. Recursive evaluation with tensor decomposition at fixed rank makes the cost linear also in the number of basis functions, graph layers and body order, and the cost of chemistry is constant in practice. For each of these reductions we know of essentially one solution, and an efficient model has to adopt all of them. This is why the requirements of efficiency leave only a narrow design space and why the majority of efficient MLIPs follow a single strategy.

The energy is invariant under translation, rotation, inversion and the exchange of atoms of the same chemical species, and we built these symmetries into the expansion exactly, which saves expansion coefficients and training data. The cost of the Clebsch-Gordan coupling has been identified as a bottleneck and has motivated giving up exact rotational symmetry. However, the coupling can be contracted from $l_{\max}^5$ to $l_{\max}^3$ operations~\cite{bochkarevFastContractedClebschGordan2026}, and efficiency therefore gives no reason to give up exact rotational symmetry or spherical harmonics.

The single-bond and single-atom basis functions of the expansion are adapted to the physics of bond formation and build on decades of electronic structure theory. We showed that overlap and Hamiltonian matrix elements take the form of single-bond basis functions, and that, following the moments theorem of Cyrot-Lackmann, the moments of the electronic density of states unfold into closed walks that reduce to tree graphs. Through the moments, the bond energy becomes an explicit function of the graph cluster basis functions, and the small number of independent two-center integrals indicates why moderate numbers of basis functions suffice.

Next to architecture and truncation, the implementation determines the efficiency. On current hardware the cost is often set by memory access rather than by arithmetic, which is why the Clebsch-Gordan coupling is recast into dense matrix multiplications and which explains part of the adoption of attention in MLIPs. Platform-specific code, which AI-assisted programming makes practical, evaluates GRACE models about five times faster than the TensorFlow implementation at unchanged accuracy, Sec.~\ref{sec:runtime}.

In the complete basis expansion, machine learning plays two roles. Non-linear functions of the basis expansion, from the square root of the Finnis-Sinclair potential to neural networks, converge faster than the linear expansion, and machine learning provides efficient non-linear functions, which we related explicitly to the basis expansion. Transformer models applied to atomic interactions, for example, can be represented within GRACE, with or without built-in rotational symmetry. Machine learning further provides loss functions, optimizers and autodifferentiation for training models with millions of parameters. Early MLIPs such as the HDNNP used a neural network to generate many-atom interactions from two- and three-body descriptors. ACE and GRACE instead construct the many-body basis functions explicitly, and non-linear functions of the expansion improve its convergence. Both routes represent the energy in the same complete expansion, and they differ in how the many-body terms are generated and in how many parameters they need for a given accuracy.

Because efficiency is meaningful only at a given accuracy, we compared accuracy and cost. On the Matbench Discovery leaderboard~\cite{Leaderboard,riebesellMatbenchDiscovery2025}, the ten best models have three to ten layers and five million to more than seven hundred million parameters, Fig.~\ref{fig:matbench}. The model with the fewest parameters has three layers and scores about two percent below the best model, which has about twenty times as many. The benchmark therefore provides no evidence that large atomic regions or many parameters are required for a CPS within two percent of the best model. However, Matbench Discovery tests bulk inorganic crystals, and the required size of the atomic region may differ for molecules, interfaces, liquids or long-range interactions. For training on the same datasets, the GRACE models in Fig.~\ref{fig:warp} show that one basis spans more than two orders of magnitude in GPU time, so that for these datasets the choice between a fast and an accurate model is a choice of truncation.

Independent of the architecture, the accuracy of an MLIP is bounded by the diversity and quality of its training data. Generation strategies such as SMAX~\cite{bochkarevExploringExtremesAtomic2026} and ASSYST~\cite{poulAutomatedGenerationStructure2025}, datasets such as MAD~\cite{mazitovPETMADLightweightUniversal2025a}, and reference data beyond DFT, up to coupled cluster accuracy, raise this bound.

Beyond the scope of this review, long-range electrostatics, charge transfer, magnetism and excited states are active fields of research. The expansion includes additional degrees of freedom by extending its single-particle functions, Sec.~\ref{sec:extendability}, and first models are available, such as charge-constrained ACE~\cite{rinaldiChargeconstrainedAtomicCluster2025}, magnetic ACE~\cite{rinaldiNoncollinearMagneticAtomic2024} and latent Ewald summation~\cite{chengLatentEwaldSummation2025,kimUniversalAugmentationFramework2025}.

\begin{acknowledgments}
We acknowledge funding from the Deutsche Forschungsgemeinschaft (DFG, German Research Foundation) through SFB 1394, project number 409476157, subproject A06, and through SFB 1625, project number 506711657, subproject A03. The funders played no role in the design of the study, the analysis and interpretation of data, or the writing of this manuscript.
\end{acknowledgments}

\subsection*{DATA AVAILABILITY}

No new training data were generated in this work. The Matbench Discovery data in Fig.~\ref{fig:matbench} are publicly available from the leaderboard~\cite{Leaderboard}. The GRACE-3L model of Sec.~\ref{sec:Relations} is publicly available at \url{https://huggingface.co/AMS-ICAMS-RUB/grace-foundation-models}. The sources of the entries in Table~\ref{tab:mapping} are listed in the Supplementary Information. The data underlying Figs.~\ref{fig:matbench} and \ref{fig:warp}, including the layer counts of the leaderboard models and their sources, are available from the corresponding author upon reasonable request.

\subsection*{CODE AVAILABILITY}

The GRACE and ACE models discussed in this work are implemented in gracemaker, \url{https://github.com/ICAMS/grace-tensorpotential}, and pacemaker, \url{https://github.com/ICAMS/python-ace}. The grace-runtime engine of Sec.~\ref{sec:runtime} is available at \url{https://github.com/ICAMS/grace-runtime}.

\subsection*{COMPETING INTERESTS}

Y.L., A.B. and R.D. hold equity interests in ACEworks GmbH, a company that provides services related to ACE and GRACE interatomic potentials. The authors declare no other competing interests.

\subsection*{AUTHOR CONTRIBUTIONS}

R.D. wrote the original draft. A.B. and Y.L. carried out the parameterization of the three-layer GRACE model, GRACE-3L. All authors contributed to reviewing and editing the manuscript. 

\subsection*{USE OF AI TOOLS}

The authors used the large language models Claude Opus 5.5 and Claude Fable 5.2 (Anthropic) to check the text for consistency, to check citations against the cited sources and to suggest formulations. The authors reviewed and edited all suggestions and take full responsibility for the content of the manuscript.

\bibliography{references}

\end{document}